\documentclass[twocolumn]{aastex631}
\usepackage{amsmath}
\usepackage{upgreek}
\usepackage{CJK}
\usepackage{graphicx}
\usepackage{subfigure}
\usepackage{enumerate}

\newcommand{\RNum}[1]{\uppercase\expandafter{\romannumeral #1\relax}}

\submitjournal{AAS Journals}

\shorttitle{Evolution of planetary obliquity}
\shortauthors{Huang X.M. \& Ji J., et al.}

\begin{document}
\begin{CJK*}{UTF8}{gbsn}

\title{Evolution of the Planetary Obliquity: The Eccentric Kozai-Lidov Mechanism Coupled with Tide}

\correspondingauthor{Jianghui Ji}
\email{jijh@pmo.ac.cn}

\author{Xiumin Huang}
\affiliation{CAS Key Laboratory of Planetary Sciences, Purple Mountain Observatory, Chinese Academy of Sciences, Nanjing 210023, China}
\affiliation{School of Astronomy and Space Science, University of Science and Technology of China, Hefei 230026, China}

\author{Jianghui Ji}
\affiliation{CAS Key Laboratory of Planetary Sciences, Purple Mountain Observatory, Chinese Academy of Sciences, Nanjing 210023, China}
\affiliation{School of Astronomy and Space Science, University of Science and Technology of China, Hefei 230026, China}
\affiliation{CAS Center for Excellence in Comparative Planetology, Hefei 230026, China}

\author{Shangfei Liu}
\affiliation{School of Physics and Astronomy, Sun Yat-sen University, Zhuhai, China}

\author{Ruobing Dong}
\affiliation{University of Victoria, Victoria, British Columbia, Canada}

\author{Su Wang}
\affiliation{CAS Key Laboratory of Planetary Sciences, Purple Mountain Observatory, Chinese Academy of Sciences, Nanjing 210023, China}
\affiliation{CAS Center for Excellence in Comparative Planetology, Hefei 230026, China}

\begin{abstract}
The planetary obliquity plays a significant role in determining physical properties of planetary surfaces and climate. As direct detection is constrained due to the present observation accuracy, kinetic theories are helpful to predict the evolution of the planetary obliquity.  {Here the coupling effect between the eccentric Kozai-Lidov (EKL) effect }and the equilibrium tide is extensively investigated, the planetary obliquity performs to follow two kinds of secular evolution paths, based on the conservation of total angular momentum. The equilibrium timescale of the planetary obliquity $t_{\mathrm{eq}}$ varies along with $r_{t}$, which is defined as the initial timescale ratio of  {the tidal dissipation and  secular perturbation}. We numerically derive the linear relationship between $t_{\mathrm{eq}}$ and $r_{t}$ with the maximum likelihood method. The spin-axis orientation of S-type terrestrials orbiting M-dwarfs reverses over $90^\circ$ when $r_{t} > 100$, then enter the quasi-equilibrium state between $40^\circ$ and $60^\circ$, while the maximum obliquity can reach $130^\circ$  when $r_{t} > 10^4 $. Numerical simulations show that the  maximum obliquity increases with the semi-major axis ratio $a_1$/$a_2$, but is not so sensitive to the eccentricity $e_2$. The likelihood of obliquity flip for S-type terrestrials in general systems with $a_2 < 45$ AU is closely related to $m_1$. The observed potential oblique S-type planets HD 42936 b, GJ 86 Ab and $\tau$ Boot Ab are explored to have a great possibility to be head-down over the secular evolution of spin.
\end{abstract}


\section{Introduction}\label{sec:1}
The planetary obliquity is simply defined as the angle between the planetary spin and normal orientation of the orbital plane. In our solar system, Uranus and Venus are well-known as retrograde planets with extremely high planetary obliquity, where the obliquity of Uranus is $97^\circ$, whereas that of Venus is approximate to $178^\circ$ \citep{Goldreich1970, Dobrovolskis1980}. The primordial planetary obliquity provides a key clue to the understanding of evolution of our solar system, for example, the four inner terrestrial planets may have experienced secular chaotic scenario of spin-axis orientation \citep{Laskar1993}. For exoplanets, the evolution of planetary obliquity will directly affect the temperature distribution on  planetary surface, thereby leading to a variable seasonality of terrestrial planets \citep{Gaidos2004}. Moreover, as to hot-Jupiters, high planetary obliquity can give rise to a severe atmosphere loss rate \citep{Nikolov2015}.  {The obliquity can greatly influence the atmospheric escape rate of Earth-like planets orbiting late-type M-dwarfs \citep{Dong2019}, which plays a vital role in regulating surficial habitability.}

As of today, 2M0122b is the only derived oblique exoplanet, which is believed to be a planetary mass companion around the 120 Myr host star at an orbital distance of 50 AU. Thus, the diversity of planetary obliquity should be clearly understood due to planetary formation. \citet{Bryan2020} measured three inclinations referring to the line-of sight, where $i_o$ for the planet orbit, $i_p$ for the planetary spin and $i_*$ for the stellar spin. The longitude of ascending node of the planet's equatorial plane on its orbital plane is defined as $\lambda_p$. According to the law of cosines, the most possible planet obliquity is estimated to be greater than $50^\circ$. So naturally arises a question: how does such high obliquity origin?

Aside from indirect calculations through the spatial geometric configurations, a wide variety of observational methods were carried out to reveal the planetary obliquity. \citet{Kawahara2016} analyzed the frequency modulation of the scattered light curve of a directly imaged exoplanet, which is induced by the planetary obliquity and orbital inclination. The modulation factor changes with difference attitudes of the spin rotations. \citet{Schwartz2016} further predicted that the planet's spin axis affects the time-resolved photometry, and east-west albedo contrast plays a vital role in constraining obliquity. \citet{Nikolov2015} investigated the Rossiter-McLaughlin effect during secondary eclipse (RMse), and showed that for transiting exoplanets, the ratio of the ingress-to-egress radial velocity amplitudes are subjected to the planetary rotational rate and axial tilt. The synthetic near-infrared spectroscopic data were utilized to estimate the sky-projected spin axis orientation and equatorial velocity of the planet. Currently, detection of planetary obliquity is limitedly known due to the spectroscopy accuracy. RMse is only effective when the host stars are brighter than $K \sim 6$ mag and large aperture telescopes (i.e., $\sim40~$m) \citep{Nikolov2015}. In this sense, kinetic theories are very helpful to understand the evolution of planetary obliquity in advance and its predictions will be further examined by future high-precision measurements.

{\citet{Colombo1966} proposed to describe the evolution of obliquity with the existence of a companion in the system, when the spin precession rate and orbital precession rate become comparable, a resonance can occur to excite a larger obliquity. The spin axis motion is characterized as precessing around its orbital axis and the total angular momentum vector. The system configuration is specified by Euler angles ($\theta$, $\phi$, $\psi$). The equilibrium solutions of the canonical equations of motion (Equation (6) in \citet{Fabrycky2007}) are defined as Cassini states \citep{Colombo1966, Peale1969}, where the planetary equatorial plane maintains constant inclination to the plane of the ecliptic, and the spin axis remains coplanar with the normal to its orbit and the normal to the ecliptic plane. Total four Cassini states may exist for a given system, each with a preferred obliquity.}

Cassini states was adopted to explain moderately oblique exoplanet induced by planet-disk interaction \citep{Su2020} and the planet-planet interaction in multi-planetary systems \citep{Su2022b}. The planetary obliquities can be maintained between $60^\circ$ and $80^\circ$ throughout the entire lifetime of Earth-like planets in the habitable zone of M-dwarf stars \citep{Valente2022}, which provides friendly temperate environment and conditions for habitability.

{In addition to the axial tilt of exoplanet, the spin-orbit alignment between planet and star plays a crucial part in the dynamical evolution of system. \citet{Storch2014b} intensively explored spin-orbit coupling effect between stellar spin and planet orbit, which leads to chaotic evolution of the stellar spin axis during Kozai cycles. \citet{Storch2015} identified the origin of the chaos to be secular spin-orbit resonances overlaps. Furthermore, \citet{Campante2016} empirically constrained the angle between a planet's orbital axis and its parent star's spin axis with the asteroseismology. \citet{Vervoort2022} indicated that the period and amplitude of the tilt of a planet's rotational axis relates directly to long-term habitability of Earth-like planets.}

In this work, we primarily aim to investigate the planetary obliquity evolution of  S-type terrestrials orbiting M-dwarfs in binary systems. The planets that simply revolve a single star in binary are referred to S-type orbits, whereas those move around both stars are P-type orbits.  As of May 2022, 154 binary star systems with 217 planets were discovered, among which more than 70 are S-type planets. For S-type systems, Kozai-Lidov model \citep{Kozai1962, Lidov1962, vonZeipel1910} is suitable for a hierarchical three-body system with more intensive orbital precession when the perturbing object is more massive than the planet. In the secular evolution of a hierarchical triple body system, the perturbation of the third body exerts on a much longer timescale than its orbital period. The Kozai-Lidov mechanism for hierarchical triple systems was studied under a wide variety of circumstances \citep{Harrington1968, Lee2003, Naoz2013, Teyssandier2013, Tan2020}. \citet{Li2014} explored Kozai-Lidov mechanism and characterized the parameter space that allowed large amplitude oscillations in eccentricity and inclination under the test particle limit. For a system with an eccentric outer orbit,  {the octupole level terms in the perturbation Hamiltonian become significant, thereby triggering EKL effect.} The planetary eccentricity increases approximately one, and the flips of orbital inclination can lead to a retrograde orbit \citep{Naoz2013}. Furthermore, the orbital flips of S-type planets were found to origin from EKL \citep{Huang2022}. Here we will explore similar flips of planetary spin-axis due to EKL.

Tidal dissipation and the dynamical torque from the adjacent planets or disk both play critical roles in the spin evolution. \citet{Wang2019} adopted a comprehensive scenario in the restricted three-body system, including the equilibrium tide and the third body perturbation, and showed that the obliquity librates for a long time or decays slowly down to zero. The planetary angular momentum, along with a positive dynamical loop consisting of orbital migration and obliquity tides, can well produce Ultra-Short-Period planets \citep{Millholland2020}. Formation of oblique super-Earths and several hot Jupiters can be further explained with tidal dissipation and spin-orbit capture \citep{Su2022a}.

In the real situation of a hierarchical system when the stellar companion is more massive than the perturbed body and the secondary's eccentricity is non-negligible, the non-restricted EKL mechanism will work to trigger the planet's evolution. Based on the conservation of angular momentum, in this work the mixed scenario between EKL and equilibrium tide is adopted to reveal a diverse secular evolution paths for planetary obliquity. The comprehensive evolution timescale of the obliquity depends on the masses, orbital elements of all bodies, and the tidal factors of the planet throughout the secular evolution. The angular momentum transferred from the distant companion to the perturbed body can sustain or excite the planetary obliquity and contend with the tidal decay effect.

To better understand the nature of planetary obliquity of exoplanets, we utilize N-body package \textit{MERCURY-T} \citep{Bolmont2015} to conduct more than 2000 numerical simulations that each run evolves at a timescale of roughly $10^7$ yr to obtain two typical evolution paths of planetary obliquity corresponding two regions in the Hamiltonian level curves. Subsequently, we derive the relationship of equilibrium timescale versus the planetary mass $m_1$, the  initial orbit $a_1$ and the timescale ratio $r_{t}$. In the parameter space of semi-major axis, we present the diagram of maximum obliquity to estimate  {the flip ratio of obliquity}. Here we further investigate reversal conditions of planetary spin axis involved in the EKL and tide effect simultaneously, which are employed to address the evolution of S-type planets, such as HD 42936 Ab, GJ 86 Ab and $\tau$ Boot Ab.

\begin{figure*}
\begin{center}
        \includegraphics[width=1.8\columnwidth,height=8.5cm]{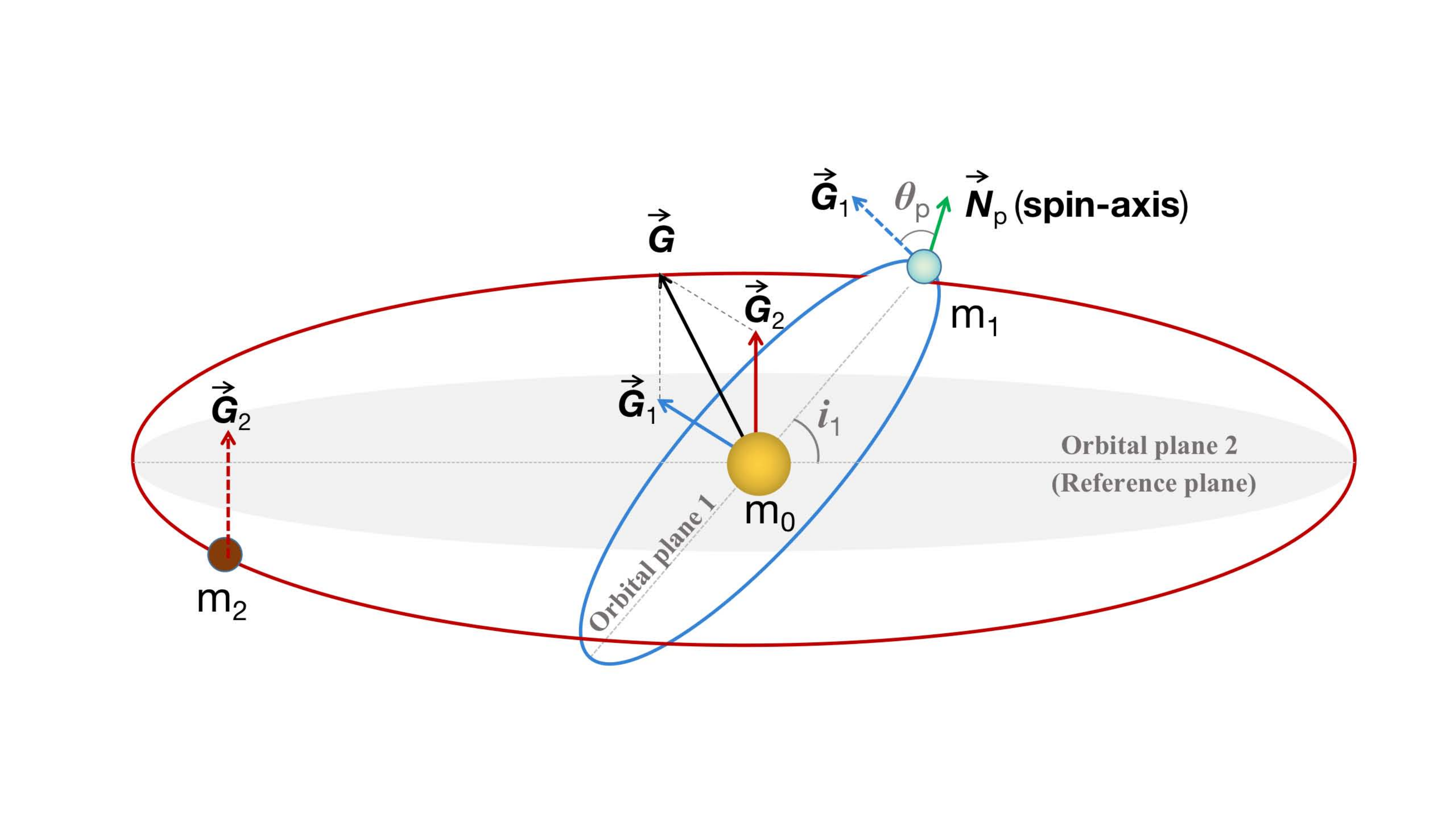}
    \caption{Geometry of the hierarchical system. $\overrightarrow{\boldsymbol{G}}_1$  and $\overrightarrow{\boldsymbol{G}}_2$ are angular momentum vectors of the inner and the outer orbits. $\overrightarrow{\boldsymbol{G}}$ is the total angular momentum vector of the orbital motion. $\overrightarrow{\boldsymbol{N}}_p$ is the rotation angular momentum.}
    \label{fig:geometry}
\end{center}
\end{figure*}

This work is structured as follows. In Section \ref{sec:2}, the theory of the mixed mechanism is described under the frame of central body equatorial coordinate, including theoretical analysis of the relationship between initial timescale ratio $r_{t}$ and the equilibrium timescale. Section \ref{sec:3} describes evolution paths for general S-type terrestrials, and provides the relationship between the maximum planetary obliquity, the equilibrium timescale and $r_{t}$.  {The obliquity flip ratio} is further calculated for a wide variety of initial conditions. Section \ref{sec:4} presents the secular evolution of several potential oblique S-type planets, and predict the possible value of the equilibrium obliquity. In final, we summarize major outcomes.

\section{The mixed scenarios}\label{sec:2}
In the secular evolution of triple body systems, the perturbation arising from the third body acts on a time scale much longer than the orbital period. If the perturbation of a circular outer orbit works, the Hamiltonian of this system can be expanded to the quadrupole level, which is referred to Kozai-Lidov mechanism as aforementioned. Here we adopt the Kozai-Lidov mechanism and the equilibrium tide to characterize the total evolution for S-type planets. It is interesting that the rotation angular momentum, the planetary orbital angular momentum and the stellar companion's angular momentum are all participating to transfer to each other. The geometry configuration of the system is plotted in Figure \ref{fig:geometry}, where the reference plane is the equatorial plane of the host star $m_0$, $m_1$ and $m_2$ are the perturbed planet and the outer stellar companion. $\overrightarrow{\boldsymbol{G}}_1$ and $\overrightarrow{\boldsymbol{G}}_2$ are angular momentum vectors of the inner and the outer orbits. $\overrightarrow{\boldsymbol{G}}$ is the total angular momentum vector of the orbital motion. $\overrightarrow{\boldsymbol{N}}_p$ is the angular momentum of planetary spin.

\subsection{Quadrupole Force vs Tide}\label{subsec:2.1}
\begin{figure*}
\begin{center}
\subfigure[]{\includegraphics[width=0.9\columnwidth,height=6.5cm]{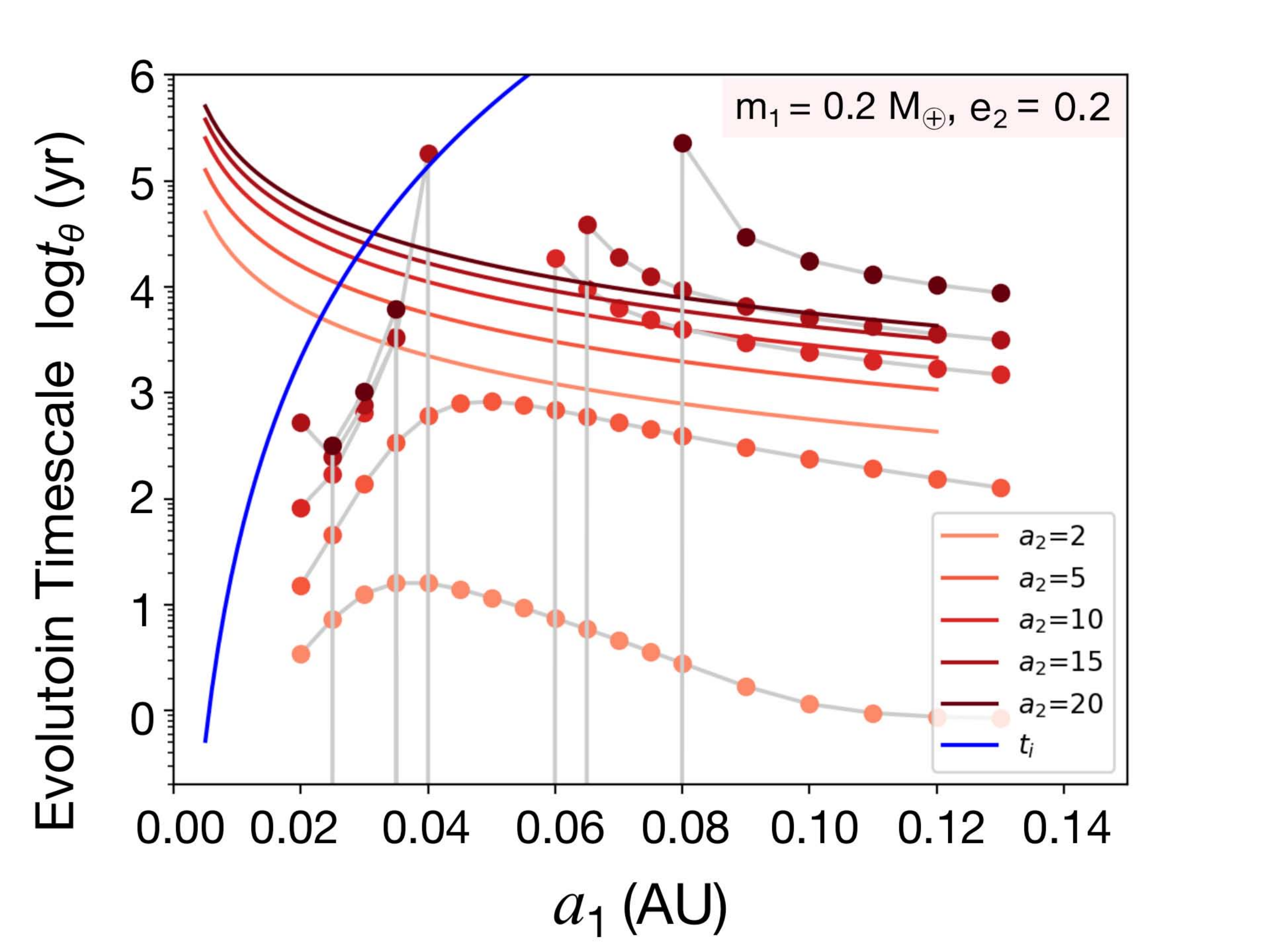}}
\subfigure[]{\includegraphics[width=0.9\columnwidth,height=6.5cm]{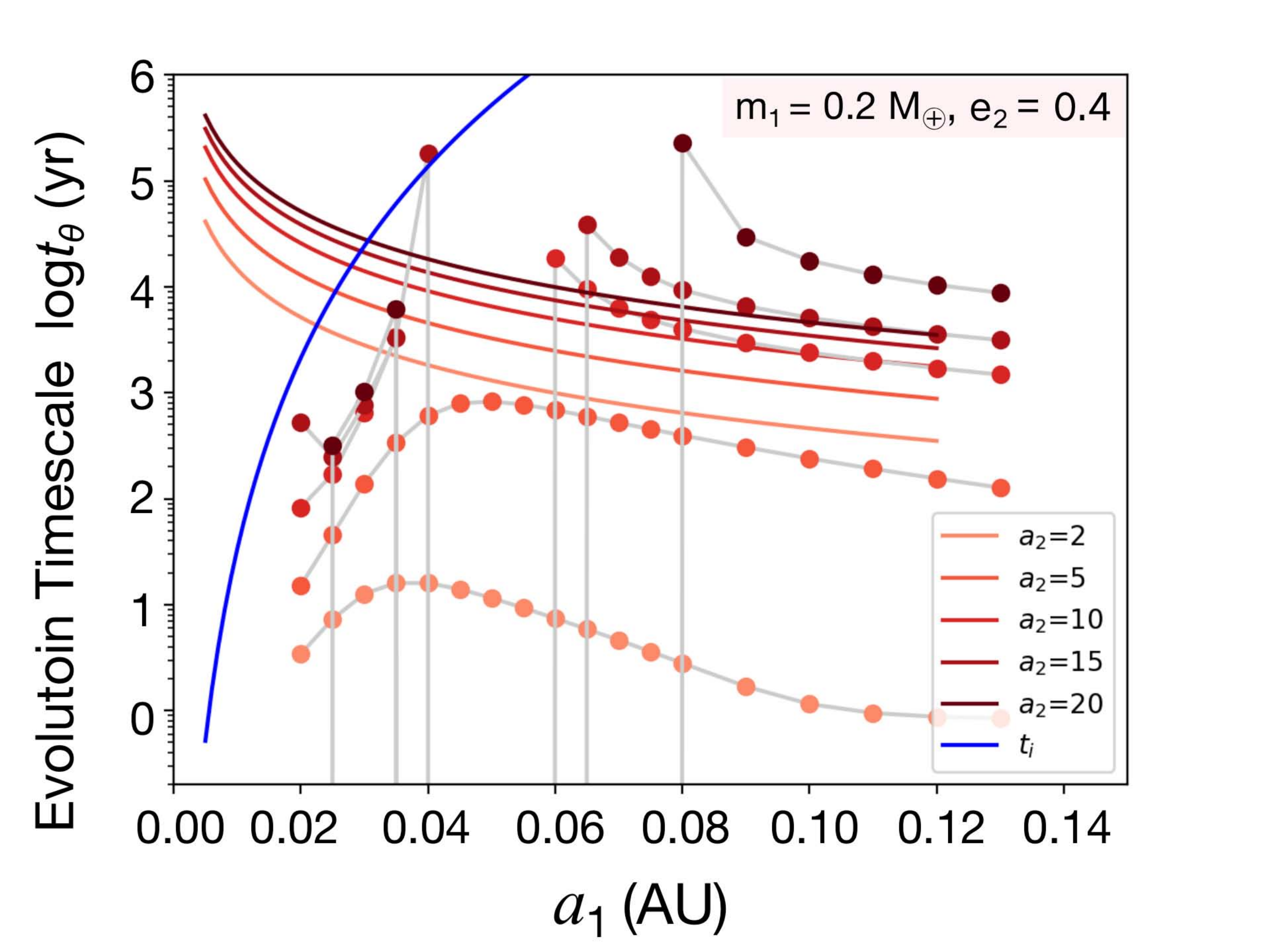}}\\
\subfigure[]{\includegraphics[width=0.9\columnwidth,height=6.5cm]{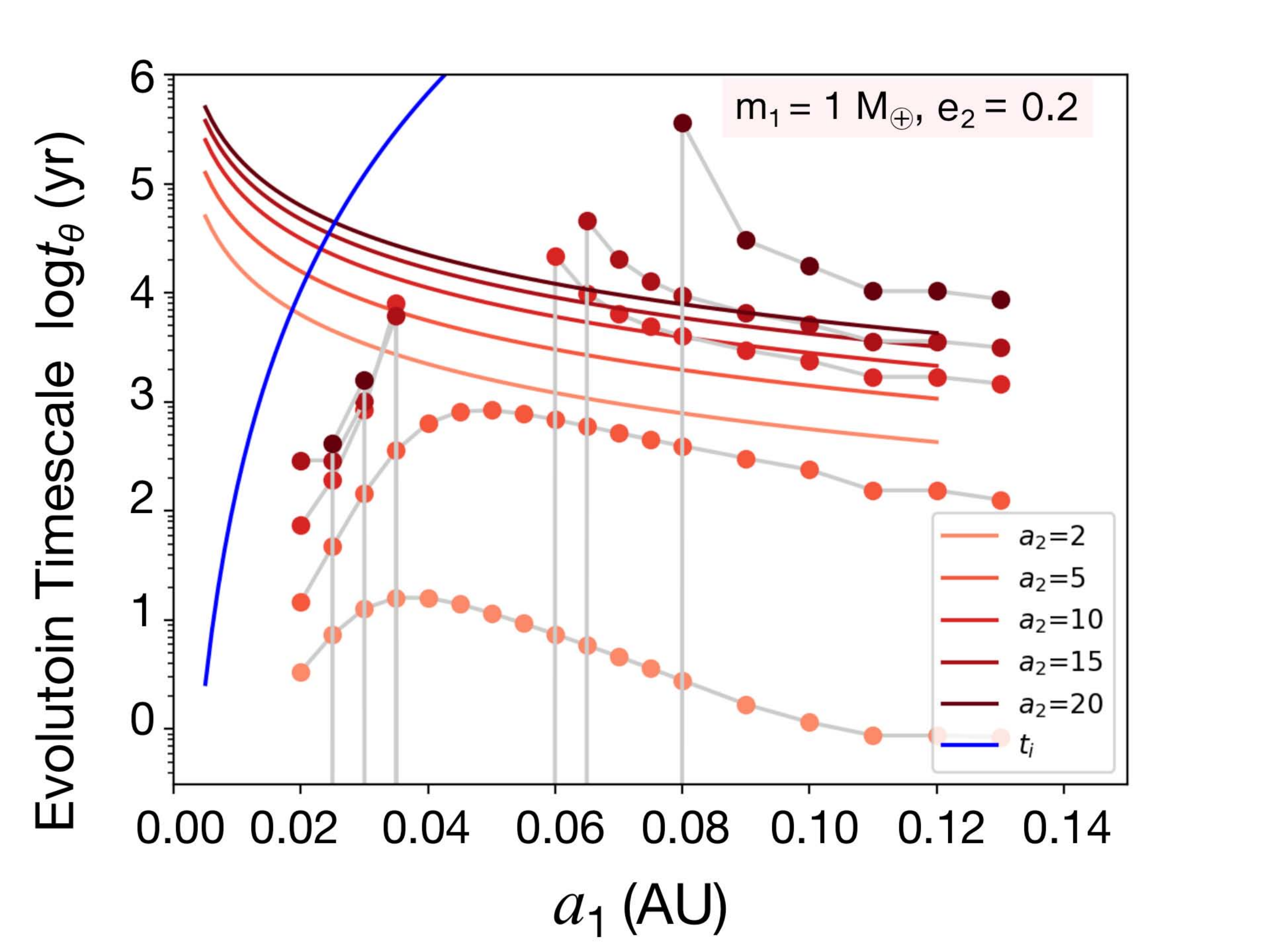}}
\subfigure[]{\includegraphics[width=0.9\columnwidth,height=6.5cm]{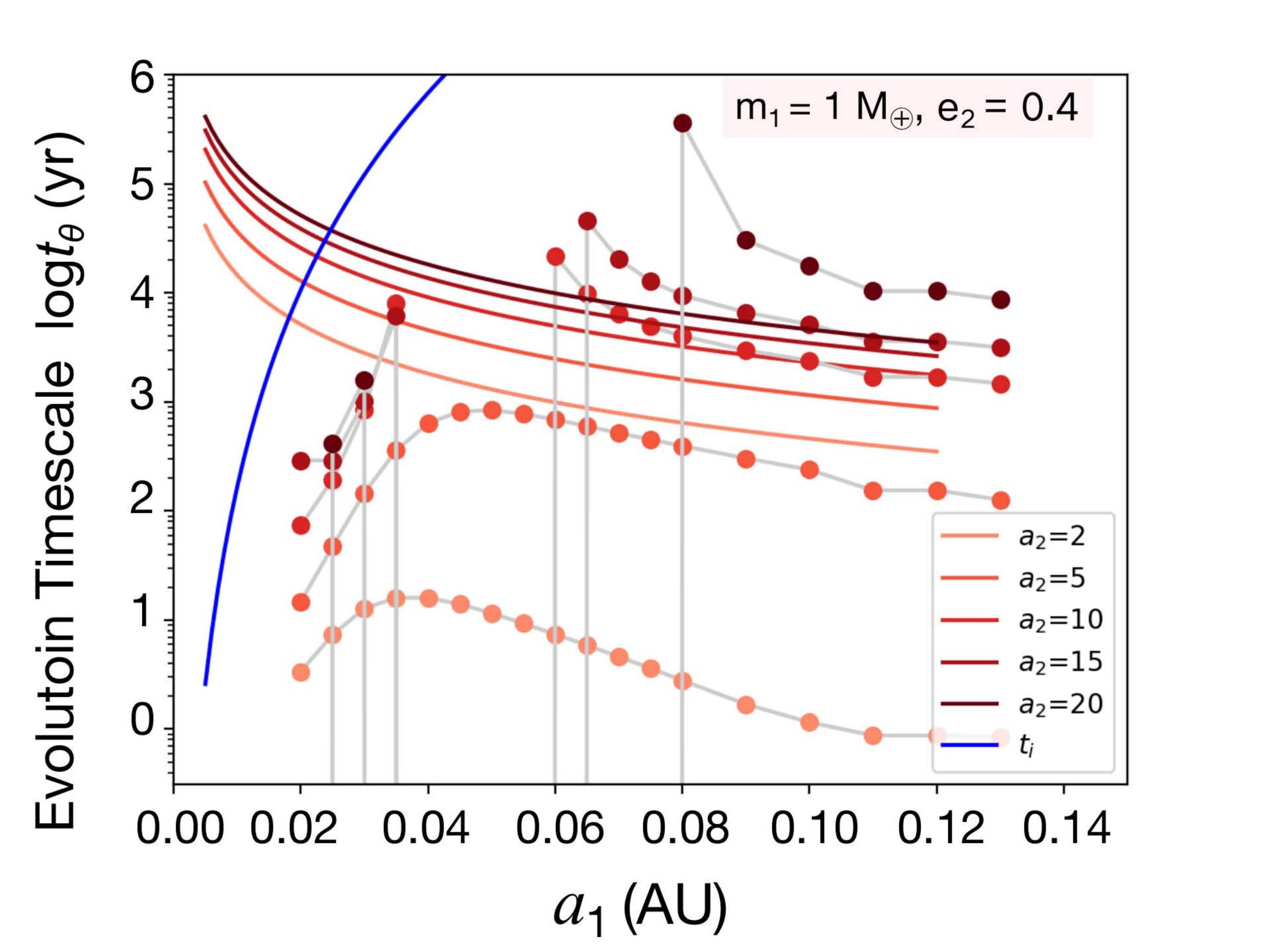}}\\
\caption{The relationship between $t_{\mathrm{I}}$ and $t_{\mathrm{kl}}$ obtained from the theoretical equations. In panel (a)-(b), $m_1$=0.2 $M_{\oplus}$, in (c)-(d), $m_1$=1 $M_{\oplus}$. $e_2$ is set to be 0.2 in panels (a)-(c) and $e_2$ =0.4 in (b)-(d).  {Dotted lines are the comprehensive evolution timescale of the planetary obliquity during the first output step of simulations. The comprehensive evolution timescale of the obliquity locally decreases when $t_{\mathrm{I}} > t_{\mathrm{kl}}$  for  $a_2 > 10$ AU and $a_1$ = 0.02 -- 0.08 AU under EKL.}}
\label{fig:timescale}
 \end{center}
\end{figure*}

The classical Kozai-Lidov mechanism shows that the inclination and eccentricity of inner test particle oscillate over secular evolution in a hierarchical system \citep{vonZeipel1910, Kozai1962, Lidov1962}. With the secular approximation, the inner and the outer orbits only exchange angular momentum, thus the semi-major axis of orbits do not change. When the SMA ratio $\alpha_r$ = $a_1$/$a_2$ is a small parameter, the perturbation term of full Hamiltonian can be expanded as a power series in $\alpha_r$ \citep{Naoz2016}:
\begin{equation}
\begin{aligned}
\label{equ:Hal}
\mathcal{H}&=\frac{k^{2} m_{0} m_{1}}{2 a_{1}}+\frac{k^{2} m_{2}\left(m_{0}+m_{1}\right)}{2 a_{2}} \\
&+\frac{k^{2}}{a_{2}} \sum_{j=2}^{\infty} \alpha^{j}_r M_{j}\left(\frac{r_{1}}{a_{1}}\right)^{j}\left(\frac{a_{2}}{r_{2}}\right)^{j+1}
P_{j}(\cos \Phi) ,\\
M_{j}&=m_{0} m_{1} m_{2} \frac{m_{0}^{j-1}-\left(-m_{1}\right)^{j-1}}{\left(m_{0}+m_{1}\right)^{j}} ,
\end{aligned}
\end{equation}
where $k^2$ is the gravitational constant (with the mass unit of $M_{\oplus}$ and the length unit of au), $r_1$ is the distance between $m_0$ and $m_1$, $r_2$ is the distance between the center of mass of the inner binary and $m_2$, $P_j$ is the Legendre polynomial,  $\Phi$ is the angle between the vectors $r_1$ and $r_2$, and the subscripts $j$ = 1, 2 represent the inner and outer orbits, respectively.

In the non-test particle approximation under the classical Kozai-Lidov mechanism, the eccentricity and the inclination oscillate regularly in a well-defined timescale $t_{\mathrm{kl}}$ \citep{Antognini2015}:
\begin{equation}
\label{equ:tquad}
t_{\mathrm{kl}} \sim \frac{16}{15} \frac{a_{2}^{3}\left(1-e_{2}^{2}\right)^{3 / 2} \sqrt{m_{0}+m_{1}}}{a_{1}^{3 / 2} m_{2} k}.
\end{equation}
This relationship was derived under the consideration of the equation of motion of $\omega$, argument of perihelion, by integrating between the maximum and minimum eccentricities.

{Meanwhile, we employ the constant time-lag equilibrium tide at arbitrary rotation \citep{Hut1981,Leconte2011,Wang2019} from the start of integration in the present work. Because the planetary eccentricity will be triggered to an extremely large value by Kozai-Lidov effect, which may result in unaccepable outcomes for the low-order expansion of eccentricity in the constant phase delay model. When the orbit plane does not coincide with the equatorial plane of the planet, \citet{Wang2019} used the Oxyz coordinate system to describe the equilibrium tide at arbitrary rotation and the equations of motion for the constant time-lag tidal model.}

{We choose a central body coordinate system, and the stellar equatorial plane is treated as the reference plane. The origin is the host star, where x-y plane runs parallel to the equatorial plane of the star, z-axis is positively aligned with the rotation axis of the star. The Oxyz coordinate system is in line with the frame illustrated in Figure \ref{fig:geometry}. When the tidal potential function expands to order 2 \citep{Kozai1965}, the tidal acceleration is derived as \citep{Wang2019}: }
\begin{equation}
\begin{aligned}
\label{equ:Rtides}
\vec{F}_{\mathrm{tide}}=\frac{m_0+m_1}{m_0 m_1}(\mathrm{F}_r \hat{r}+\mathrm{F}_\theta \hat{\theta}+\mathrm{F}_w \hat{w}),
\end{aligned}
\end{equation}

\begin{equation}
\begin{aligned}
\label{equ:Rtides}
\left\{\begin{array}{l}
\mathrm{F}_r=-\frac{G m_0 m_1}{r^2}[3 \frac{m_0}{m_1}\left(\frac{R_\mathrm{p}}{r}\right)^5 k_2\left(1+3 \frac{\dot{r}}{r} \tau\right)] \\
F_\theta=3 \frac{G m_0^2}{r^2}\left(\frac{R_\mathrm{p}}{r}\right)^5 k_2\left(\Omega_p A-\dot{\theta}\right) \tau \\
\mathrm{F}_w=-3 \frac{G m_0^2}{r^2}\left(\frac{R_\mathrm{p}}{r}\right)^5 k_2 \Omega_p \tau[B \cos (f+\omega)+C \sin (f+\omega)],
\end{array}\right.
\end{aligned}
\end{equation}

\begin{equation}
\begin{aligned}
\label{equ:Rtides}
\left\{\begin{array}{l}
A=\cos I \cos i+ \sin I \sin i \cos(\Omega-\Theta) \\
B=\cos I \sin i- \sin I \cos i \cos(\Omega-\Theta) \\
C=\sin I \sin (\Omega-\Theta) \\
D=\sin i \sin(\Omega-\Theta),
\end{array}\right.
\end{aligned}
\end{equation}
{where $\hat{w}$ is the unit vector of orbital angular momentum,  $\hat{r}$ is the vector from the star to the center of mass of the planet, and $\hat{\theta}$ is the unit vector along the orbital velocity. $G$ is the gravitational constant, $r$ is the distance from the planet to the center of the star, $k_2$ is the 2th order Love number, $\dot{\theta}$ is the instantaneous orbital angular velocity. $\tau$ denotes the time delay factor in the constant time lag tidal model \citep{Hut1981}. $\Omega$ is the longitude of ascending node of the planetary orbit, and $i$ is the orbital inclination of the planet's orbit relative to the Oxyz system. $\Omega_{\mathrm{p}}$ is the planetary rotational velocities. $I$ and $\Theta$, refer to the inclination and the longitude of ascending node of the planet's equatorial plane with respect to the Oxyz system, respectively.}

{We can calculate equations of rotational motion based on the total angular momentum conservation:
 \begin{equation}
\begin{aligned}
\label{equ:Rtides}
H_1 \widehat{w}_1+H_2 \widehat{w}_2+I_p \Omega_{\mathrm{p}} \widehat{N}=\vec{C},
\end{aligned}
\end{equation}
where $\vec{C}$ is the total angular momentum vector, and the moment of inertia $I_{\mathrm{p}}$ = $m_1 R_{\mathrm{p}}^2 r_g^2$, with $r_g^2$ the square of radius of gyration (constant). Total angular momentum $\vec{C}$ and orbital angular momentum $H_1$, $H_2$ can be calculated by the initial orbital elements: $a_0$, $e_0$, $i_0$, $\Omega_0$, $I_{\mathrm{p}}$ , $\Omega_{\mathrm{p,0}}$, $I_{\mathrm{0}}$, $\Theta_{0}$. In the evolution, the rotational velocity of planet $\Omega_{\mathrm{p}}$ can be estimated with orbital parameters according to angular momentum conservation. Hence, $I$ and $\Theta$ can be further derived through three components of the planetary spin vector.}

The evolutionary timescale of $I$ is defined as \citep{Wang2019}:
\begin{equation}
\begin{aligned}
\label{equ:ti}
t_{\mathrm{I}}&=\frac{2I_p T_t m_1 a_1^6}{3 k_2 m_0^2 R_p^8}(1-e_1^2)^{\frac{9}{2}}\\
     &=\frac{2 m_1 r_g^2 a_1^6}{3 G \tau_{\oplus} k_2 m_0^2 R_p^3}(1-e_1^2)^{\frac{9}{2}},
\end{aligned}
\end{equation}
where $I_p$ is the momentum inertia of the planet, $T_t$ is a typical tidal dissipation timescale, which keeps constant and can be calculated by:
\begin{equation}
\label{equ:Tt}
T_t=\frac{R_p^3}{G m_1 \tau_{\oplus}},
\end{equation}
$\tau_{\oplus}$ is the time delay factor of Earth. In the following, we define the planetary obliquity as $\theta_p$, which is the angle between the orbital angular momentum vector and the rotational angular momentum vector (Figure \ref{fig:geometry}).

{Note that Equation (\ref{equ:Hal}) and the secular perturbation timescale $t_{\mathrm{kl}}$ in Equation (\ref{equ:tquad}) are defined in the Jacobian coordinate system \citep{Naoz2016}. }For a host star with  $m_0 = 0.1~M_{\odot}$ and the terrestrial planet with a mass of 1 $M_{\oplus}$, the center of mass of them will be inside the host star if the semi-major axis $a_1 < 0.2 $ AU. The difference of coordinates between the Jacobian frame and the central body frame can be neglected.  {Here we employ the astrocentric coordinate as the system frame of reference.}

According to \citet{Colombo1966}, the spin precession rate and orbital precession rate is comparable to excite planetary obliquity. As the tidal dissipation timescale $t_{\mathrm{I}}$ is much greater than $t_{\mathrm{kl}}$ induced by the Kozai-Lidov mechanism.  {Blue and red solid lines in Figure \ref{fig:timescale} simply focus on the comparison of $t_{\mathrm{kl}}$ and $t_{\mathrm{I}}$ for a wide variety of initials $m_1$, $e_2$, $a_1$ and $a_2$ according to the theoretical equations, which is also considered to be a pair of initial conditions.} Figure \ref{fig:timescale} indicates that when $a_1$ increases, the timescale $t_{\mathrm{I}}$ can rise but $t_{\mathrm{kl}}$ decreases. By contrast, $t_{\mathrm{kl}}$ grows when $a_2$ increases. For $a_1$ = 0.02 -- 0.04 AU, the line of $t_{\mathrm{I}}$ crosses $t_{\mathrm{kl}}$ at roughly $10^4$ yr. The tidal dissipation timescale goes up along with the planetary mass, and $t_{\mathrm{kl}}$ is not so sensitive to the eccentricity $e_2$ of the secondary. In addition, Figure \ref{fig:timescale} exhibits that $t_{\mathrm{kl}}$ remains larger when $e_2 = 0.2$, $t_{\mathrm{I}}$ and $t_{\mathrm{kl}}$ are equivalent (the intersections of two curves) at $5\times10^3$ -- $3 \times 10^4$ yr. When $e_2$ = 0.4, the EKL timescale is smaller, and the intersection range of $t_{\mathrm{I}}$ and $t_{\mathrm{kl}}$ is $4\times10^3$ -- $3\times10^4$ yr. For $m_1 = 1$ $M_{\oplus}$, the timescale of tide is larger in comparison with that of $m_1 = 0.2$ $M_{\oplus}$, and the crossing point of $t_{\mathrm{I}}$ and $t_{\mathrm{kl}}$ varies from $7\times10^3$ to $5\times10^4$ yr.

\subsection{Octupole Force vs Tide}\label{subsec:2.2}
Unlike the improved tidal triple body model \citep{Wang2019}, here we consider the combined effects of the octupole level perturbation and the equilibrium tide for non-restricted hierarchical systems. In such model, the total angular momentum is conserved, which consists of the orbital angular momentum of planet, that of the secondary star and the angular momentum of planet' spin. The three components of angular momentum are mutually coupling over secular evolution. However, for the restricted model, the total angular momentum of the planet is assumed to be constant.

As \citet{Naoz2016} pointed out, the EKL secular evolution timescale is sensitive to $e_2$ of the secondary star, which is difficult to be directly calculated.  {Thus we numerically investigate secular evolution of the systems with N-body simulations.}

{In Figure \ref{fig:timescale}, red dots exhibit the comprehensive evolution timescale of planetary obliquity corresponding to different comparison between $t_{\mathrm{I}}$ and $t_{\mathrm{kl}}$. Vertical coordinates of the red dots represent the mean evolution timescale of the planetary obliquity during the first integration output step from t = 0 to t = $\delta t$. In the runs, the output step $\delta t$ is set to be 100 yr. When the vertical coordinates of the dots fall down to be negative (gray lines) , the planetary obliquity goes down or evolves to be retrograde.} For a relatively distant secondary star, $\theta_p$ will decrease initially with $a_2 > 10$ AU and $a_1$ = 0.02 -- 0.08 AU.

\section{Secular evolution of the obliquity}\label{sec:3}
Here the principal goal for this work is mainly to explore the spin evolution of terrestrial planets, as the planetary obliquity plays a crucial role in habitability, therefore our study will cast light on spin evolution for habitable planets. Over two hundred S-type planetary systems were discovered so far, however, there are only 10 binaries that own eccentricities, which those of the secondary stars are detected to be greater than 0.2. Note that the octupole force frequency is proportional to the eccentricity $e_2$ of the perturber under EKL scenario.

\subsection{Numerical setup}\label{subsec:3.1}
{To extensively investigate the spin evolution for terrestrial planets, here we adopt the standard equilibrium tidal model and constant time lag model given by \textit{MERCURY-T} \citep{Bolmont2015} with numerical simulations. Here all runs refer to the N-body simulations. The constant time lag model consists on the assumption that the bodies under consideration are made of a weakly viscous fluid \citep{Alexander1973}. \textit{MERCURY-T}  models the tidal forces between the star and the planets but we neglect the tidal interaction between planets, then adds the forces in the acceleration of orbital motion and the spin angular momentum. The tidal torque exerted by planet and the star is explicitly written in \citet{Bolmont2015}.}

{Here host stars are considered to be non-evolutionary in the simulations. Mass of the secondary star is set to be 0.2 $M_{\odot}$. The initials of $a_1$ are in the range 0.02 -- 0.13 AU, which keeps the orbital periastron of a rocky planet outside of the Roche limit of a host star \citep{Liu2013}. For rocky planets, the lower value of Roche limit is 1.58 $R_{\odot}$. The separation of the binaries $a_2$ are given in [2, 5, 10, 15, 20] AU. According to $t_{\mathrm{I}}$ and $t_{\mathrm{kl}}$ in Equation (\ref{equ:tquad}) and (\ref{equ:ti}), $t_{\mathrm{I}}$ can always be comparable with $t_{\mathrm{kl}}$ for each set of $a_2$.  To maintain the system stability, we choose $e_1$ = 0.2, $e_2$ = [ 0.2, 0.4] for terrestrial planets orbiting M-dwarfs, and set $e_2$ = [0.6, 0.8] for solar-type host star, respectively. The initial inclinations are given to be $i_1$ = $50^{\circ}$ and $i_2$ = $0^{\circ}$ with $\Omega_1$ = $\Omega_2$ = $0^{\circ}$. }

{Recently, the stellar spin was proved to has large implications on the dynamics of planetary systems \citep{Becker2020, Chen2022, Faridani2023}, our calculations demonstrate that the radius of M-dwarf with age of 700 Myr and a mass of 0.1 $M_{\odot}$ will decrease from 0.1238 $R_{\odot}$ to 0.1236 $R_{\odot}$. The terrestrial planet's orbital angular momentum variation due to the stellar spin evolution over $2\times10^7$ yr is less than $4\times10^{-6}$, and the final planetary obliquity differs from that of the non-evolving model by $\sim 1^{\circ}$. Thus we adopt the non-evolving M-dwarf host model, which will not bring about significant influence to our results.}

{In the non-evolving host body model, the time delay factor in the constant time lag tidal model is set as $\tau_{\oplus}$ = 698 s \citep{Bolmont2015}. For terrestrials, we select the love number $k_2$ = 0.305, the radius of gyration $r_g^2$ = 0.3308, and the planetary radius $R_{\mathrm{p}}$ = $R_{\oplus}$. The initial planet rotation period is set to be 24 hr, and the initial value of $\theta_p$ is assumed to be $10^{\circ}$ in the simulations. }

{The former study revealed that general relativity (GR) may have influence on triples' evolution, which may result in resonant-like dynamics \citep{Naoz2013, Liu2015} or destabilize the existing resonance \citep{Hansen2020}. We conduct additional runs by considering GR effect  with the same initials as Figure \ref{fig:terrestrials}. For a terrestrial planet at $a_1$ = 0.04 AU with the GR and non-GR models, the difference in final rotation period is 0.5 hours and the final planetary orbital angular momentum differs by $6\times10^{-5}$. To speed up the integration, here we simply adopt the non-GR model in the simulations.}

{In this work, we conduct over 2000 simulations that each evolves for 10 Myr ($\sim 10^4~ t_{\mathrm{kl}}$) with \textit{MERCURY-T}. When all runs reach $10^7$ yr, the simulation is ceased once the equilibrium state of planetary obliquity occurs after $10^7$ yr, which is the stopping condition. }

\subsection{Typical Evolution Paths}\label{subsec:3.2}
\begin{figure*}
\begin{center}
\subfigure[]{\includegraphics[width=0.67\columnwidth,height=5.2cm]{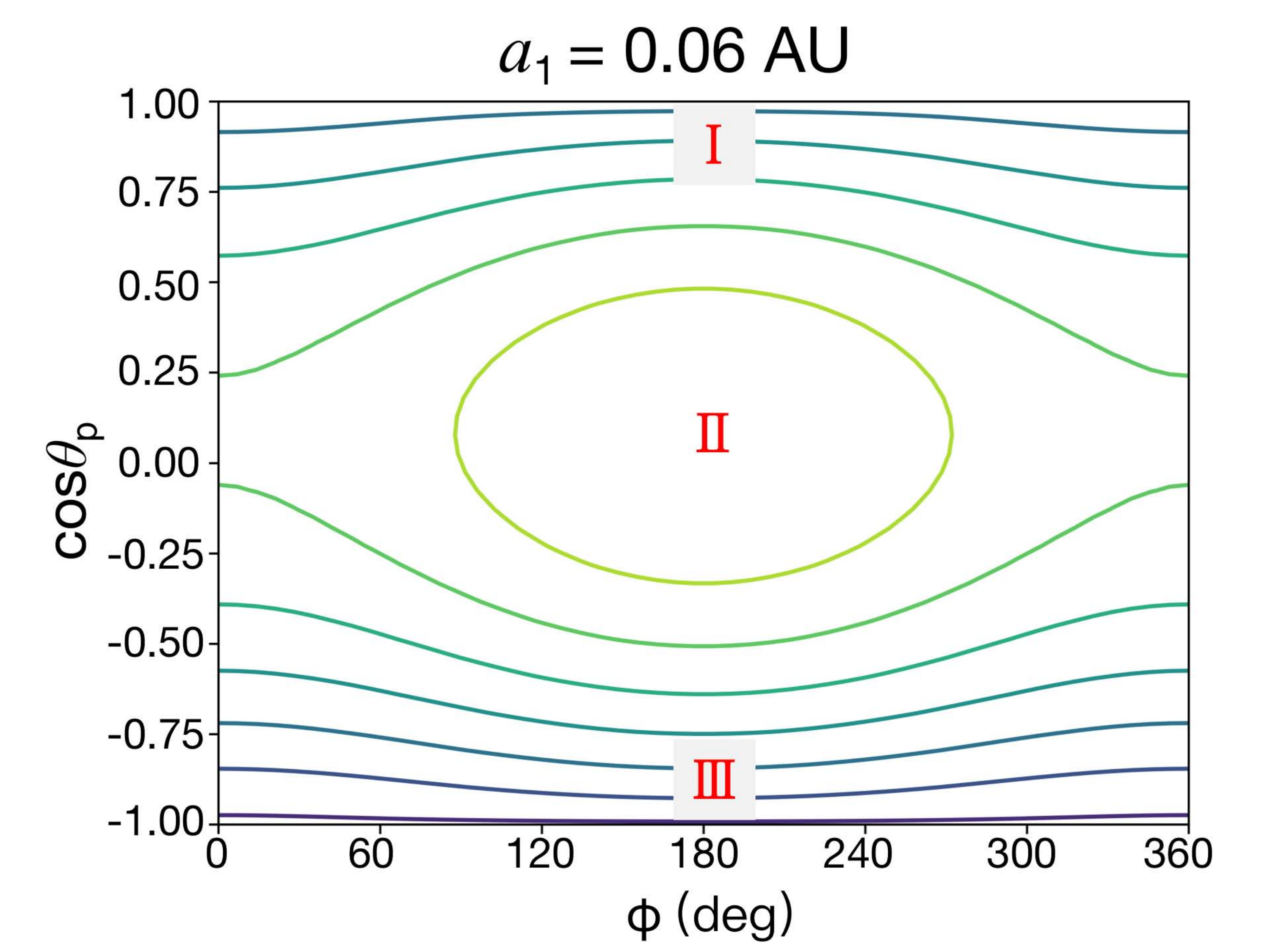}}
\subfigure[]{\includegraphics[width=0.67\columnwidth,height=5.2cm]{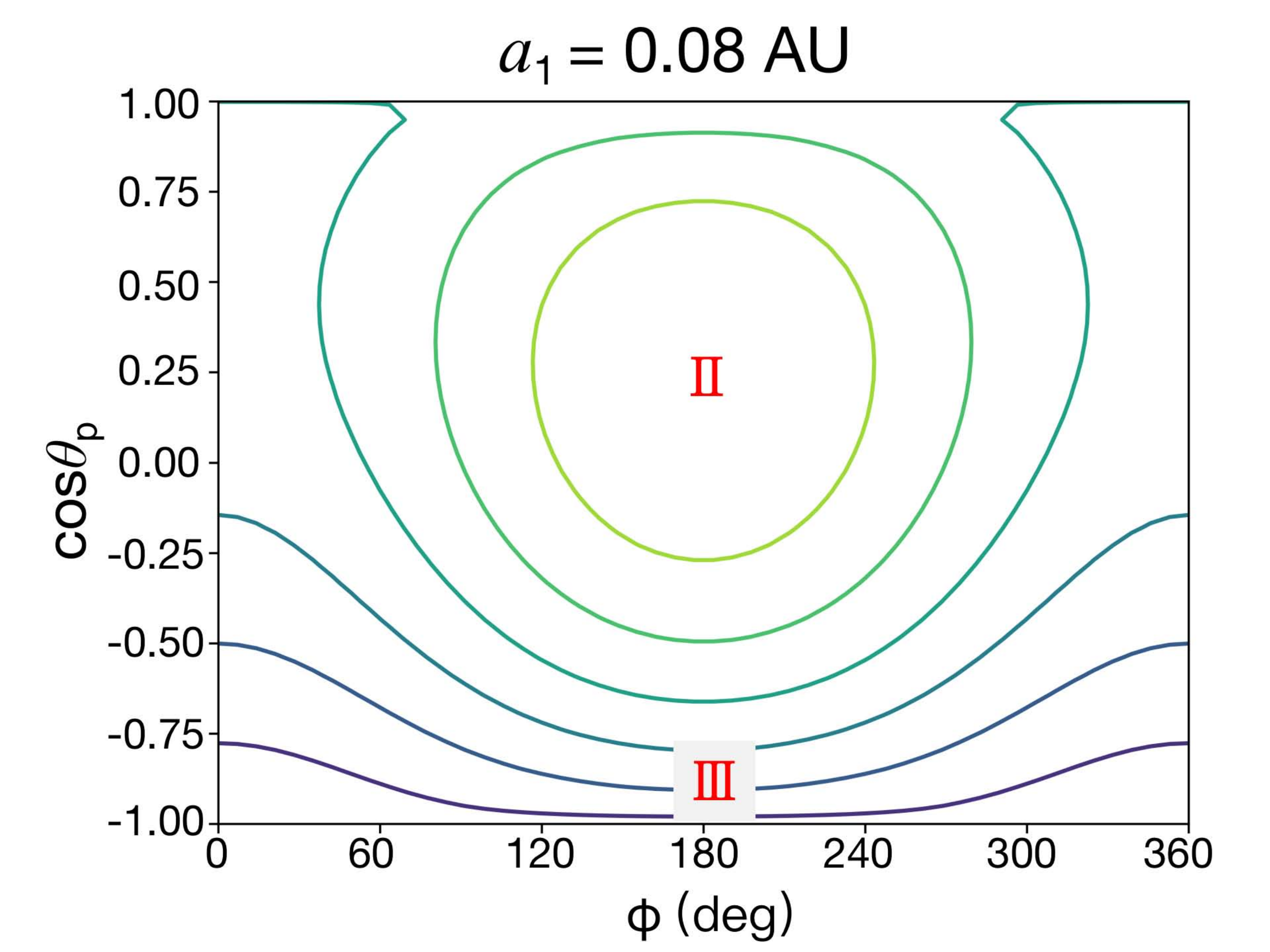}}
\subfigure[]{\includegraphics[width=0.67\columnwidth,height=5.2cm]{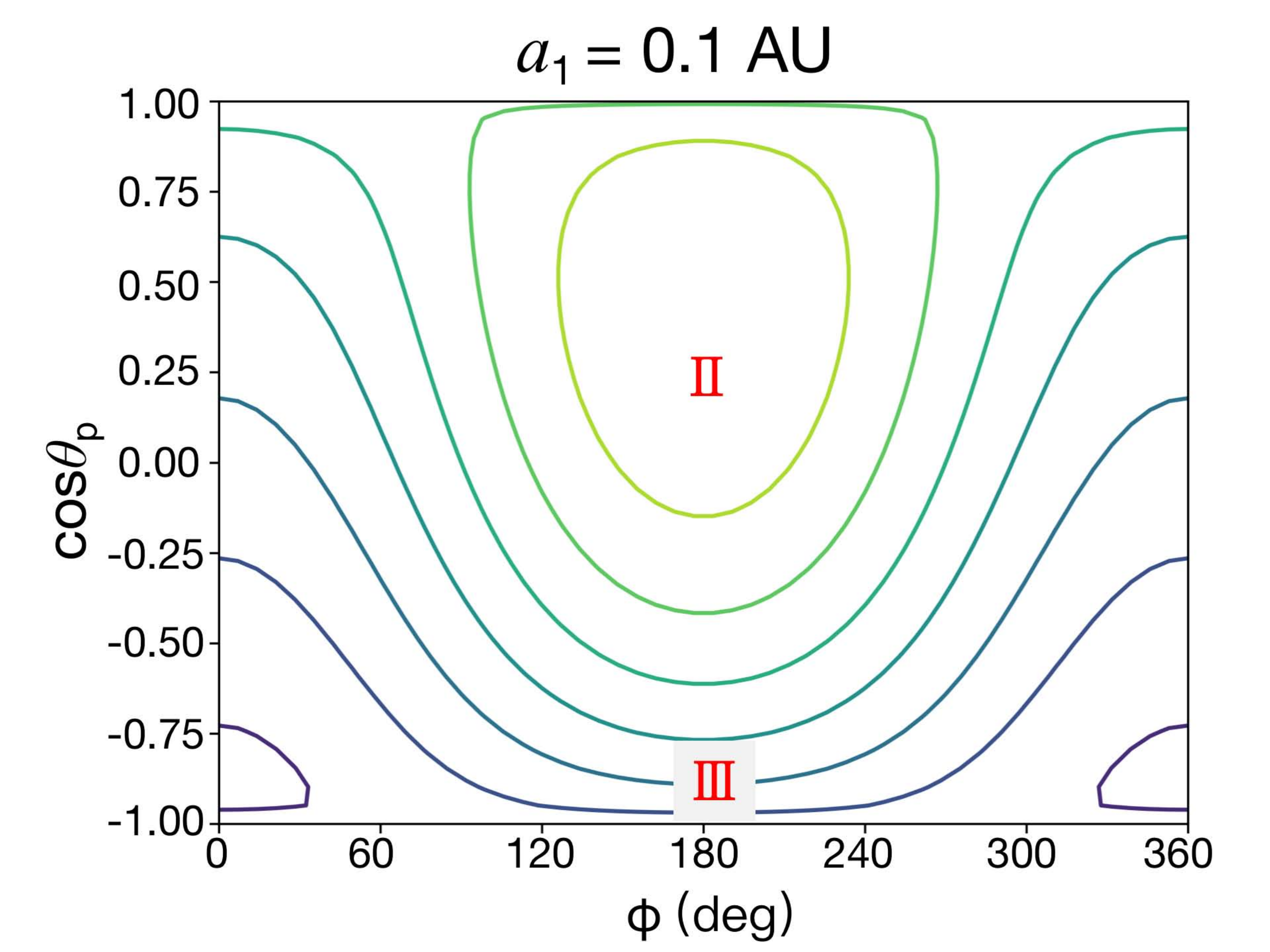}}
\caption{Three examples of Hamiltonian level curves for the spin evolution of terrestrial planets around the M-dwarf. The planetary semi-major axis ranges from 0.05 AU to 0.1 AU, and the precession frequency ratio noted as $\eta$ by \citep{Su2020}  is selected as [0.12, 0.45, 1.2] respectively. The different Cassini states are marked out with labels \RNum{1}, \RNum{2}, \RNum{3}.}
    \label{fig:level_curves}
 \end{center}
\end{figure*}

{According to the origin of Cassini States described in this work and Equation (8) in \citet{Su2022a}, we first plot the Hamiltonian  level curves in the phase space of cos$~\theta_\mathrm{p}-\phi$ to theoretically describe the spin evolution feature, $\phi$ is the angle to represent the precessional phase of spin vector about the orbital angular momentum vector. We find the  dynamical structures similar to \citet{Su2022a} as shown in Figure \ref{fig:level_curves}, including up to three kinds of typical oscillation modes, which refer to three kinds of Cassini States. For region \RNum{1} in Figure \ref{fig:level_curves}, the obliquity suffers excitation below 90$^{\circ}$. $\theta_p$ experiences a larger oscillation amplitude and the spin-axis flip with $\theta_p > 90^{\circ}$ in region \RNum{2}. After flip, $\theta_p$ stably librates between 40$^{\circ}$ and 60$^{\circ}$. In region \RNum{3}, the planetary obliquity is initially stirred up to $\theta_p  >  90^{\circ}$, finally evolves to the stable retrograde spin orientation.}

{The Hamiltonian level curves indicate that the planet spin evolution dynamics is greatly affected by the initial conditions, the equilibrium obliquity corresponding to Cassini State \RNum{2} that decreases with the precession ratio of the orbit and spin-axis. Based on the theoretical results from the Hamiltonian level curves, we then mainly investigate the typical oscillation cases of region \RNum{1} and \RNum{2}, to summarize the relationship between initial conditions and the final evolution feature.}

 \begin{figure*}
\begin{center}
        \includegraphics[width=2.0\columnwidth,height=10.5cm]{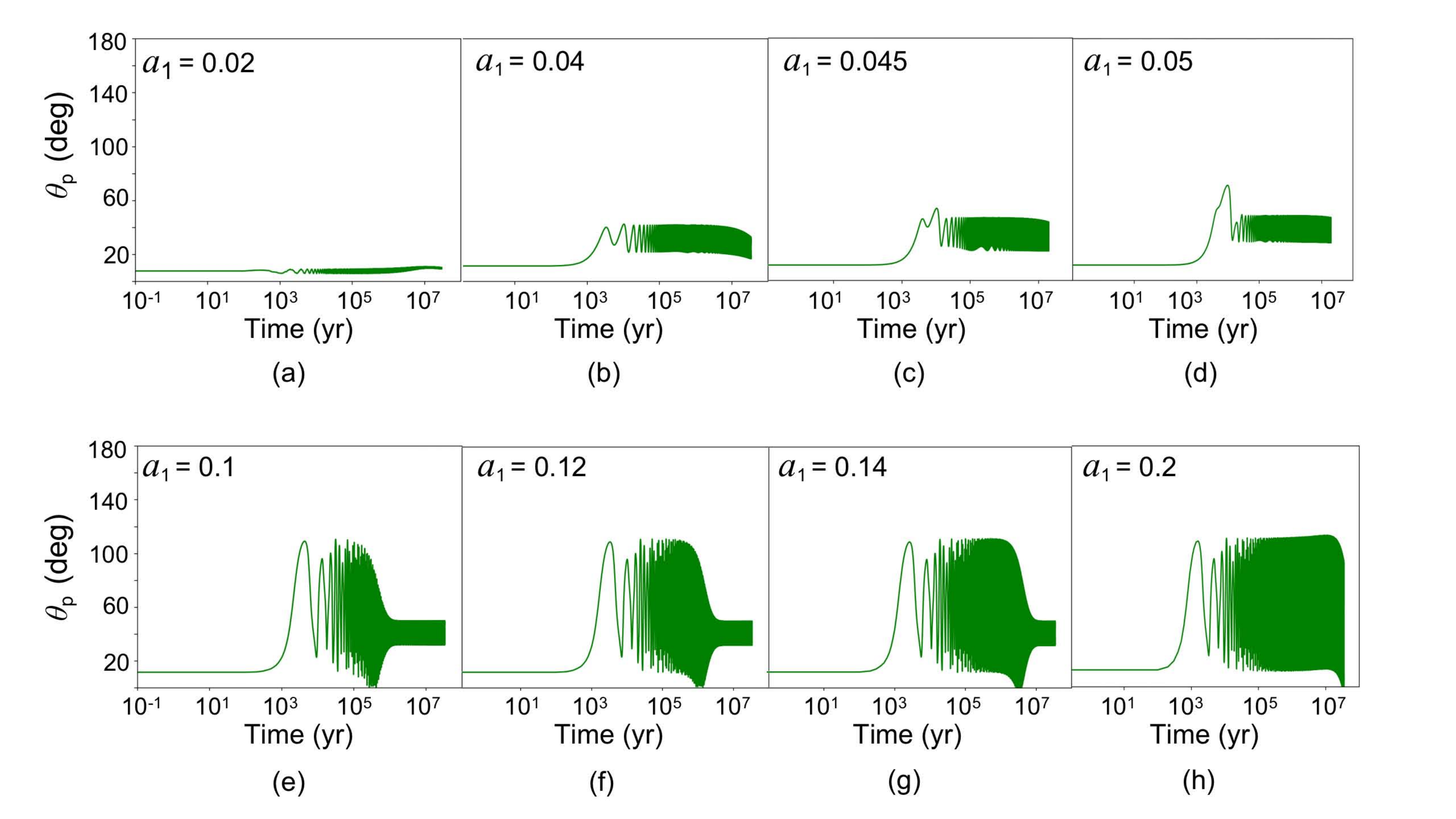}
    \caption{Evolution of the planetary obliquity or S-type terrestrials with a mass of 1 $M_{\oplus}$ and semi-major axis smaller than 0.2 AU, other initial conditions are set to be $m_0$ = 0.1 $M_{\odot}$, $m_1$ = 1  $M_{\oplus}$, $m_2$ = 0.1 $M_{\odot}$, $a_1$ = [0.02, 0.04, 0.045, 0.05, 0.1, 0.12, 0.14, 0.2] AU, $a_2$ = 5 AU, $e_1$ = 0.4, $e_2$ = 0.2, $i_1$ = 50 deg, $i_2$ = 0 deg. The obliquity can be triggered to retrograde and the time to reach the obliquity equilibrium state is much sensitive to the planetary initial position.}
    \label{fig:terrestrials}
\end{center}
\end{figure*}

\begin{figure*}
\begin{center}
        \includegraphics[width=2.0\columnwidth,height=10.5cm]{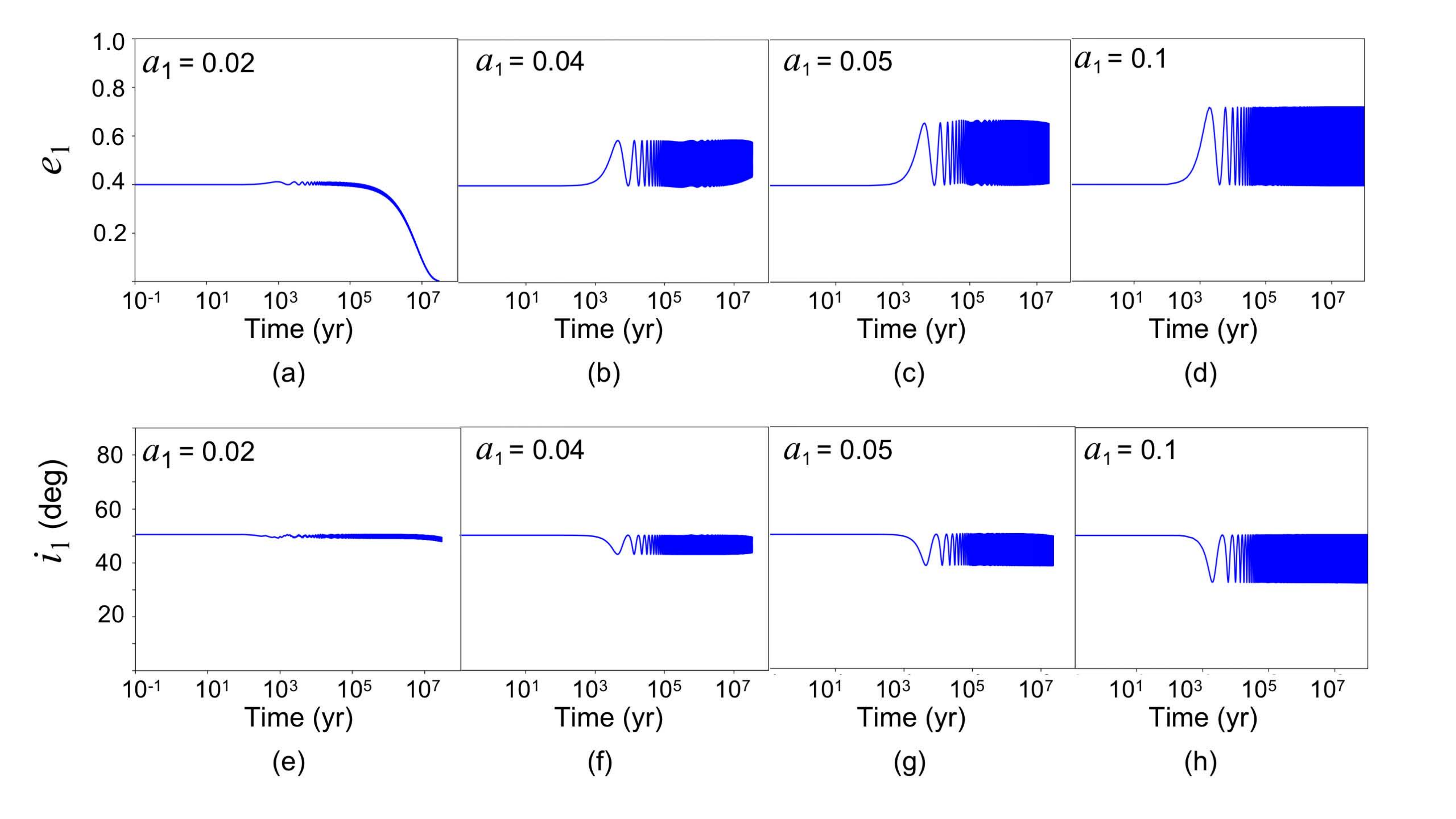}
    \caption{ {Evolution of the eccentricity and inclination for S-type terrestrials with a mass of 1 $M_{\oplus}$ and semi-major axis smaller than 0.1 AU, other initials same as Figure \ref{fig:terrestrials}. The eccentricity can be excited to 0.7 and the inclination decreases to below $40^{\circ}$ when $a_1$ = 0.1 AU. The oscillation amplitude of $e_1$ and $i_1$ are increasing with $a_1$. When $a_1 > 0.1$ AU, the upper and lower limit $e_1$ and $i_1$ in the evolution are not sensitive to $a_1$ any more.}}
    \label{fig:terrestrials_ei}
\end{center}
\end{figure*}

{Figure \ref{fig:terrestrials} presents a typical set of obliquity evolution with respect to diverse initial orbits for terrestrials, where $a_1$ of the planet varies from 0.02 to 0.2 AU. Panel (a)-(d) follow the evolution path of region \RNum{1} in Figure \ref{fig:level_curves}, and panel (e)-(f) follow the region \RNum{2}. We also perform the time-series evolution of inclination and eccentricity in Figure \ref{fig:terrestrials_ei} to show the effect of the EKL strength, and the excitation effect of EKL is greater when the planet and the secondary star are closer. The eccentricity can be excited to 0.7 and the orbital inclination decreases to lower than $40^{\circ}$ when $a_1$ = 0.1 AU. The oscillation amplitude of $e_1$ and $i_1$ are increasing with $a_1$. The variation of semi-major axis is too little during the evolution to capture the orbit decay, thus we could not see the noticeable circulation.}

\begin{figure*}
\begin{center}
\subfigure[]{\includegraphics[width=0.9\columnwidth,height=6.5cm]{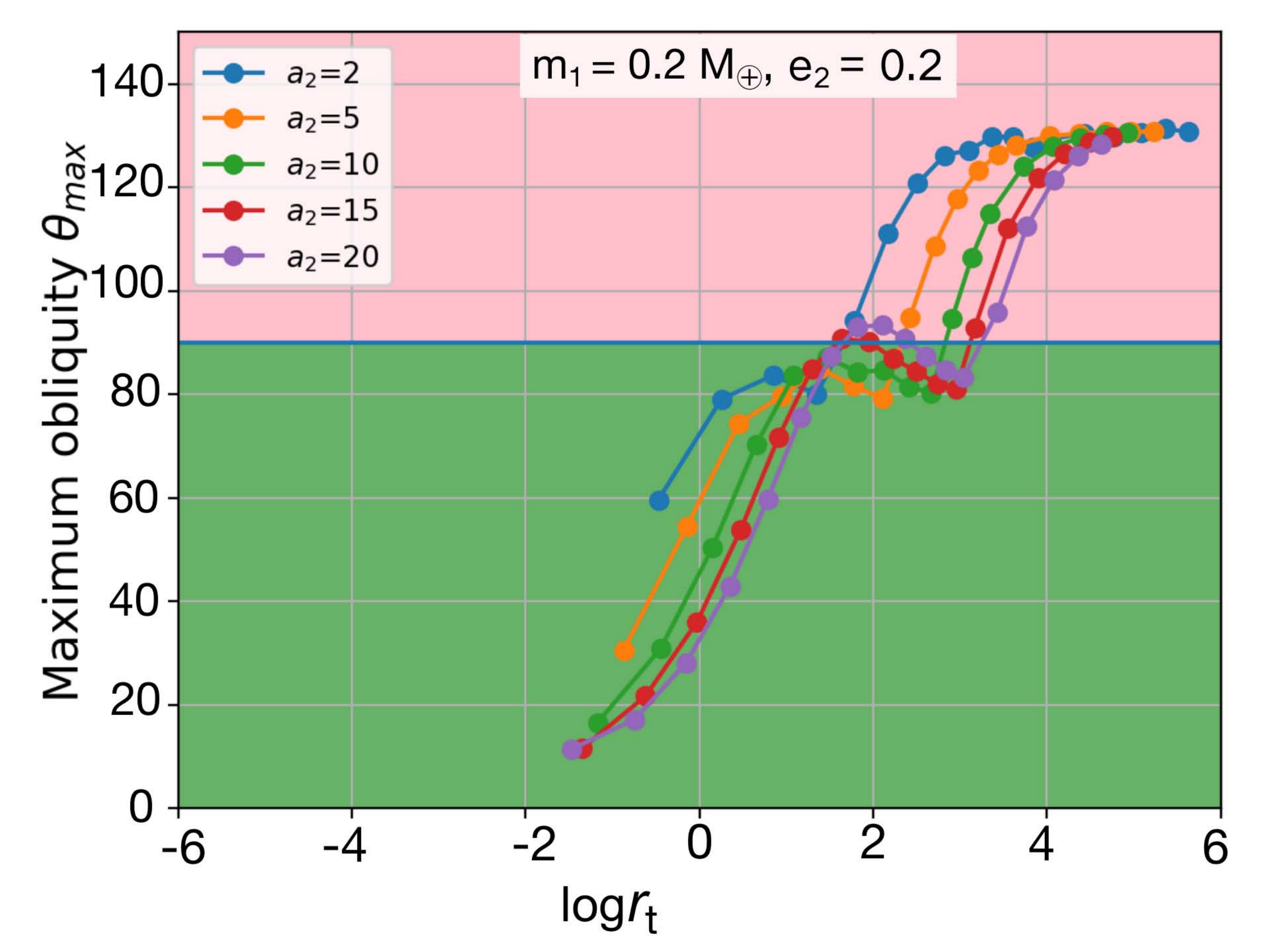}}
\subfigure[]{\includegraphics[width=0.9\columnwidth,height=6.5cm]{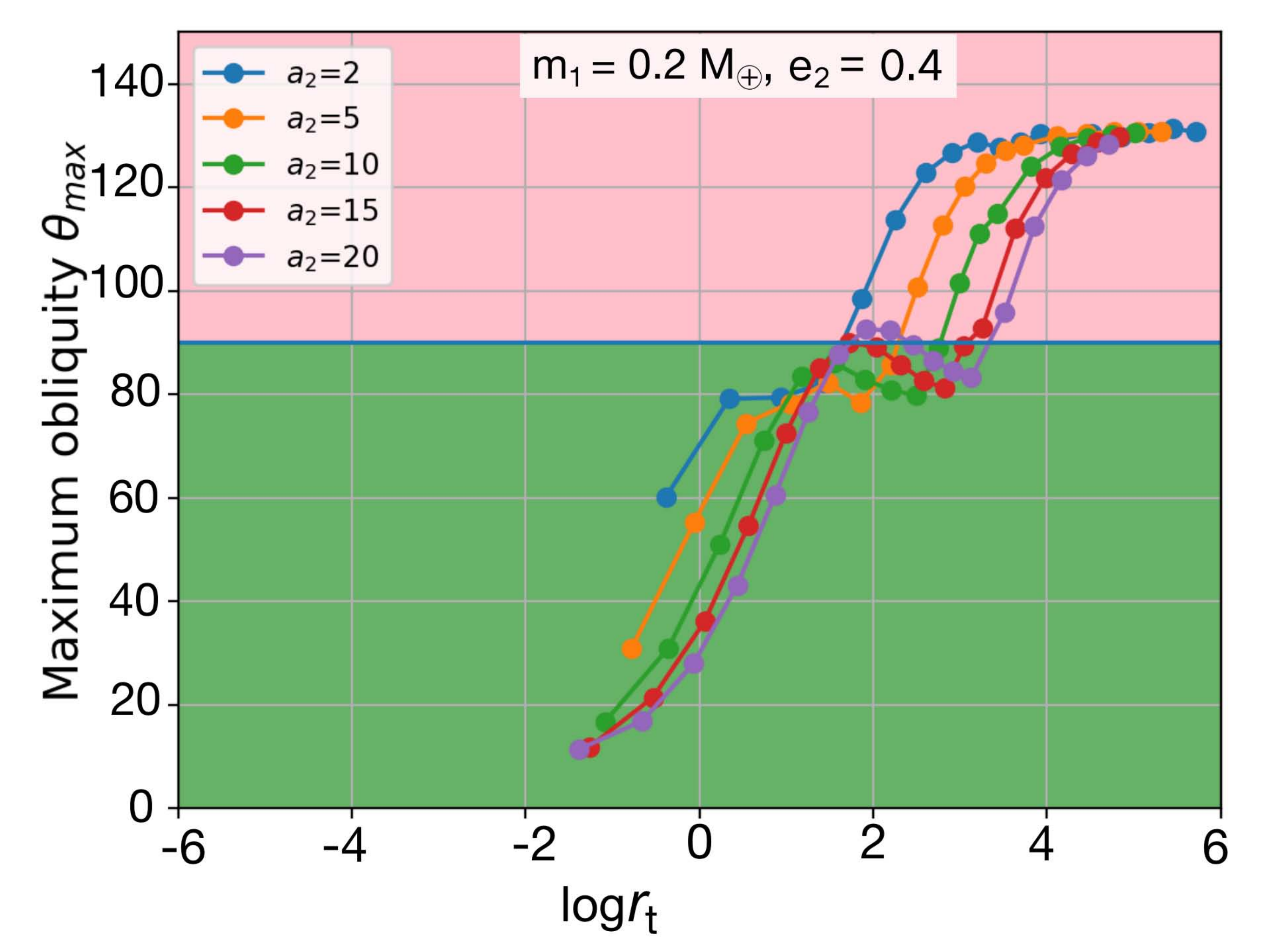}}\\
\subfigure[]{\includegraphics[width=0.9\columnwidth,height=6.5cm]{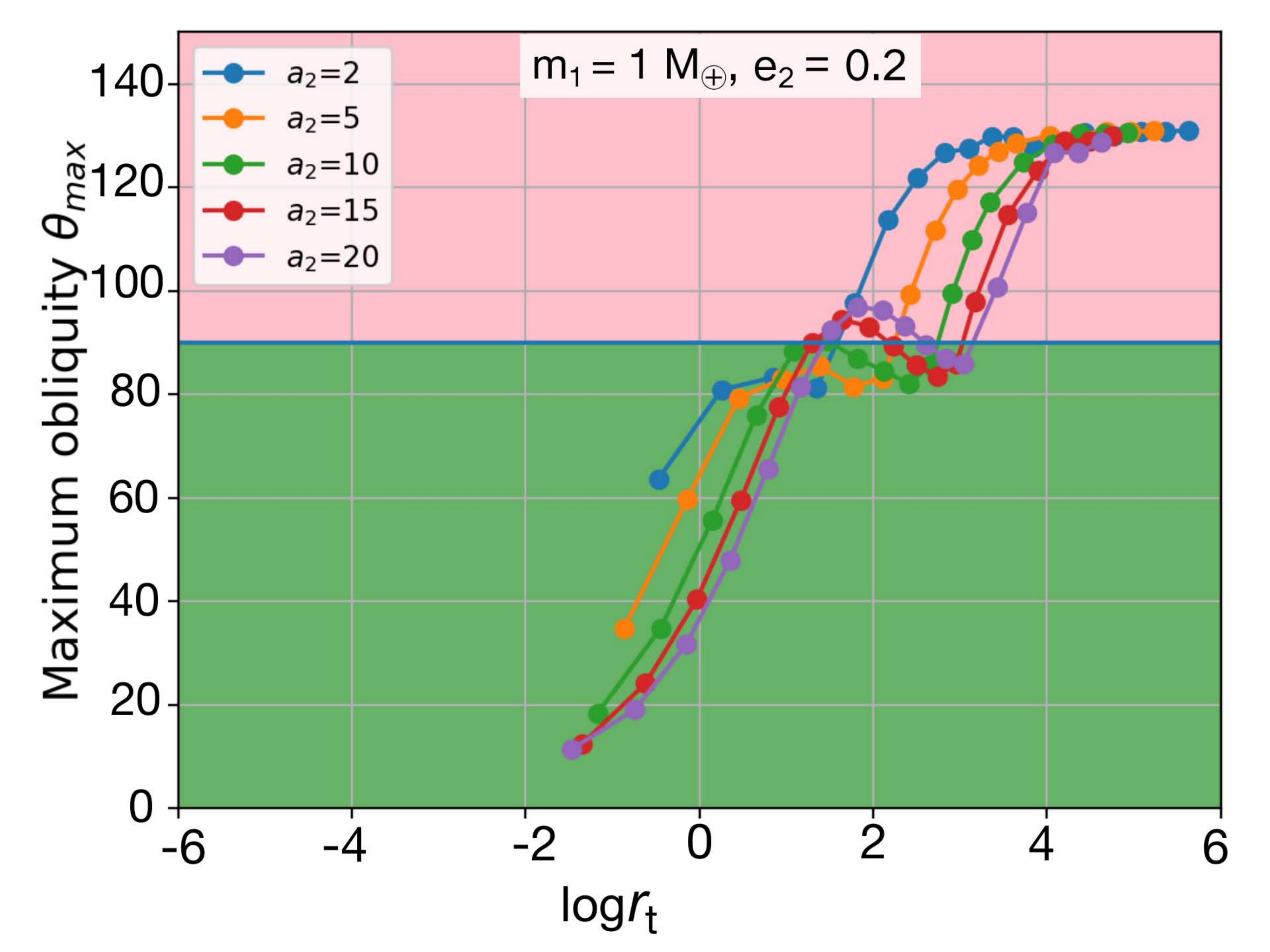}}
\subfigure[]{\includegraphics[width=0.9\columnwidth,height=6.5cm]{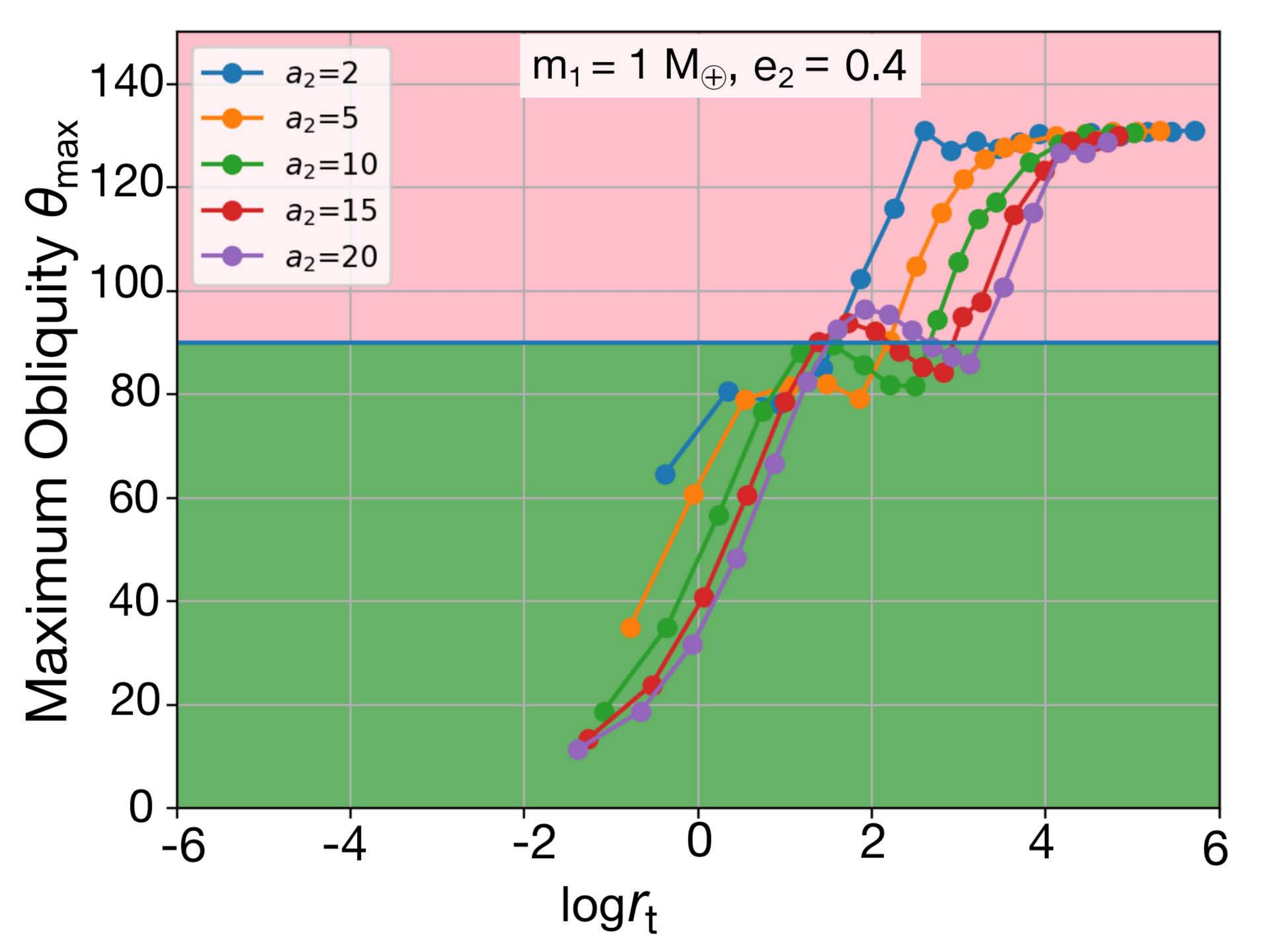}}\\
\caption{Distribution of the maximum obliquity $\theta_{max}$ in the dimension of timescale ratio $r_{t}$. The semi-major axis $a_2$ ranges from 2 to 20 AU. Regions with or without flip are also filled with pink and green respectively, with 90$^{\circ}$ as horizontal boundaries.}
    \label{fig:obl_max_tr}
 \end{center}
\end{figure*}

{The evolution results can be further distinguished by $r_{t}$, as shown in Figure \ref{fig:obl_max_tr}. Here we perform the numerical results with respect to four groups of $m_1 = 0.2$ $M_{\oplus}$, 1.0 $M_{\oplus}$ and $e_2$ = 0.2, 0.4. With the increase of $r_{t}$, the maximum obliquity also increases. Figure \ref{fig:obl_max_tr} shows that the planetary obliquity without flip when $r_{t} < 1$,  while the obliquity of Earth-mass planet could flip when $r_{t}$ ranges from 1 to $10^3$, and $\theta_p$ will always flip when $r_{t}$ spans from $10^3$ to $10^6$.}

{It is noteworthy that when $10 < r_{t} < 10^3$, $\theta_{max}$ goes up to a local bulge and then declines to a valley. It is speculated that the extreme planetary eccentricity originates from EKL mechanism when $r_{t}$ gets larger, and the tidal effect is dramatically enhanced owing to a much closer pericenter to the host star. Under the combined scenarios, the planetary obliquity undergoes an inhibited growth. On the other hand, we further compare the obliquity evolution due to $e_2$. When $e_2$ rises to 0.4, the more obvious obliquity excitation occurs, and the maximum $\theta_p$ enters the flip region with smaller log$r_{\mathrm{t}}$. For $r_{t} > 10^4$, the maximum obliquity converges to $\sim$ 130$^{\circ}$.}

In total numerical simulations, we calculate the evolution paths for terrestrial planets with $m_1 = 0.2 \sim 8~M_{\oplus}$, where the general evolution trend is consistent. When $m_1$ grows  to 1 $M_{\oplus}$ , the local maximum obliquity of the bulge ascends to 100$^{\circ}$ with $a_2 = 20$ AU. Additional simulation results reveal that when $m_1$ increases to 4 $M_{\oplus}$ and 8 $M_{\oplus}$, the local maximum obliquity can arrive at 105$^{\circ}$ and 110$^{\circ}$, respectively. It can be inferred that more runs may locate at the region of flip with larger planetary masses for terrestrials around M-dwarfs.

{We carry out additional N-body simulations with \textit{MERCURY-T}  to study more general S-type systems with an extremely higher $e_2$.} Currently, there are over five S-type binary systems that are observed to have an eccentric outer companion with $e_2 >$ 0.5, thus we can further investigate the evolution of $\theta_p$ in these cases. The simulations show that the terrestrials orbiting M-dwarfs can be scattered or captured by the secondary with $e_2 = 0.8$. Furthermore, we conduct a couple of runs to explore terrestrials orbiting solar-type stars in comparison with those about M-dwarfs,  {where 55$\%$ of them can survive under the comprehensive effect of EKL and tidal dissipation.}

{Aside from typical oscillation cases presented in Figure \ref{fig:terrestrials}, we also find chaotic evolution behaviour. \citet{Storch2014b} demonstrated the stellar spin-orbit angle can have a chaotic region, and similar chaotic behaviour of the planetary obliquity exist in our work. These chaotic cases always occur when  $a_2$ = 2 AU and  $e_2 > 0.6$, the extremely excitation of eccentricity leads to the instability of the system. Comparing with the stellar spin-orbit angle chaotic behaviour, the chaotic results of planetary obliquity in our work origin from spin-orbit coupling effect between the planet spin and the planet orbit, rather than between the stellar spin and the planet orbit.}

\subsection{The Equilibrium Timescale}\label{subsec:3.3}
\begin{figure*}
\begin{center}
\label{fig:eqtmap}
\subfigure[]{\includegraphics[width=0.9\columnwidth,height=6cm]{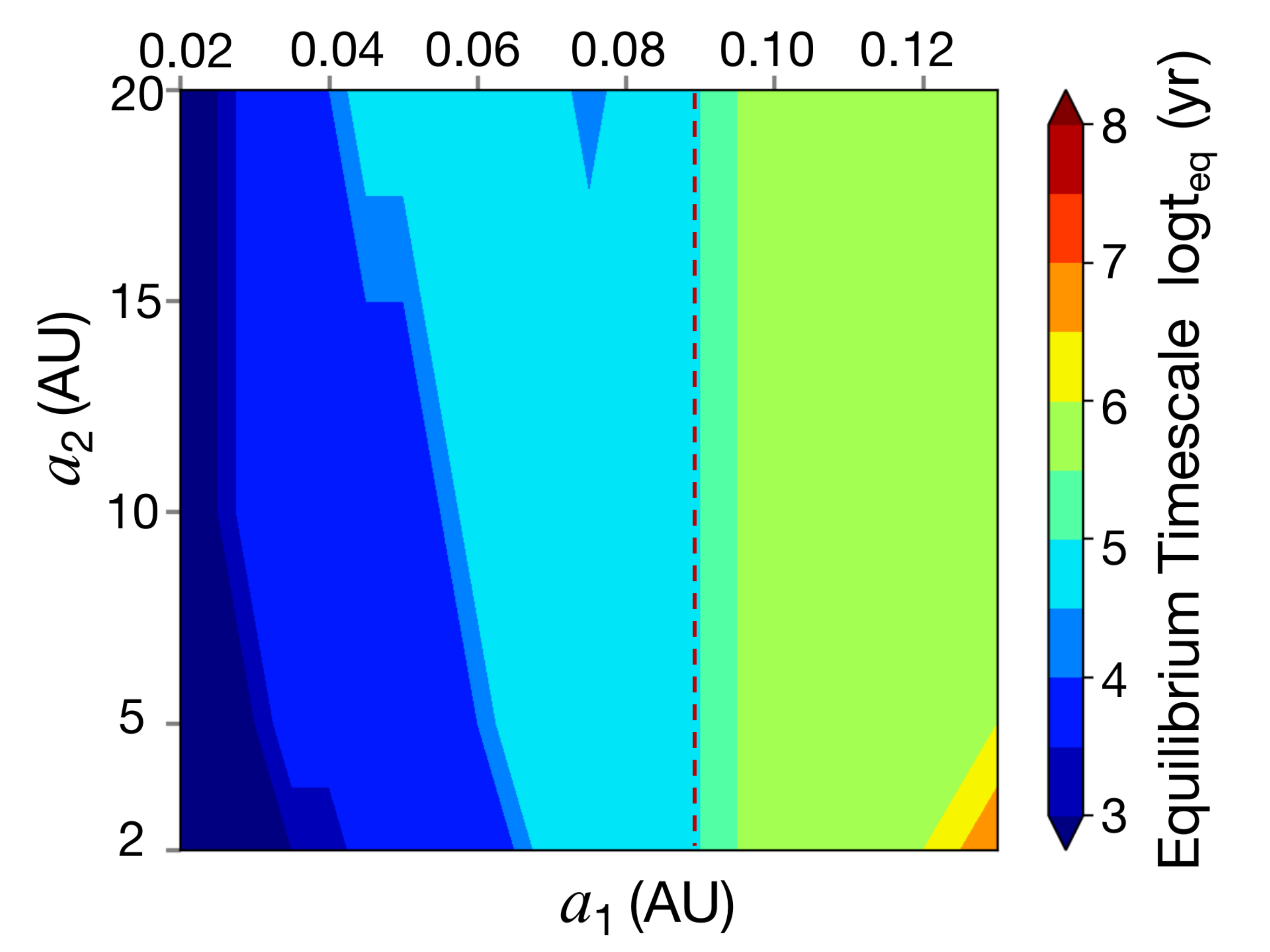}}
\subfigure[]{\includegraphics[width=0.9\columnwidth,height=6cm]{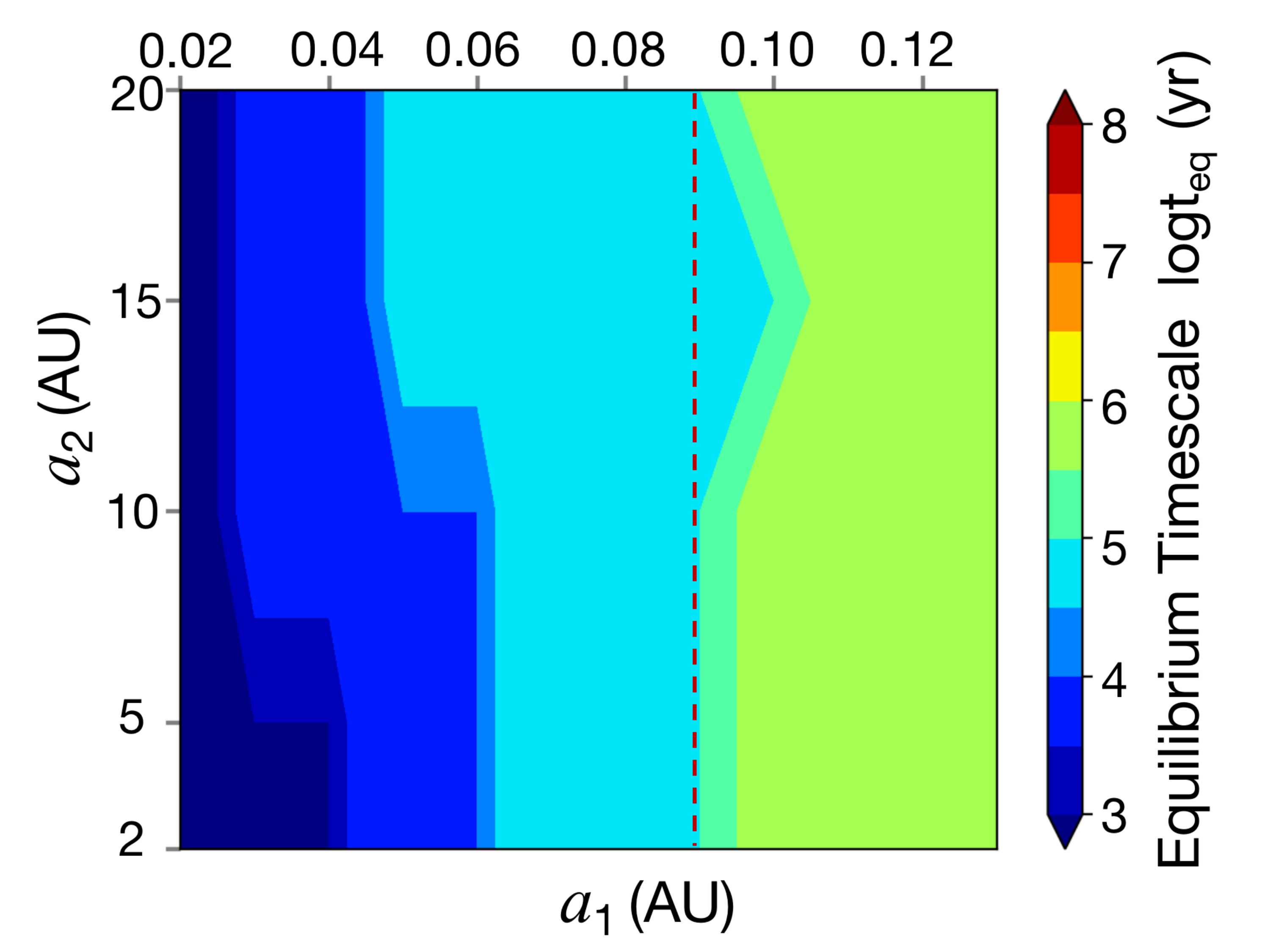}}\\
\subfigure[]{\includegraphics[width=0.9\columnwidth,height=6cm]{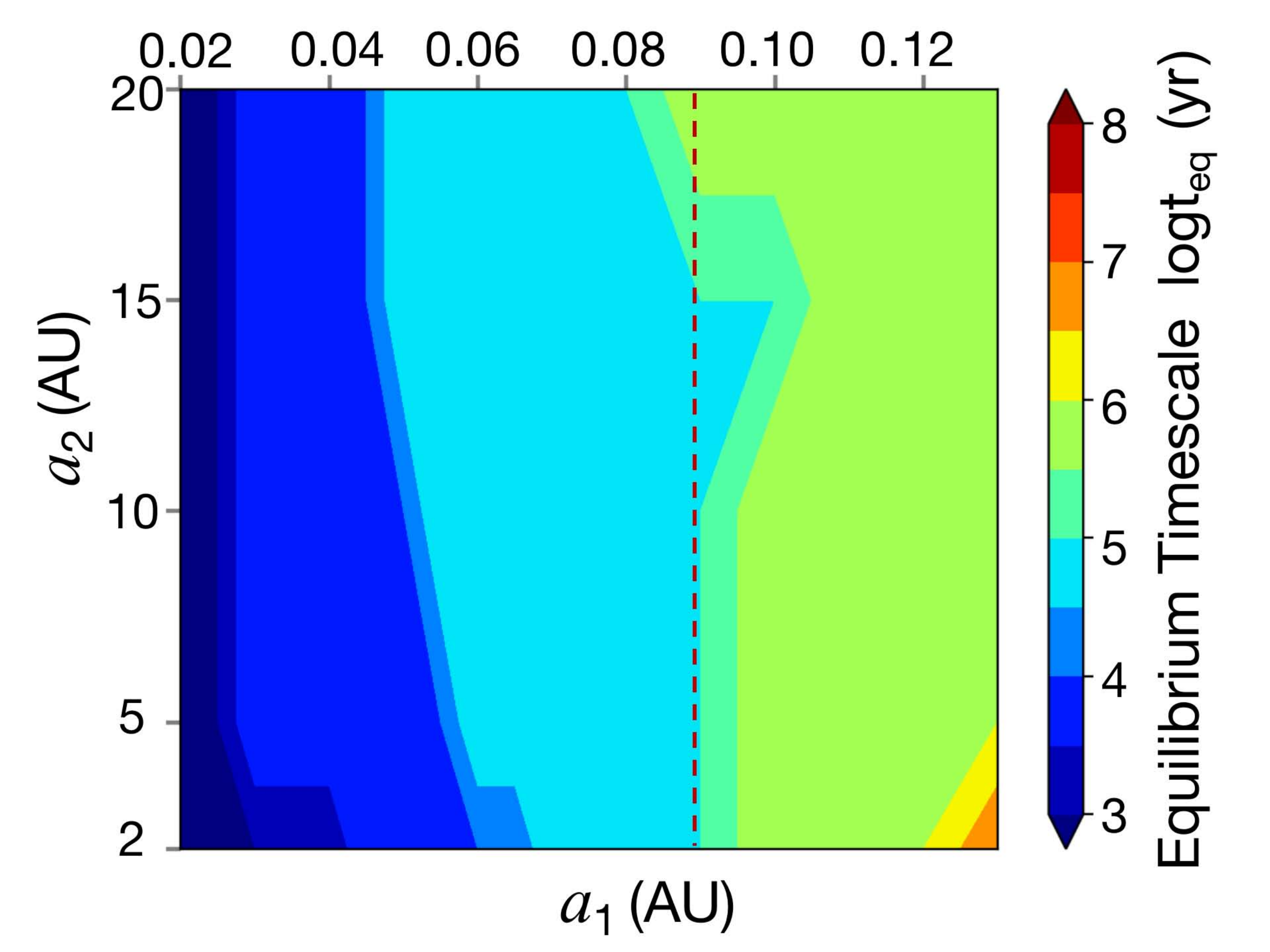}}
\subfigure[]{\includegraphics[width=0.9\columnwidth,height=6cm]{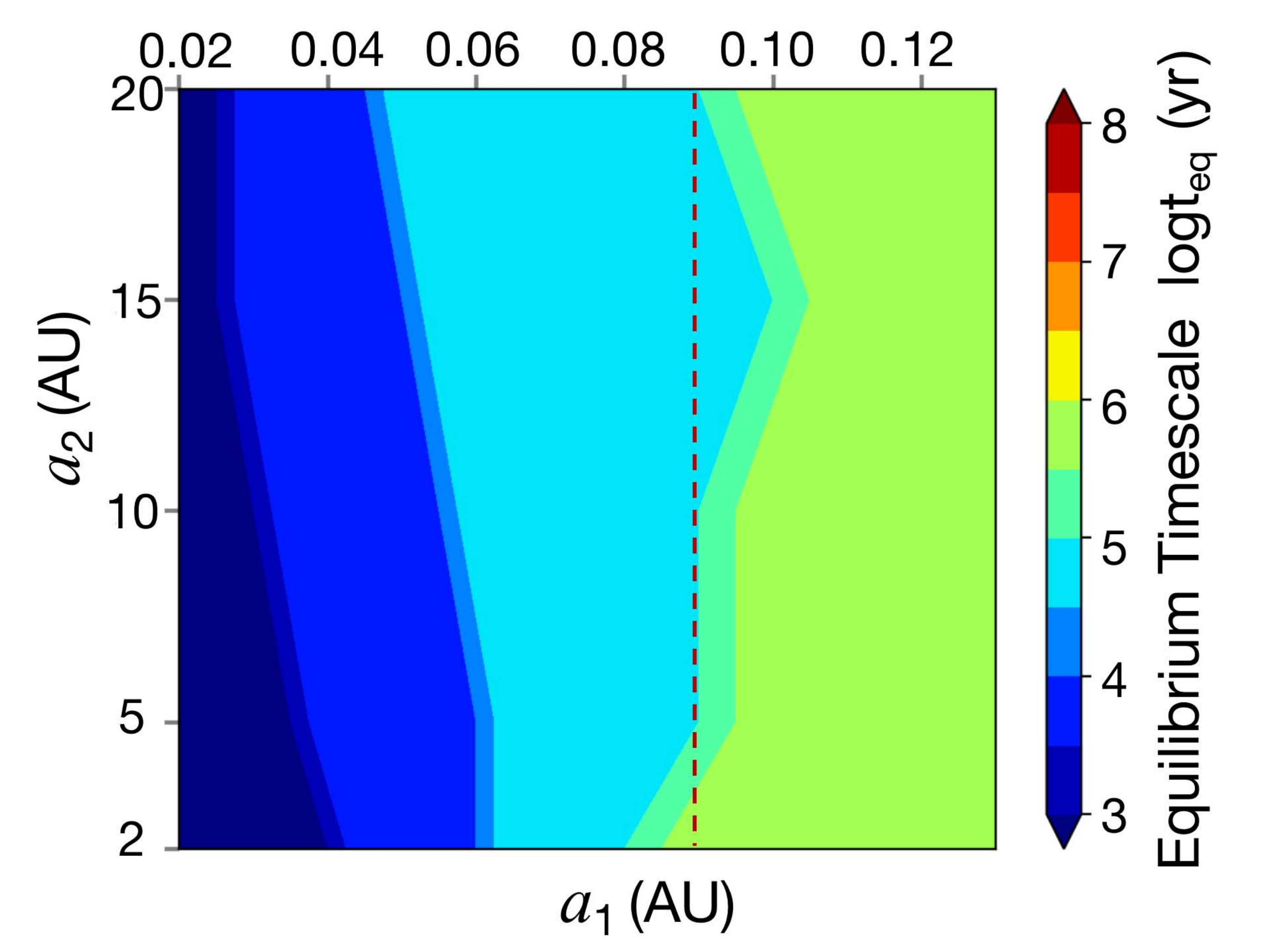}}
\caption{The equilibrium timescale map in the parameter spaces of semi-major axis. In panel (a)-(b), the planetary mass $m_1=0.2 M_{\oplus}$, $e_2 = 0.2$ and 0.4, in panel (c)-(d), the planetary mass changes to $m_1=0.2 M_{\oplus}$, and $e_2$ is same as panel (c)-(d).}
    \label{fig:eqtmap}
 \end{center}
\end{figure*}

\begin{figure*}
\begin{center}
\subfigure[]{\includegraphics[width=0.9\columnwidth,height=6.5cm]{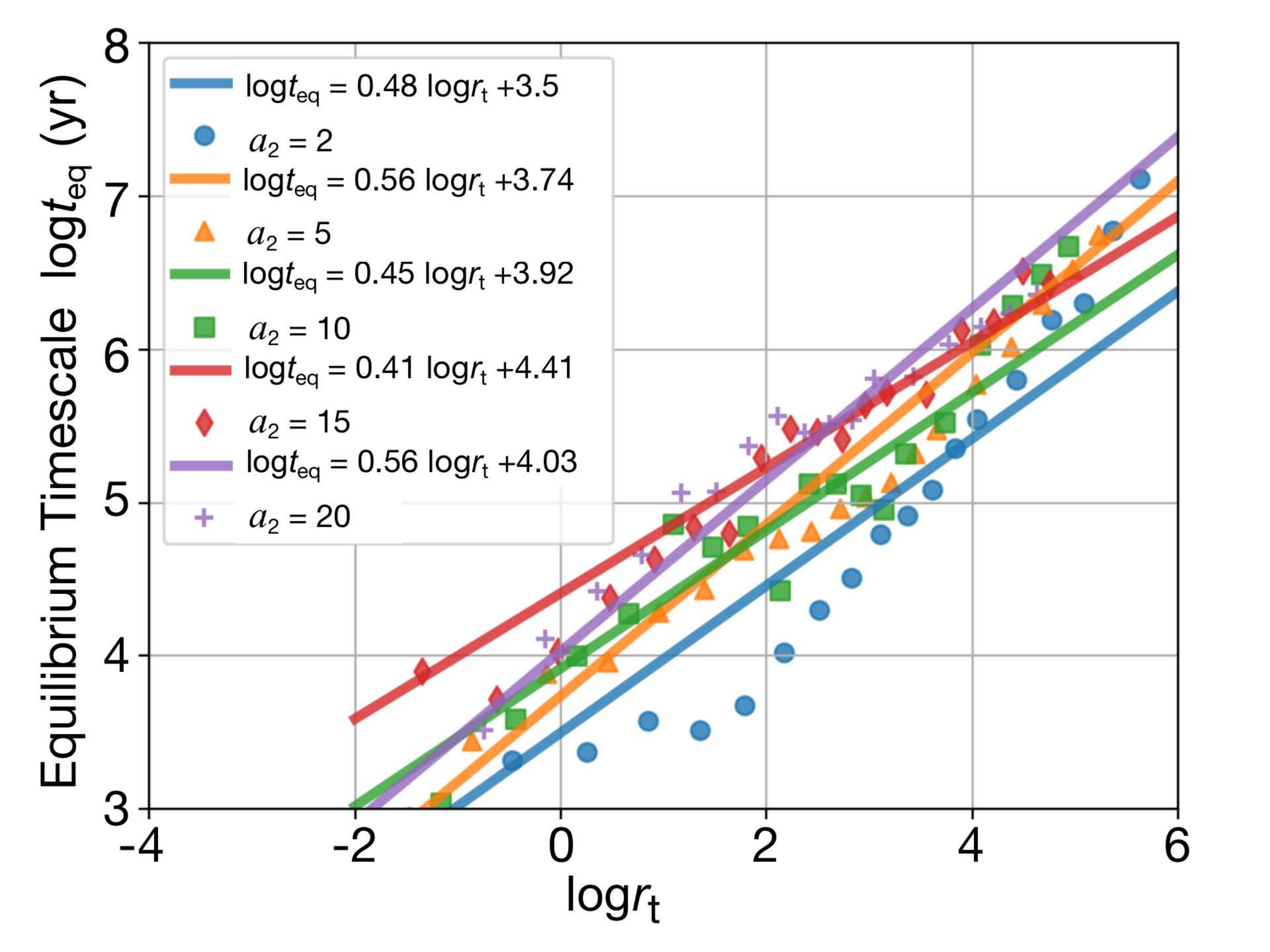}}
\subfigure[]{\includegraphics[width=0.9\columnwidth,height=6.5cm]{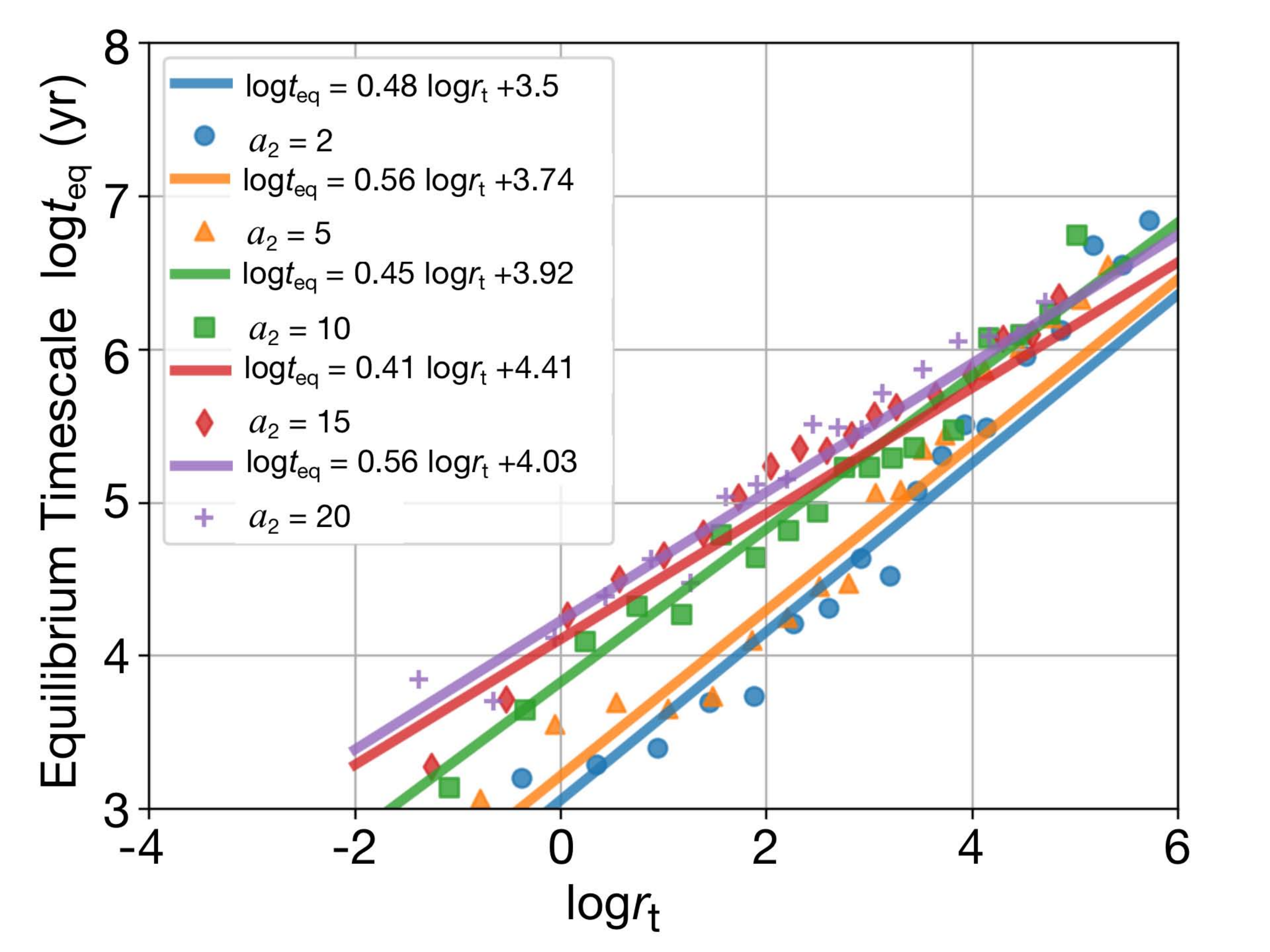}}\\
\subfigure[]{\includegraphics[width=0.9\columnwidth,height=6.5cm]{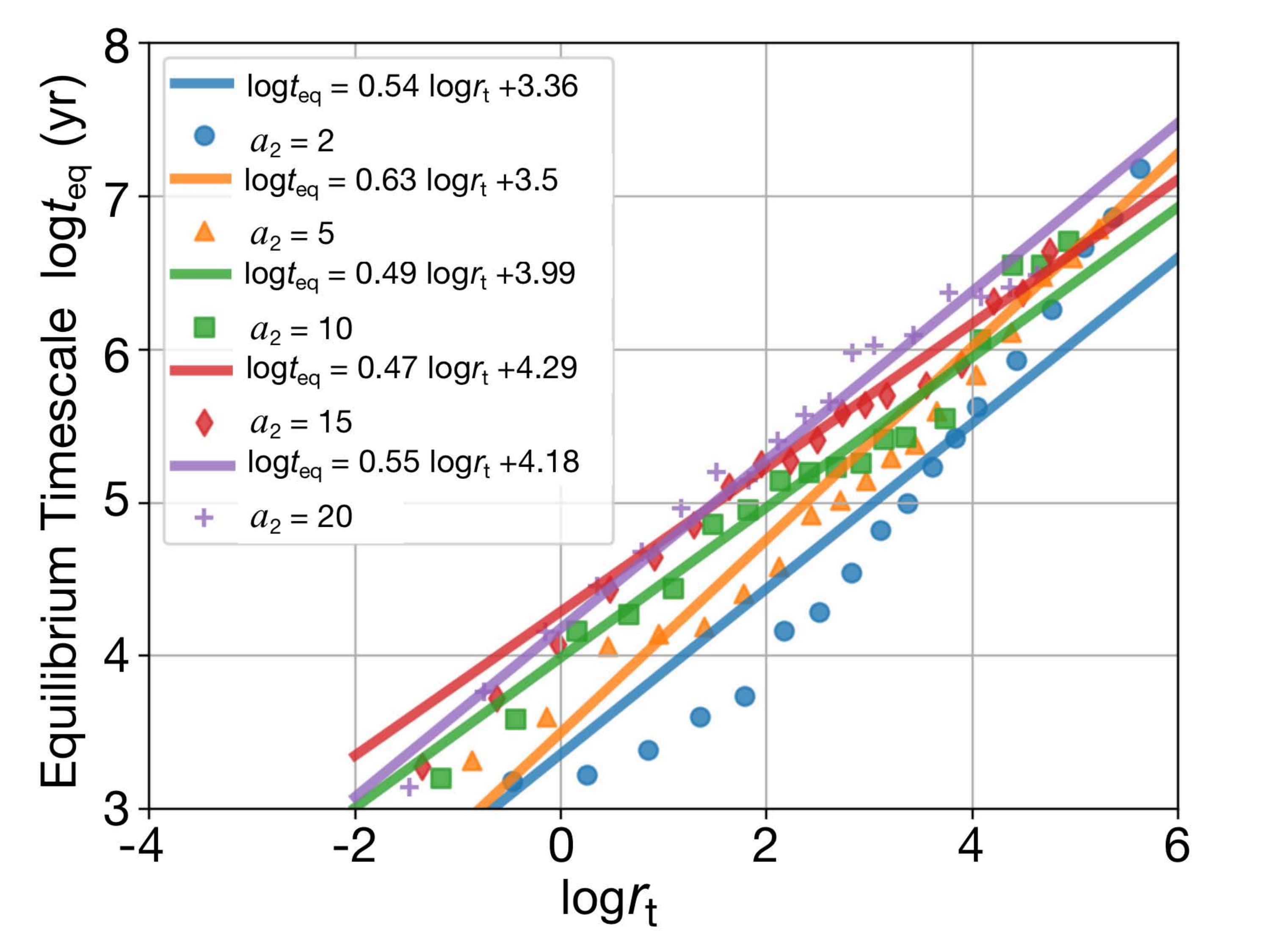}}
\subfigure[]{\includegraphics[width=0.9\columnwidth,height=6.5cm]{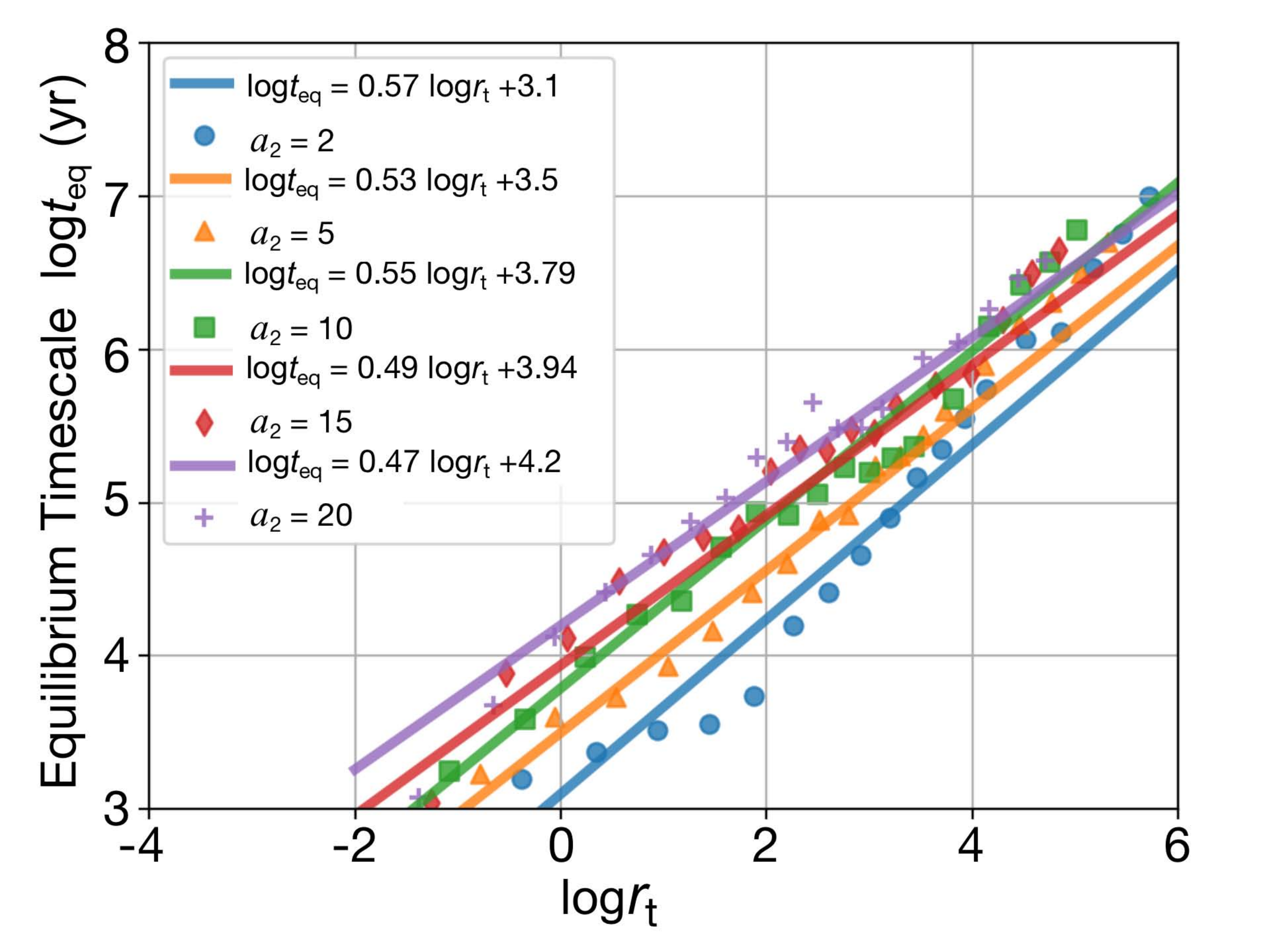}}\\
\caption{The relationship between $t_{\mathrm{eq}}$ and the timescale ratio $r_{t}$. Panel (a): $m_1=0.2M_{\oplus}$, $e_2=0.2$; Panel (b): $m_1=0.2M_{\oplus}$, $e_2=0.4$. Panel (c): $m_1=1 M_{\oplus}$, $e_2=0.2$; Panel (d): $m_1= 1 M_{\oplus}$, $e_2=0.4$. In each plot, the initial semi-major axis of the secondary star ranges from 2 AU to 20 AU. Dots plotted with different markers are numerical simulations results. The solid lines are the estimated relationships fitted by the maximum likelihood method.}
    \label{fig:eqt}
 \end{center}
\end{figure*}

In Section \ref{subsec:3.2}, $\theta_p$ remains a quasi-equilibrium state for over $10^7$ yr prior to a decrease to zero, and the timescale of entering the equilibrium obliquity is denoted as $t_{\mathrm{eq}}$. We then calculate the ratio between the orbital period and rotation period $P_r$  over the entire evolution.  For all terrestrial cases with $m_1$ = 1 $M_{\oplus}$ and $a_1 = 0.1 \sim 0.2 $ AU, $P_r$ will fall down to 6 at the time of $t_{\mathrm{eq}}$,  {which seems to be in a high-order spin-orbit resonant-like state. }The total angular momentum of the planet, the summation of $\overrightarrow{\boldsymbol{G}}_1$  and $\overrightarrow{\boldsymbol{N}}_p$, also drops down to a constant.  {We infer that the planetary angular momentum is trapped in the high-order spin-orbit resonant-like state.}

\citet{Su2022b} showed that the equilibrium obliquity ranges from $-90^{\circ}$ to $90^{\circ}$ for Cassini State \RNum{1}, \RNum{2} and \RNum{4}, whereas the obliquity varies from $160^{\circ}$ to $180^{\circ}$ for Cassini State \RNum{3}. The related precession ratio of spin vector and orbital angular momentum vector is in ($10^{-2}$ -- $10^{1}$). { Here the equilibrium $\theta_p$ performs the oscillation state consisting of Cassini State \RNum{2}-- evolves between $40^{\circ}$ and $60^{\circ}$, which agrees with the typically evolution in our simulations. \citet{Winn2005} addressed that Cassini state \RNum{2} is the most favorable configuration to maintain a significant obliquity.} In this study, we adopt a non-restricted model, the initial planetary mass is set to be 0.2 -- 8 $M_{\oplus}$, then we perform in a larger parameter space of $r_\mathrm{t}$ spanning from $10^{-2}$ to $10^{7}$. Our findings unveil a new equilibrium value of the maximum obliquity $\theta_{\mathrm{max}}$ close to $130^{\circ}$ when $r_\mathrm{t} > 10^{4}$ (Figure \ref{fig:obl_max_tr}).

In the equilibrium stage, the initial planetary orbit residing in the habitable zone of M-dwarf is not shrunk to approach the Roche-limit, thus the spin-axis orientation remains relatively stable, thereby helping maintaining the planetary habitability.

Figure \ref{fig:eqtmap} shows the contours of equilibrium timescale $t_{\mathrm{eq}}$ in the parameter space $a_1$-$a_2$, where $t_{\mathrm{eq}}$ evenly distributes in the dimension of $a_1$  but varies with $a_2$ and $e_2$. Next, we observe that $t_{\mathrm{eq}}$ is always less than $10^5$ yr for $a_1 = 0.09$ AU (see the dashed lines in each panel). For $a_2 < 20$ AU, $t_{\mathrm{eq}}$ is approximately proportional to the initial orbit of planet, which primarily governs tidal timescale. For $a_1 = 0.02$ AU and $a_1 > 0.1$ AU, the obliquity equilibrium timescale keeps unchanged for the variations $a_2$ at same $a_1$.  {When the semi-major axis meets 0.04 $< a_1 < 0.08$ AU, $a_2 = 20$ AU, the tidal timescale is equivalent to the secular perturbation timescale.}

We further investigate the relationship between $t_{\mathrm{eq}}$ and the timescale ratio $r_{t}$ as shown in Figure \ref{fig:eqt}, where Figure \ref{fig:eqt} (a)-(b) shows the planetary mass of $m_1=0.2$ $M_{\oplus}$ while $m_1=1 M_{\oplus}$ in Figure \ref{fig:eqt} (c)-(d), with respect to $e_2=0.2$ and 0.4, respectively, for each panel. We then utilize a maximum likelihood (ML) function and numerically optimize the relation. The fitting model is $\mathrm{log} t_{\mathrm{eq}}$ = $m_{ML} \times (\mathrm{log} r_{t}) + b_{ML}$. The fitting results of $t_{\mathrm{eq}}$ and $t_{\mathrm{I}}/t_{\mathrm{kl}}$ are provided in the panel legend.  {When $a_2$ = 2 or 5 AU, the tidal torque and the EKL effect are  intensive for $(t_{\mathrm{I}}/t_{\mathrm{kl}}) < 10^2 $,} the equilibrium timescale of the obliquity much arises from the tidal dissipation, thus the blue dots deviate to the lower end.

\subsection{The Obliquity Flip Ratio $R_\mathrm{flip}$}\label{subsec:3.4}
\begin{figure*}
\begin{center}
\subfigure[]{\includegraphics[width=0.95\columnwidth,height=6cm]{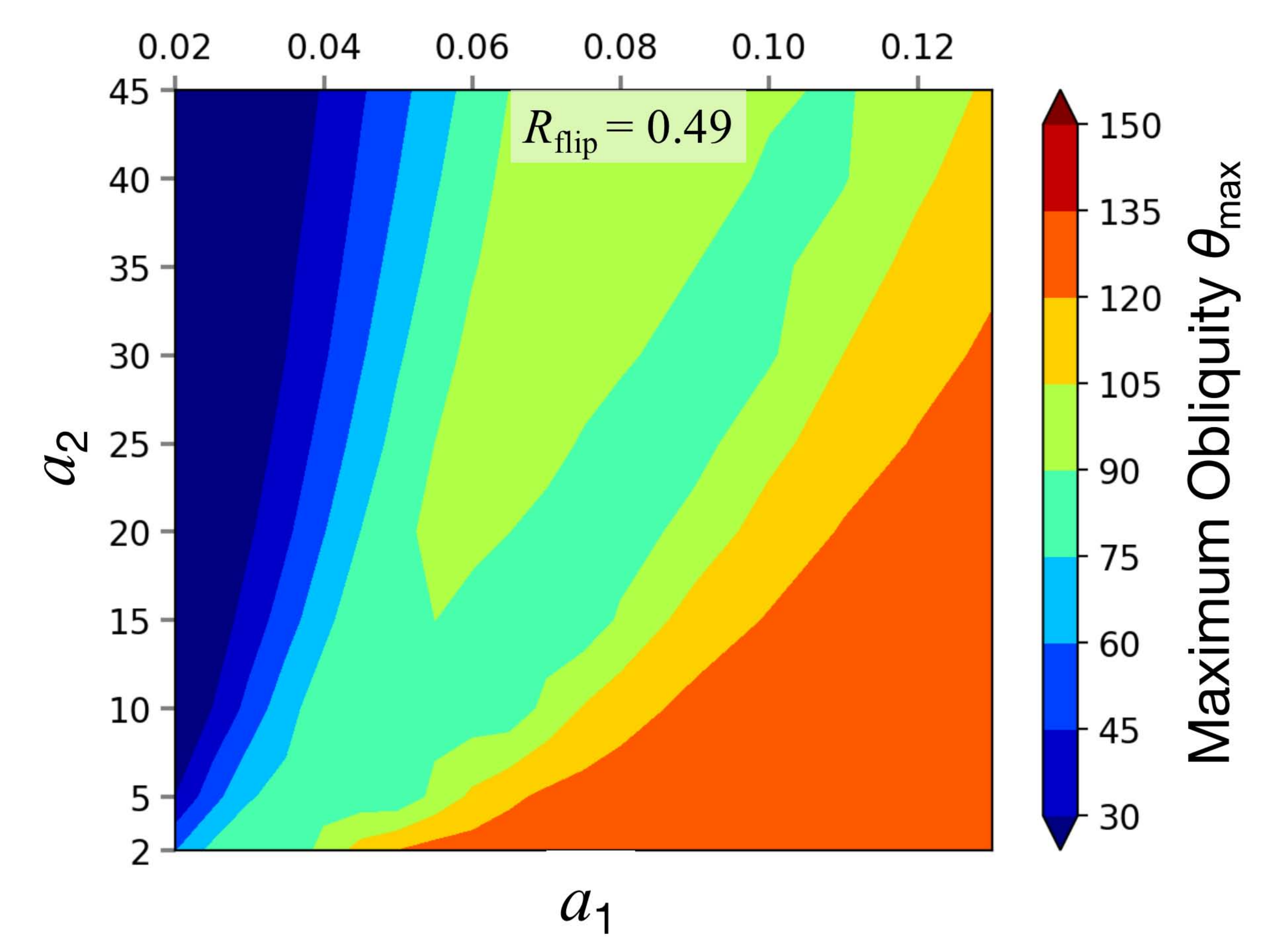}}
\subfigure[]{\includegraphics[width=0.95\columnwidth,height=6cm]{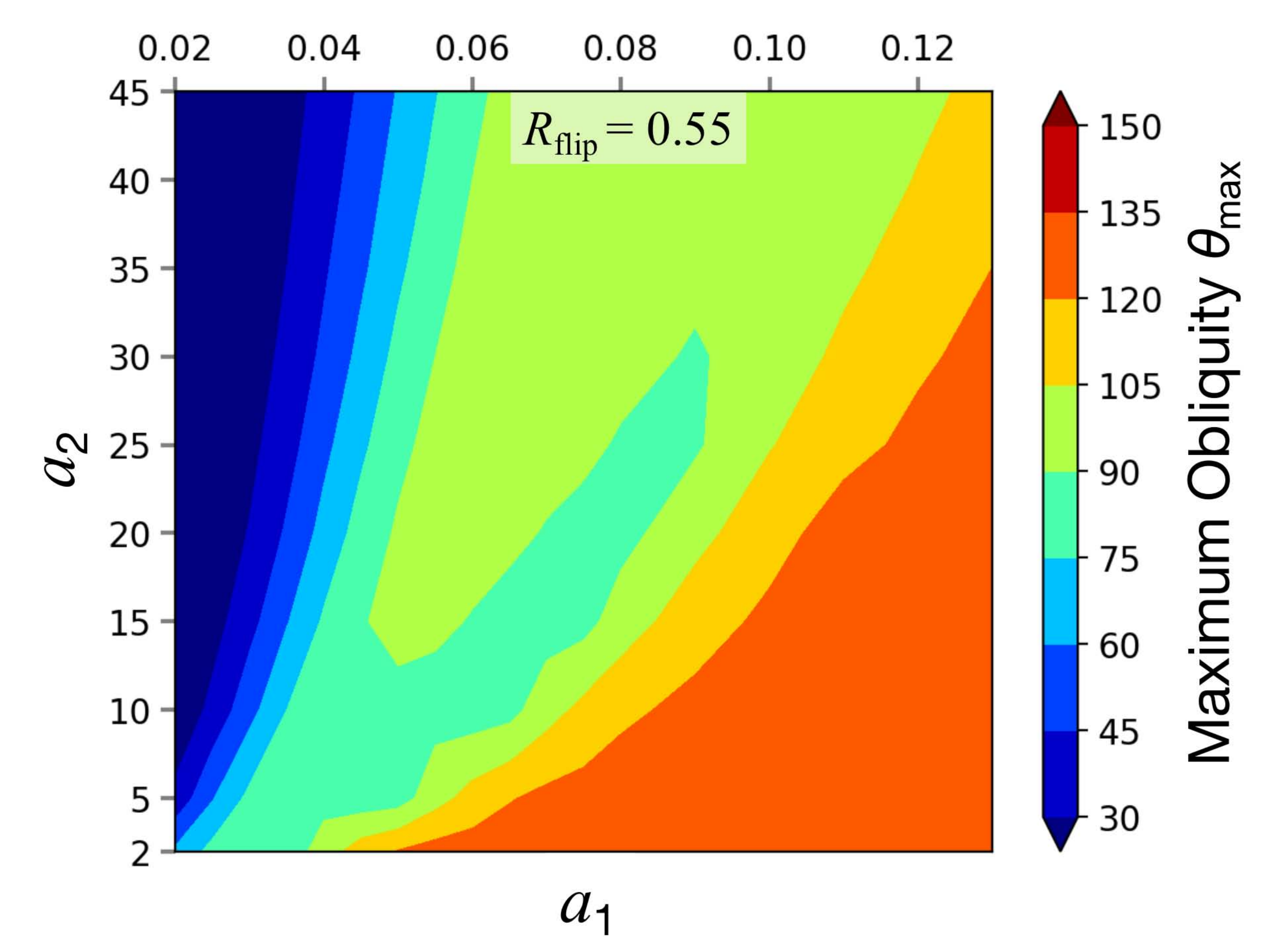}}\\
\subfigure[]{\includegraphics[width=0.95\columnwidth,height=6cm]{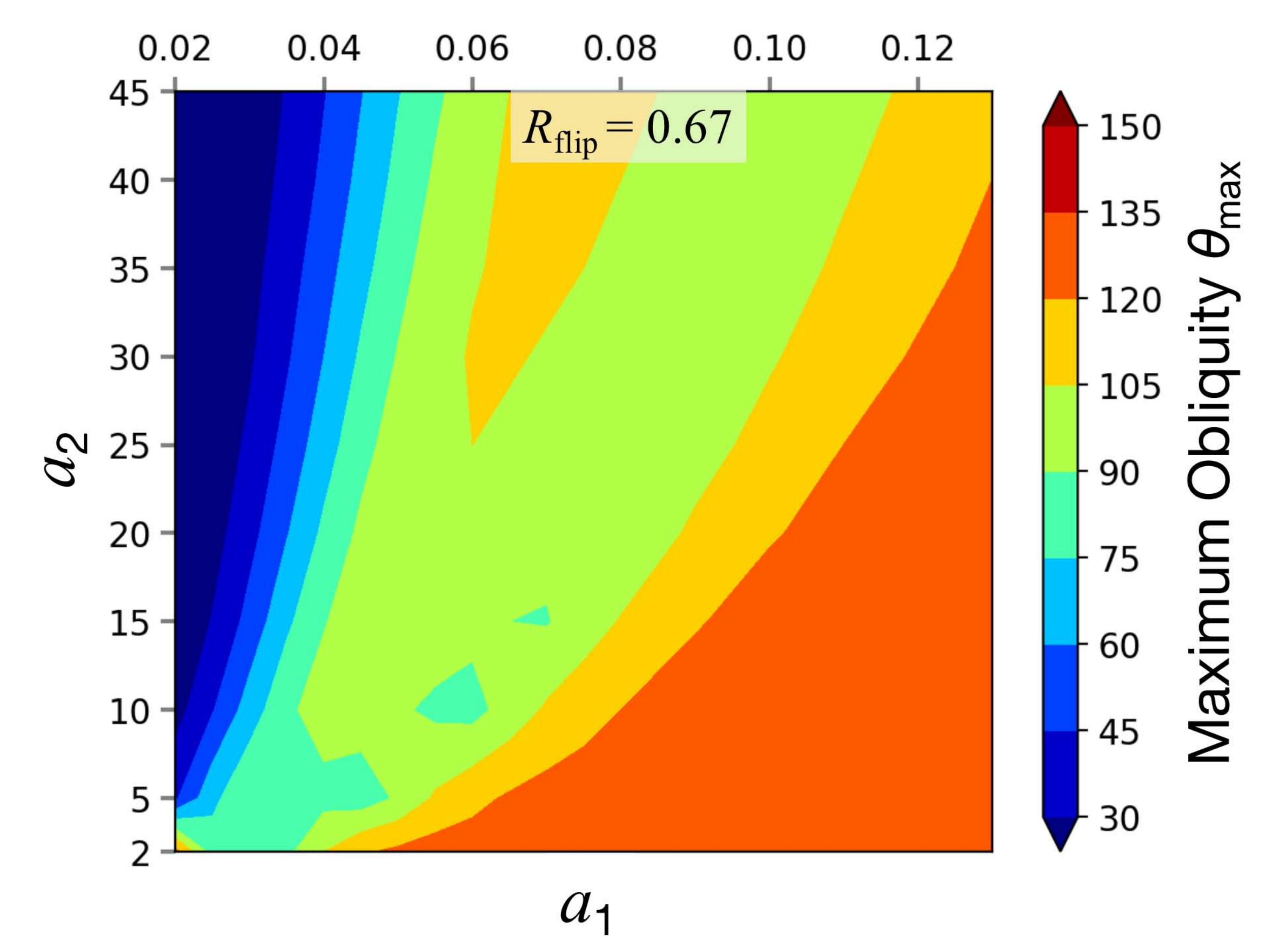}}
\subfigure[]{\includegraphics[width=0.95\columnwidth,height=6cm]{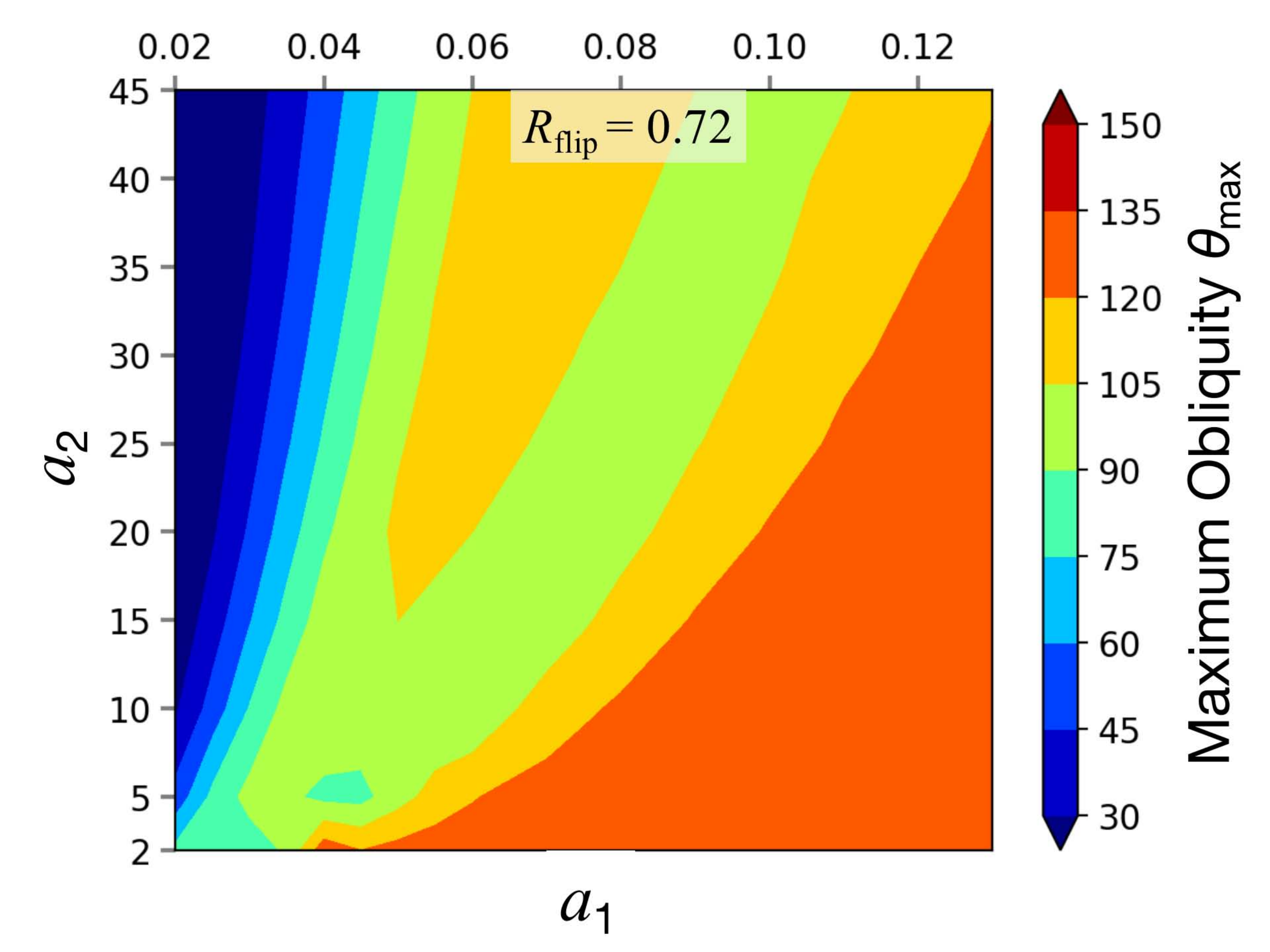}}
\caption{The spin-axis flip diagram. Each panel has specific initial parameters $m_1$ and $e_2$, (a): $m_1 = 0.2 M_{\oplus}$, $e_2 = 0.2$, (b): $m_1 = 0.2 M_{\oplus}$, $e_2 = 0.4$, (c): $m_1 = 1 M_{\oplus}$, $e_2 = 0.2$, (d): $m_1 = 1 M_{\oplus}$, $e_2 = 0.4$. The colour bar represents the range of the maximum value of $\theta_{\mathrm{p}}$ over 100 Myr, whereas those in blue denote cases without spin-axis flip. The flipping regions are filled with yellow-green and orange.}
    \label{fig:omax}
 \end{center}
\end{figure*}

In Section \ref{subsec:3.1}, we describe two typical evolution paths of obliquity, and cases follow region \RNum{2} can always trigger the flip of spin-axis. We extend the distance of the second star to 45 AU in this section, and investigate  {the obliquity flip ratio} in the parameter space of $a_1$-$a_2$. In Figure \ref{fig:omax}, over 700 simulations are conducted for S-type terrestrials with $m_1 = 0.2 \sim 8$ $M_{\oplus}$, accompanied with a constant perturber of 0.2 solar mass, in which each panel is plotted with a grid resolution of $10\times18$. The flip ratio is calculated by the fraction of the simulations that undergoes flip.

{The flip ratio} in Figure \ref{fig:omax}(a)-(b) are 0.49, 0.55, 0.67, 0.72, respectively. We conclude that the reverse of the spin axis can be very common for S-type terrestrials with a distance companion. When the planetary mass increases,  {the flip ratio} increases with the same initial semi-major axis, thus the flips get easier with higher $m_1/m_2$.  This will provide the clues to predict what kind of  exoplanets can evolve to be head-down, like Venus in our solar system. We also change the eccentricity of the companion, but the flip region of obliquity is not much sensitive to $e_2$. For the Earth-mass planet,  {the flip ratio} spans from 0.52 to 0.56 when $e_2 $= 0.2 -- 0.6.

\section{Applications to Potential Oblique S-type systems}\label{sec:4}
In this section, several S-type planets are reported to figure out whether they rotate with the oblique attitude under the combined scenarios of EKL and equilibrium tide. Considering the timescale of secular evolution, we select the terrestrial planet HD 42936 b and the Jupiter-like GJ 86 Ab and $\tau$ Boot Ab as case study. In Table \ref{tab:stars} and \ref{tab:planets}, we summarize essential parameters for three potential oblique S-type systems.

\begin{deluxetable}{ccccc} \label{tab:stars}
\tablecaption{Parameters for binary stars in three potential oblique S-type systems}
 \tablehead{ & $\text{HD } 42936$ & $\text { GJ } 86$ & $\tau \text { Boot}$}
\startdata
$a_{\text {AB}} (\text{AU}) $& 1.22& 20 & 45\\
$e_{\text{AB}}$ & 0.594 & 0.4 & 0.91\\
$m_{\text {A}} (M_{\odot})$ & 0.87 & 0.8 & 1.4 \\
$m_{\text{B}} (M_{\odot})$ & 0.08 &  0.49 & 0.4 \\
\enddata
\end{deluxetable}

\begin{deluxetable}{ccccc} \label{tab:planets}
\tablecaption{Physical parameters for  planets in three potential oblique S-type systems.}
 \tablehead{ & $\text{HD} 42936 \text{b}$& $\text { GJ } 86 \text{ Ab}$ & $\tau \text { Boot Ab}$}
\startdata
 $a_{\text{p}} (\text{AU})$ & 0.066 & 0.110 & 0.046\\
 $P (\text{day})$ & 6.67 & 15.77 & 3.31\\
$e_{\text{p}}$ & 0.140 & 0.046 & 0.079\\
$i_{\text{p}} (\text{deg})$ & -- & -- & 45\\
$\omega_{\text{p}} (\text{deg})$ & 80.6 & 266.0 & 218.4\\
$m_{\text{p}} (M_{\text{J}})$ & 0.008 & 4.01 & 4.13 \\
\enddata
\end{deluxetable}

\subsection{HD 42936 b}\label{sec:4.1}
\begin{figure}
\includegraphics[width=\columnwidth,height=7cm]{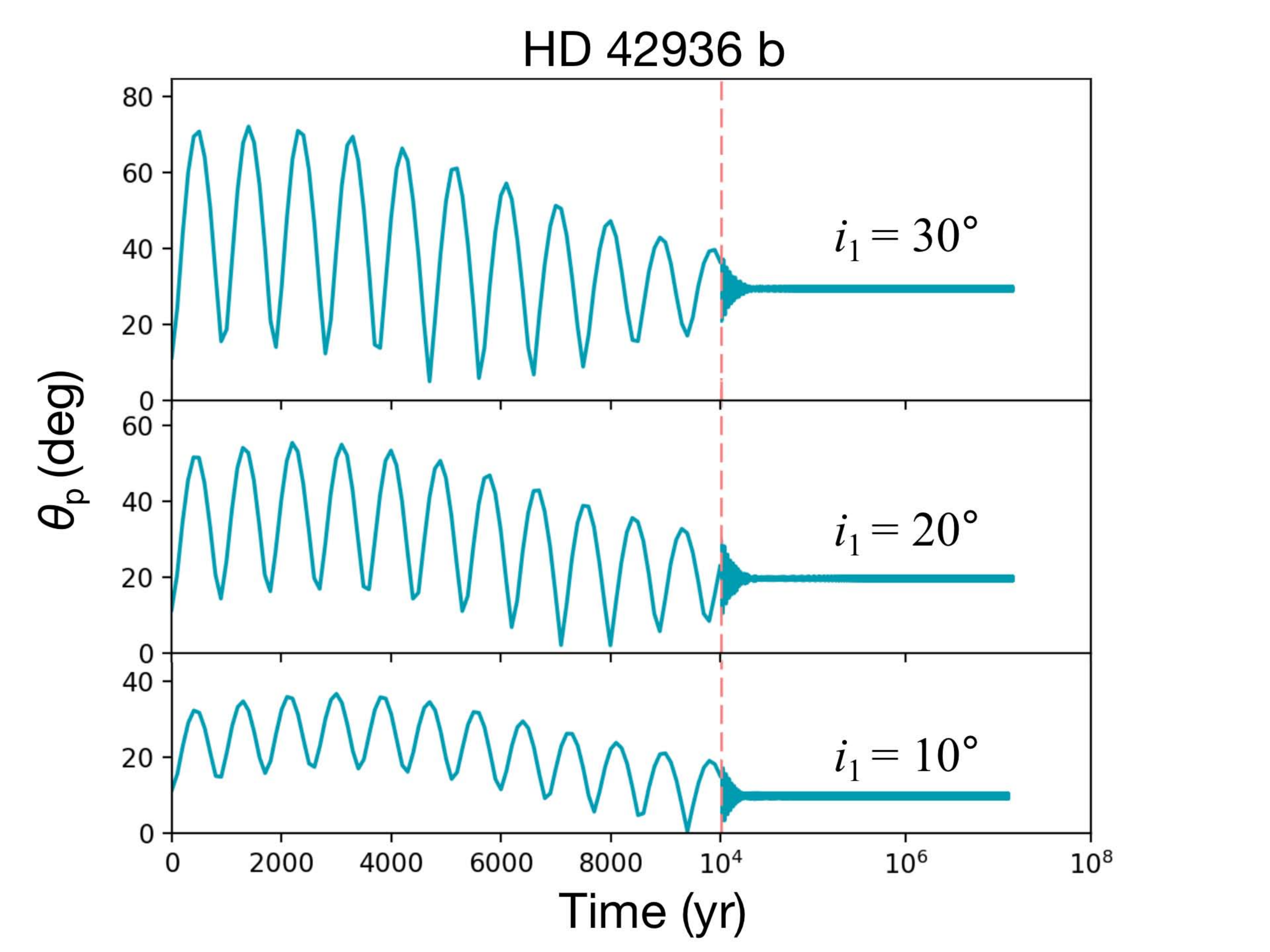}
\caption{The obliquity evolution for HD 42936 b of the orbital inclination $i_1$=$10^\circ$, $20^\circ$, $30^\circ$.}
\label{fig:HD42936b1}
\end{figure}

\begin{figure}
\label{fig:HD42936b2}
\includegraphics[width=\columnwidth,height=7cm]{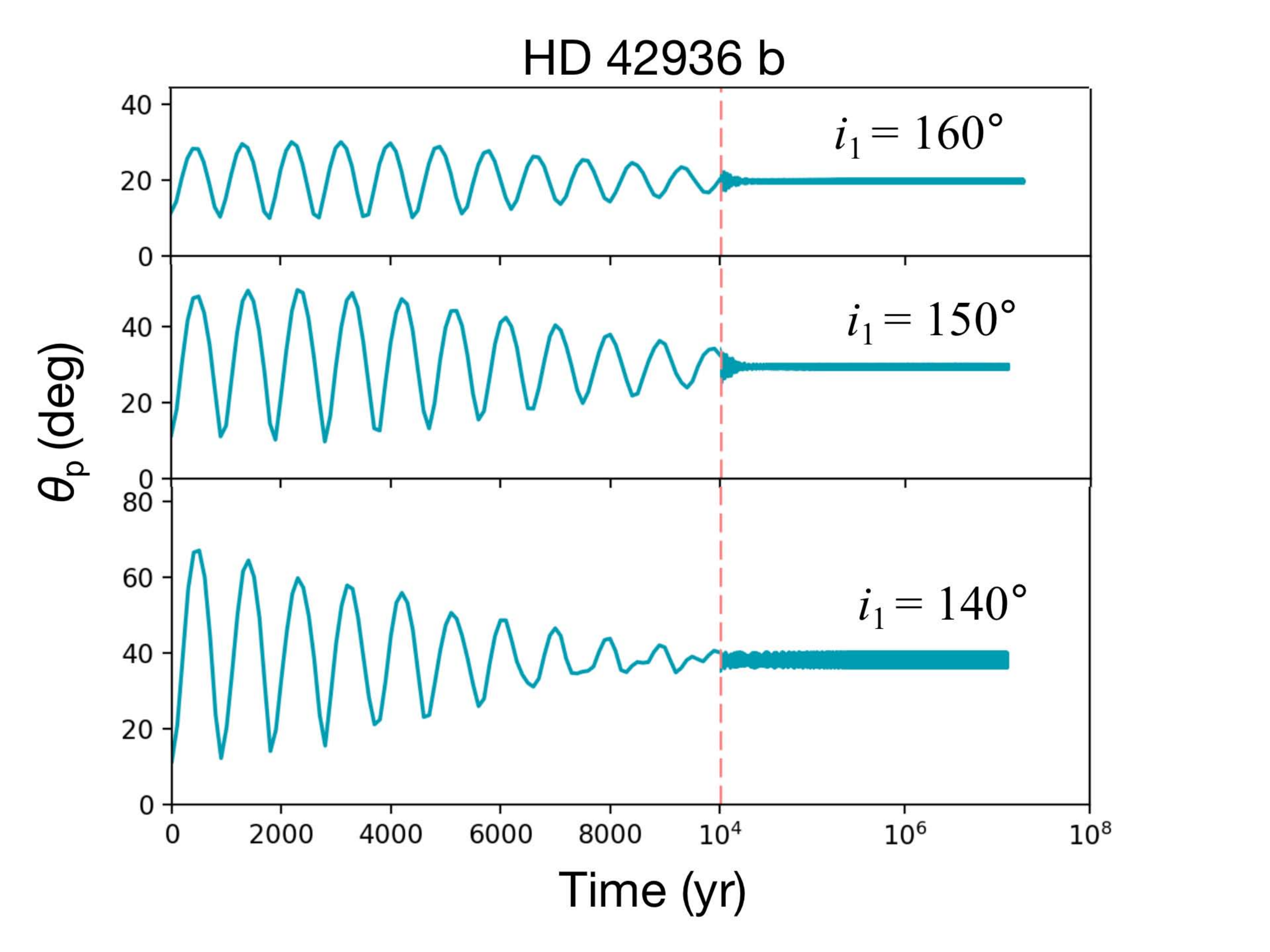}
\caption{The obliquity evolution for HD 42936 b of the orbital inclination $i_1$=$140^\circ$, $150^\circ$, $160^\circ$.}
\label{fig:HD42936b2}
\end{figure}

HD 42936 (DMPP-3) is the closest binary harboring a S-type super-Earth HD 42936 b \citep{Barnes2020}. The primary star HD 42936 is a solar-like star with mass of 0.87 $M_{\odot}$ at a distance of $48.9 \pm~0.6$ pc, while the semi-major axis of the secondary star is simply $1.22 \pm~0.02$ AU, and the eccentricity is relatively high with $e_2$ = 0.594. Such a compact and hierarchical system naturally brings secular perturbation scenario and tidal evolution into our mind. \citet{Barnes2020} concluded that the EKL mechanism, as one of the most likely formation models, can drive HD 42936 b move to current orbital configuration, in which the eccentricity of planet can be initially stirred up, and then shrink to a circular orbit when the tidal timescale is comparable to or shorter than $t_{\mathrm{kl}}$. Hence, HD 42936 is a good example to explore the spin evolution with our mixed mechanism and reveal the dependence of the spin-axis evolution tendency on the initial conditions.

HD 42936 b was discovered by radial velocity, probably moving at an inclined orbit. Thus in this work, the initial semi-major axis of this planet is assumed to be 0.06 -- 0.12 AU, and the initial inclination ranges from $10^\circ$ to $170^\circ$. When $40^\circ$ $< i_{\text{p}} <$ $140^\circ$, HD 42936 b will collide with the primary star or be scattered out of the system, before that the obliquity $\theta_p$ can be excited to $150^\circ$. For the survivals, $\theta_p$ can be triggered to the maximum value of $80^\circ$. Figure \ref{fig:HD42936b1} shows the obliquity evolution for HD 42936 b with the initial prograde orbits, while Figure \ref{fig:HD42936b2} stands for those results for the retrograde orbits at beginning.

\subsection{GJ 86 Ab}\label{subsubsec:4.2.1}
Aside from terrestrials in binary systems, we again investigate the spin evolution of hot-Jupiters with the mass ranging from 0.3 to 8 $M_{\text{J}}$. The simulation results show that the obliquity can be triggered to retrograde, and the maximum obliquity can arrive at $160^\circ$, being indicative of that the equilibrium obliquity may be sensitive to the planetary mass.

GJ 86 is a nearby S-type system at a distance of 10.9 pc from the solar system, and hosts a giant planet GJ 86 Ab \citep{Queloz2000} at the orbital distance of 0.11 AU, the outer companion is a white dwarf in the eccentric orbit with $a_2 = 21$ AU, $e_2 = 0.4$ \citep{Mugrauer2005,Lagrange2006}.  The orbital elements and masses of both companions are measured \citep{Zeng2022} with the observations from radial velocity and high angular resolution imaging along with the absolute astrometry data from \textit{Hipparcos} \citep{Perryman1997} and \textit{Gaia} \citep{Gaia2016,Brandt2018}. The current distance between the binary is 21 AU, whereas the mutual closest approach may reach 9 AU \citep{Zeng2022}, being indicative of that the gas-giant may origin from the disk truncation. The secular perturbation also participates in the obliquity evolution due to close separation of the binary. Our simulations suggest that GJ 86 Ab have a large possibility to follow obliquity evolution path \RNum{2}, and maintain the head-down rotation attitude.

GJ 86 is selected as one of the candidates for \textit{CHES} (Closeby Habitable Exoplanet Survey) mission \citep{Ji2022}, which will observe nearby solar-type stars to hunt for terrestrial planets in the habitable zones at an ultrahigh resolution via astrometry. The future observations will extensively provide the orbital elements and true mass of GJ 86 Ab, which helps place constraints on the dynamics of spin evolution.

\subsection{$\tau$ Boot Ab}\label{subsubsec:4.3}
\begin{figure}
\includegraphics[width=\columnwidth,height=7cm]{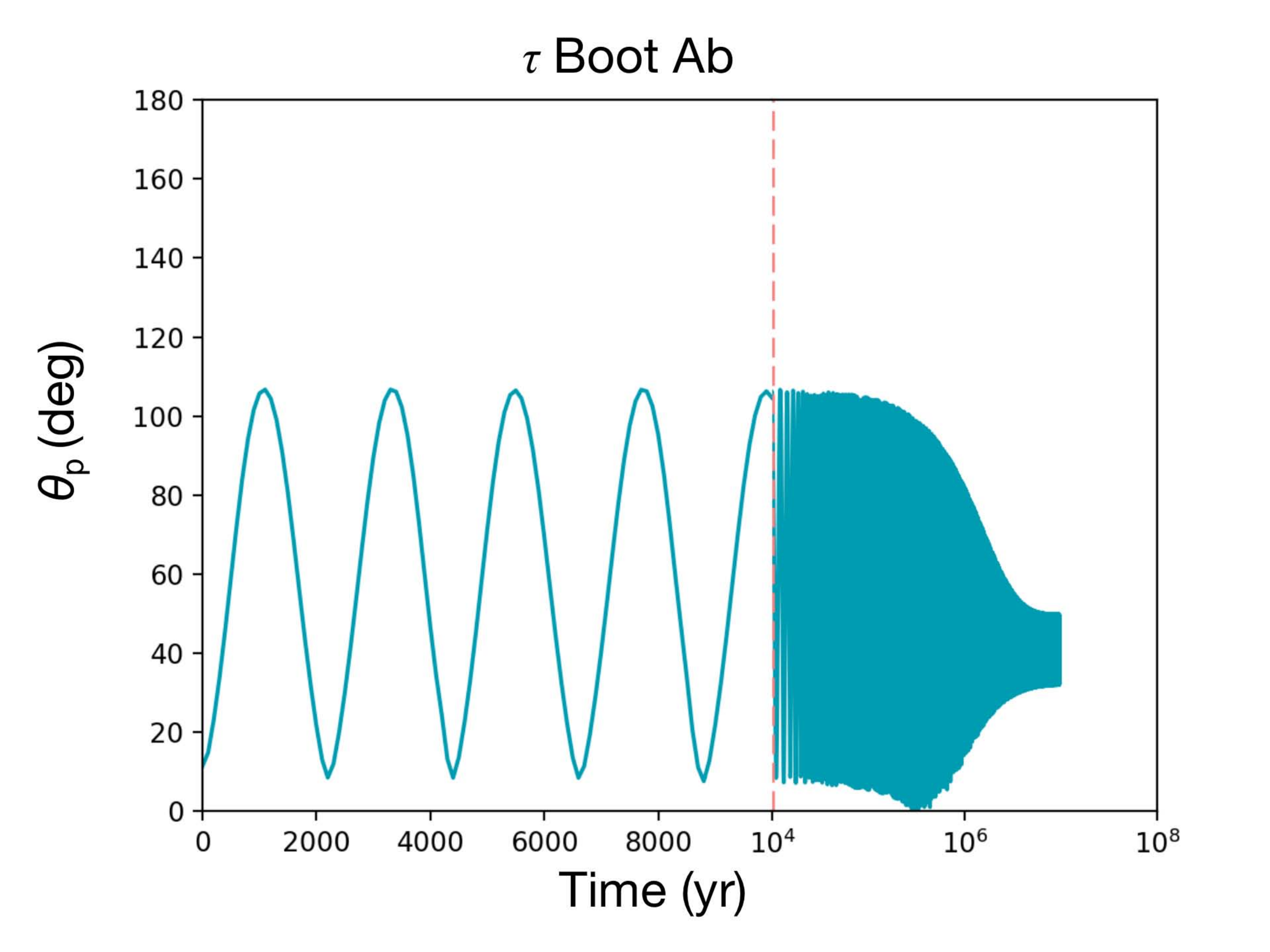}
\caption{The obliquity evolution for $\tau$ Boot Ab of the orbital inclination $i_1$ = $45^\circ$.}
\label{fig:tboot}
\end{figure}

The giant planet orbiting $\tau$ Boot is also one of the best-known exoplanets in the nearby stars from our solar system. The planet $\tau$ Boot Ab was first discovered in 1996 \citep{Butler1997}. \citet{Brogi2012} utilized radial velocity measurements to determine an orbital inclination of $i_p = 44.5\pm1.5$ $^\circ$ and a true planetary mass of $5.95\pm0.28 $ $M_{\mathrm{J}}$. They presented the atmospheric characterization of this non-transiting planet and showed the planetary temperature distribution decreases towards higher altitudes. Table \ref{tab:stars} and \ref{tab:planets} summarize the essential parameters for $\tau$ Boot. For $a_1 = 0.046$ AU, $a_{\mathrm{AB}} = 45$ AU, we still believe the perturbation from the outer companion can excite the orbital eccentricity and spin obliquity of $\tau$ Boot Ab, as the eccentricity $e_2$ of the secondary remains extremely high. Our numerical results indicate that $\theta_p$ follows the evolution path \RNum{2}, and the maximum value can reach above $100^\circ$, then falls to the equilibrium stage before 100 Myr (Figure \ref{fig:tboot}).

\section{Conclusions and Discussion}\label{sec:5}
In this work, we explore the spin evolution of terrestrial planets as the planetary obliquity plays a vital role in habitability. As the binary systems of stars that harbor planets are very common, thus secular perturbation from the secondary may greatly lead to a diverse planetary evolution.  {Here we adopt a comprehensive model composed of the octupole level secular perturbation and equilibrium tide to study dynamical evolution of S-type planets' obliquity,} which may provide essential clues for elucidating spin evolution for habitable planets.

We first calculate the comprehensive evolution timescale of $\theta_p$, which  relates very much to the Kozai-Lidov timescale $t_{\mathrm{kl}}$ and tidal timescale $t_{\mathrm{I}}$. We further plot the Hamiltonian level curves of the spin dynamics to theoretically derive the evolution paths of terrestrials' planetary obliquity, and also conduct numerical simulations to extensively explore the evolution of planetary obliquity induced by the coupling effect due to EKL resonance and the equilibrium tide, the planetary obliquity equilibrium state varies as a function of $r_{t}$. In addition, the maximum $\theta_\mathrm{p}$ during evolution can be distinguished by $r_{t}$. Cases in region \RNum{1} presents no extreme excitation of the obliquity, and $\theta_p$ will stably oscillate between $40^\circ$ and $60^\circ$ in region \RNum{2}, the planetary obliquity has excitation with flip. When $e_2$ ascends to 0.4, more obliquity excitation emerges, and $\theta_p$ with five sets of $a_2$ enter the flip region earlier. For $r_{t} > 10^4$, the maximum planetary obliquity evolves to nearly $130^\circ$.

According to the numerical results in the parameter space of semi-major axis $a_1$ and $a_2$, we present diagrams of the equilibrium timescale and maximum obliquities. The  equilibrium timescale $t_{\mathrm{eq}}$ evenly distributes in the dimension of $a_1$, and the timescale is always smaller than $10^5$ yr when $a_1 < 0.09$ AU. For $a_2 < 20$ AU, $t_{\mathrm{eq}}$ is approximately proportional to the initial orbit of the planet, which mainly determines the tidal timescale. Then we derive the linear relationship between $t_{\mathrm{eq}}$ and the timescale ratio $r_{\mathrm{t}}$ with the maximum likelihood method, $\mathrm{log} t_{\mathrm{eq}}$ is approximately proportional to $\mathrm{log} r_{t}$ and the slope of fitted lines rises with $m_1$.

Moreover, we estimate  {the ratio of obliquity flip} in the extensive parameter space $a_2 < 45 $ AU. We conclude that the reverse of spin axis appear to be common for S-type terrestrial planets accompanied with a distant massive companion. When the planetary mass ascends, the flip  {ratio} grows with the same initial semi-major axis, thus the flips become easier for a larger $m_1/m_2$. To employ our comprehensive model in the observed systems and predict the evolution of potential oblique planets, then we further numerically investigate three cases of HD 42936 b, GJ 86 Ab and $\tau$ Boot Ab, to understand their evolution paths of $\theta_p$.

{Aside from the planetary spin-orbit flip, the flip of the star-spin-orbit angle induced by the EKL mechanism were noted in the literature previously \citep{Naoz2011,Naoz2012,Petrovich2015bb,Lai2018,Stephan2018}, and the original high stellar spin-orbit obliquity will be damped for the cooler star \citep{Lai2012,Dawson2014}. As above-mentioned,  \citet{Storch2014b} intensively studied the spin-orbit coupling effect between the stellar spin and the planet orbit, which gives rise to the chaotic evolution of stellar spin axis during Kozai cycles.}

{Effects of tidal dissipation and stellar spin-down will also influence the final distribution of spin-orbit misalignment angles of hot-Jupiters. \citet{Storch2015} explored the origin of this chaotic stellar spin behaviour, they identified secular spin-orbit resonances and the resonance overlaps are responsible for the chaos. Key parameters including the adiabaticity parameter, the ratio of the Kozai-Lidov nodal precession rate and the stellar spin precession rate are proved to dominate the degree of chaos. In our research, similar chaotic behavior of the planetary obliquity also exist in several simulation cases. It was also demonstrated that before being destroyed by their stars either Roche-limit crossing or engulfment during stellar expansion, there could be diverse results on the final stellar-spin orbit angle \citep{Petrovich2015bb, Petrovich2015aa, Stephan2018, Angelo2022}.}

{It was recently highlighted that chaotic tide plays a key role in the evolution of planetary systems \citep{Wu2018,Vick2019}. In this case, in eccentric orbits, the planet's orbital energy is converted into internal fluid energy via tidal stretching and compression at periastron passage, causing the orbit to circularize over time. For Earth-like planets, a multilayered internal structure would increase the efficiency of tidal dissipation and affect the climate and habitability of the planet. For Jupiter-mass planets, increasing the rate of planetary orbital energy loss accelerates the tidal dissipation process and faster entry into the proposed equilibrium state.}

With the improvement of observational accuracy, the space-borne telescopes, such as \textit{JWST} (James Webb Space Telescope) \citep{Gardner2006}, \textit{TESS} (Transiting Exoplanet Survey Satellite) \citep{Ricker2015}, \textit{PLATO} (Planetary Transits and Oscillations of stars) \citep{Rauer2014}, \textit{ARIEL} (Atmospheric remote-sensing infrared exoplanet large survey) \citep{Tinetti2021} and \textit{CHES} (Closeby Habitable Exoplanet Survey) \citep{Ji2022}, will release a bulk of detailed information about the planetary three dimension orbital configuration, the planetary mass, the presence of atmosphere, and  spin orientation, thereby further identifying the stable oblique planet in habitable zone. There are prospects for constraining exoplanet obliquities in the coming years, such as using high-resolution spectroscopy to obtain the projected rotational radial-velocity of the planet \citep{Snellen2014,Bryan2018} based on the Rossiter-Mclaughlin effect during the second transit eclipsing. The planetary obliquity is also an important indicator for the seasonality of habitable terrestrials and the atmosphere loss rate of hot-Jupiters. This can be further examined by upcoming observations.  {In future work, we will further explore the origin of these chaotic behaviors, the stellar spin evolution, and the connection between planetary obliquity and stellar obliquity as well.}

We thank the referee for constructive comments and suggestions to improve the manuscript. This work is financially supported by the National Natural Science Foundation of China (Grant Nos. 12033010, 11773081), the B-type Strategic Priority Program of the Chinese Academy of Sciences (Grant No. XDB41000000), Foundation of Minor Planets of the Purple Mountain Observatory.


\bibliography{refs}{}

\begin{thebibliography}{}
\expandafter\ifx\csname natexlab\endcsname\relax\def\natexlab#1{#1}\fi
\providecommand{\url}[1]{\href{#1}{#1}}
\providecommand{\dodoi}[1]{doi:~\href{http://doi.org/#1}{\nolinkurl{#1}}}
\providecommand{\doeprint}[1]{\href{http://ascl.net/#1}{\nolinkurl{http://ascl.net/#1}}}
\providecommand{\doarXiv}[1]{\href{https://arxiv.org/abs/#1}{\nolinkurl{https://arxiv.org/abs/#1}}}

\bibitem[{{Alexander}(1973)}]{Alexander1973}
{Alexander}, M.~E. 1973, \apss, 23, 459, \dodoi{10.1007/BF00645172}

\bibitem[{{Angelo} {et~al.}(2022){Angelo}, {Naoz}, {Petigura}, {MacDougall},
  {Stephan}, {Isaacson}, \& {Howard}}]{Angelo2022}
{Angelo}, I., {Naoz}, S., {Petigura}, E., {et~al.} 2022, \aj, 163, 227,
  \dodoi{10.3847/1538-3881/ac6094}

\bibitem[{{Antognini}(2015)}]{Antognini2015}
{Antognini}, J.~M.~O. 2015, \mnras, 452, 3610, \dodoi{10.1093/mnras/stv1552}

\bibitem[{{Barnes} {et~al.}(2020){Barnes}, {Haswell}, {Staab},
  {Anglada-Escud{\'e}}, {Fossati}, {Doherty}, {Cooper}, {Jenkins}, {D{\'\i}az},
  {Soto}, \& {Pe{\~n}a Rojas}}]{Barnes2020}
{Barnes}, J.~R., {Haswell}, C.~A., {Staab}, D., {et~al.} 2020, Nature
  Astronomy, 4, 419, \dodoi{10.1038/s41550-019-0972-z}

\bibitem[{{Becker} {et~al.}(2020){Becker}, {Batygin}, {Fabrycky}, {Adams},
  {Li}, {Vanderburg}, \& {Rodriguez}}]{Becker2020}
{Becker}, J., {Batygin}, K., {Fabrycky}, D., {et~al.} 2020, \aj, 160, 254,
  \dodoi{10.3847/1538-3881/abbad3}

\bibitem[{{Bolmont} {et~al.}(2015){Bolmont}, {Raymond}, {Leconte}, {Hersant},
  \& {Correia}}]{Bolmont2015}
{Bolmont}, E., {Raymond}, S.~N., {Leconte}, J., {Hersant}, F., \& {Correia}, A.
  C.~M. 2015, \aap, 583, A116, \dodoi{10.1051/0004-6361/201525909}

\bibitem[{{Brandt}(2018)}]{Brandt2018}
{Brandt}, T.~D. 2018, \apjs, 239, 31, \dodoi{10.3847/1538-4365/aaec06}

\bibitem[{{Brogi} {et~al.}(2012){Brogi}, {Snellen}, {de Kok}, {Albrecht},
  {Birkby}, \& {de Mooij}}]{Brogi2012}
{Brogi}, M., {Snellen}, I. A.~G., {de Kok}, R.~J., {et~al.} 2012, \nat, 486,
  502, \dodoi{10.1038/nature11161}

\bibitem[{{Bryan} {et~al.}(2018){Bryan}, {Benneke}, {Knutson}, {Batygin}, \&
  {Bowler}}]{Bryan2018}
{Bryan}, M.~L., {Benneke}, B., {Knutson}, H.~A., {Batygin}, K., \& {Bowler},
  B.~P. 2018, Nature Astronomy, 2, 138, \dodoi{10.1038/s41550-017-0325-8}

\bibitem[{{Bryan} {et~al.}(2020){Bryan}, {Chiang}, {Bowler}, {Morley},
  {Millholland}, {Blunt}, {Ashok}, {Nielsen}, {Ngo}, {Mawet}, \&
  {Knutson}}]{Bryan2020}
{Bryan}, M.~L., {Chiang}, E., {Bowler}, B.~P., {et~al.} 2020, \aj, 159, 181,
  \dodoi{10.3847/1538-3881/ab76c6}

\bibitem[{{Butler} {et~al.}(1997){Butler}, {Marcy}, {Williams}, {Hauser}, \&
  {Shirts}}]{Butler1997}
{Butler}, R.~P., {Marcy}, G.~W., {Williams}, E., {Hauser}, H., \& {Shirts}, P.
  1997, \apjl, 474, L115, \dodoi{10.1086/310444}

\bibitem[{{Campante} {et~al.}(2016){Campante}, {Lund}, {Kuszlewicz}, {Davies},
  {Chaplin}, {Albrecht}, {Winn}, {Bedding}, {Benomar}, {Bossini}, {Handberg},
  {Santos}, {Van Eylen}, {Basu}, {Christensen-Dalsgaard}, {Elsworth}, {Hekker},
  {Hirano}, {Huber}, {Karoff}, {Kjeldsen}, {Lundkvist}, {North}, {Silva
  Aguirre}, {Stello}, \& {White}}]{Campante2016}
{Campante}, T.~L., {Lund}, M.~N., {Kuszlewicz}, J.~S., {et~al.} 2016, \apj,
  819, 85, \dodoi{10.3847/0004-637X/819/1/85}

\bibitem[{{Chen} {et~al.}(2022){Chen}, {Li}, \& {Petrovich}}]{Chen2022}
{Chen}, C., {Li}, G., \& {Petrovich}, C. 2022, \apj, 930, 58,
  \dodoi{10.3847/1538-4357/ac6024}

\bibitem[{{Colombo}(1966)}]{Colombo1966}
{Colombo}, G. 1966, \aj, 71, 891, \dodoi{10.1086/109983}

\bibitem[{{Dawson}(2014)}]{Dawson2014}
{Dawson}, R.~I. 2014, \apjl, 790, L31, \dodoi{10.1088/2041-8205/790/2/L31}

\bibitem[{{Dobrovolskis}(1980)}]{Dobrovolskis1980}
{Dobrovolskis}, A.~R. 1980, \icarus, 41, 18,
  \dodoi{10.1016/0019-1035(80)90157-8}

\bibitem[{{Dong} {et~al.}(2019){Dong}, {Huang}, \& {Lingam}}]{Dong2019}
{Dong}, C., {Huang}, Z., \& {Lingam}, M. 2019, \apjl, 882, L16,
  \dodoi{10.3847/2041-8213/ab372c}

\bibitem[{{Fabrycky} {et~al.}(2007){Fabrycky}, {Johnson}, \&
  {Goodman}}]{Fabrycky2007}
{Fabrycky}, D.~C., {Johnson}, E.~T., \& {Goodman}, J. 2007, \apj, 665, 754,
  \dodoi{10.1086/519075}

\bibitem[{{Faridani} {et~al.}(2023){Faridani}, {Naoz}, {Li}, \&
  {Inzunza}}]{Faridani2023}
{Faridani}, T.~H., {Naoz}, S., {Li}, G., \& {Inzunza}, N. 2023, arXiv e-prints,
  arXiv:2305.03104, \dodoi{10.48550/arXiv.2305.03104}

\bibitem[{{Gaia Collaboration} {et~al.}(2016){Gaia Collaboration}, {Prusti},
  {de Bruijne}, {Brown}, {Vallenari}, {Babusiaux}, {Bailer-Jones}, {Bastian},
  {Biermann}, {Evans}, {Eyer}, {Jansen}, {Jordi}, {Klioner}, {Lammers},
  {Lindegren}, {Luri}, {Mignard}, {Milligan}, {Panem}, {Poinsignon},
  {Pourbaix}, {Randich}, {Sarri}, {Sartoretti}, {Siddiqui}, {Soubiran},
  {Valette}, {van Leeuwen}, {Walton}, {Aerts}, {Arenou}, {Cropper}, {Drimmel},
  {H{\o}g}, {Katz}, {Lattanzi}, {O'Mullane}, {Grebel}, {Holland}, {Huc},
  {Passot}, {Bramante}, {Cacciari}, {Casta{\~n}eda}, {Chaoul}, {Cheek}, {De
  Angeli}, {Fabricius}, {Guerra}, {Hern{\'a}ndez}, {Jean-Antoine-Piccolo},
  {Masana}, {Messineo}, {Mowlavi}, {Nienartowicz}, {Ord{\'o}{\~n}ez-Blanco},
  {Panuzzo}, {Portell}, {Richards}, {Riello}, {Seabroke}, {Tanga},
  {Th{\'e}venin}, {Torra}, {Els}, {Gracia-Abril}, {Comoretto},
  {Garcia-Reinaldos}, {Lock}, {Mercier}, {Altmann}, {Andrae}, {Astraatmadja},
  {Bellas-Velidis}, {Benson}, {Berthier}, {Blomme}, {Busso}, {Carry},
  {Cellino}, {Clementini}, {Cowell}, {Creevey}, {Cuypers}, {Davidson}, {De
  Ridder}, {de Torres}, {Delchambre}, {Dell'Oro}, {Ducourant}, {Fr{\'e}mat},
  {Garc{\'\i}a-Torres}, {Gosset}, {Halbwachs}, {Hambly}, {Harrison}, {Hauser},
  {Hestroffer}, {Hodgkin}, {Huckle}, {Hutton}, {Jasniewicz}, {Jordan},
  {Kontizas}, {Korn}, {Lanzafame}, {Manteiga}, {Moitinho}, {Muinonen},
  {Osinde}, {Pancino}, {Pauwels}, {Petit}, {Recio-Blanco}, {Robin}, {Sarro},
  {Siopis}, {Smith}, {Smith}, {Sozzetti}, {Thuillot}, {van Reeven}, {Viala},
  {Abbas}, {Abreu Aramburu}, {Accart}, {Aguado}, {Allan}, {Allasia},
  {Altavilla}, {{\'A}lvarez}, {Alves}, {Anderson}, {Andrei}, {Anglada Varela},
  {Antiche}, {Antoja}, {Ant{\'o}n}, {Arcay}, {Atzei}, {Ayache}, {Bach},
  {Baker}, {Balaguer-N{\'u}{\~n}ez}, {Barache}, {Barata}, {Barbier}, {Barblan},
  {Baroni}, {Barrado y Navascu{\'e}s}, {Barros}, {Barstow}, {Becciani},
  {Bellazzini}, {Bellei}, {Bello Garc{\'\i}a}, {Belokurov}, {Bendjoya},
  {Berihuete}, {Bianchi}, {Bienaym{\'e}}, {Billebaud}, {Blagorodnova},
  {Blanco-Cuaresma}, {Boch}, {Bombrun}, {Borrachero}, {Bouquillon}, {Bourda},
  {Bouy}, {Bragaglia}, {Breddels}, {Brouillet}, {Br{\"u}semeister},
  {Bucciarelli}, {Budnik}, {Burgess}, {Burgon}, {Burlacu}, {Busonero}, {Buzzi},
  {Caffau}, {Cambras}, {Campbell}, {Cancelliere}, {Cantat-Gaudin}, {Carlucci},
  {Carrasco}, {Castellani}, {Charlot}, {Charnas}, {Charvet}, {Chassat},
  {Chiavassa}, {Clotet}, {Cocozza}, {Collins}, {Collins}, {Costigan}, {Crifo},
  {Cross}, {Crosta}, {Crowley}, {Dafonte}, {Damerdji}, {Dapergolas}, {David},
  {David}, {De Cat}, {de Felice}, {de Laverny}, {De Luise}, {De March}, {de
  Martino}, {de Souza}, {Debosscher}, {del Pozo}, {Delbo}, {Delgado},
  {Delgado}, {di Marco}, {Di Matteo}, {Diakite}, {Distefano}, {Dolding}, {Dos
  Anjos}, {Drazinos}, {Dur{\'a}n}, {Dzigan}, {Ecale}, {Edvardsson}, {Enke},
  {Erdmann}, {Escolar}, {Espina}, {Evans}, {Eynard Bontemps}, {Fabre},
  {Fabrizio}, {Faigler}, {Falc{\~a}o}, {Farr{\`a}s Casas}, {Faye}, {Federici},
  {Fedorets}, {Fern{\'a}ndez-Hern{\'a}ndez}, {Fernique}, {Fienga}, {Figueras},
  {Filippi}, {Findeisen}, {Fonti}, {Fouesneau}, {Fraile}, {Fraser}, {Fuchs},
  {Furnell}, {Gai}, {Galleti}, {Galluccio}, {Garabato}, {Garc{\'\i}a-Sedano},
  {Gar{\'e}}, {Garofalo}, {Garralda}, {Gavras}, {Gerssen}, {Geyer}, {Gilmore},
  {Girona}, {Giuffrida}, {Gomes}, {Gonz{\'a}lez-Marcos},
  {Gonz{\'a}lez-N{\'u}{\~n}ez}, {Gonz{\'a}lez-Vidal}, {Granvik}, {Guerrier},
  {Guillout}, {Guiraud}, {G{\'u}rpide}, {Guti{\'e}rrez-S{\'a}nchez}, {Guy},
  {Haigron}, {Hatzidimitriou}, {Haywood}, {Heiter}, {Helmi}, {Hobbs},
  {Hofmann}, {Holl}, {Holland}, {Hunt}, {Hypki}, {Icardi}, {Irwin}, {Jevardat
  de Fombelle}, {Jofr{\'e}}, {Jonker}, {Jorissen}, {Julbe}, {Karampelas},
  {Kochoska}, {Kohley}, {Kolenberg}, {Kontizas}, {Koposov}, {Kordopatis},
  {Koubsky}, {Kowalczyk}, {Krone-Martins}, {Kudryashova}, {Kull}, {Bachchan},
  {Lacoste-Seris}, {Lanza}, {Lavigne}, {Le Poncin-Lafitte}, {Lebreton},
  {Lebzelter}, {Leccia}, {Leclerc}, {Lecoeur-Taibi}, {Lemaitre}, {Lenhardt},
  {Leroux}, {Liao}, {Licata}, {Lindstr{\o}m}, {Lister}, {Livanou}, {Lobel},
  {L{\"o}ffler}, {L{\'o}pez}, {Lopez-Lozano}, {Lorenz}, {Loureiro},
  {MacDonald}, {Magalh{\~a}es Fernandes}, {Managau}, {Mann}, {Mantelet},
  {Marchal}, {Marchant}, {Marconi}, {Marie}, {Marinoni}, {Marrese},
  {Marschalk{\'o}}, {Marshall}, {Mart{\'\i}n-Fleitas}, {Martino}, {Mary},
  {Matijevi{\v{c}}}, {Mazeh}, {McMillan}, {Messina}, {Mestre}, {Michalik},
  {Millar}, {Miranda}, {Molina}, {Molinaro}, {Molinaro}, {Moln{\'a}r},
  {Moniez}, {Montegriffo}, {Monteiro}, {Mor}, {Mora}, {Morbidelli}, {Morel},
  {Morgenthaler}, {Morley}, {Morris}, {Mulone}, {Muraveva}, {Musella},
  {Narbonne}, {Nelemans}, {Nicastro}, {Noval}, {Ord{\'e}novic},
  {Ordieres-Mer{\'e}}, {Osborne}, {Pagani}, {Pagano}, {Pailler}, {Palacin},
  {Palaversa}, {Parsons}, {Paulsen}, {Pecoraro}, {Pedrosa}, {Pentik{\"a}inen},
  {Pereira}, {Pichon}, {Piersimoni}, {Pineau}, {Plachy}, {Plum}, {Poujoulet},
  {Pr{\v{s}}a}, {Pulone}, {Ragaini}, {Rago}, {Rambaux}, {Ramos-Lerate},
  {Ranalli}, {Rauw}, {Read}, {Regibo}, {Renk}, {Reyl{\'e}}, {Ribeiro},
  {Rimoldini}, {Ripepi}, {Riva}, {Rixon}, {Roelens}, {Romero-G{\'o}mez},
  {Rowell}, {Royer}, {Rudolph}, {Ruiz-Dern}, {Sadowski}, {Sagrist{\`a}
  Sell{\'e}s}, {Sahlmann}, {Salgado}, {Salguero}, {Sarasso}, {Savietto},
  {Schnorhk}, {Schultheis}, {Sciacca}, {Segol}, {Segovia}, {Segransan},
  {Serpell}, {Shih}, {Smareglia}, {Smart}, {Smith}, {Solano}, {Solitro},
  {Sordo}, {Soria Nieto}, {Souchay}, {Spagna}, {Spoto}, {Stampa}, {Steele},
  {Steidelm{\"u}ller}, {Stephenson}, {Stoev}, {Suess}, {S{\"u}veges}, {Surdej},
  {Szabados}, {Szegedi-Elek}, {Tapiador}, {Taris}, {Tauran}, {Taylor},
  {Teixeira}, {Terrett}, {Tingley}, {Trager}, {Turon}, {Ulla}, {Utrilla},
  {Valentini}, {van Elteren}, {Van Hemelryck}, {van Leeuwen}, {Varadi},
  {Vecchiato}, {Veljanoski}, {Via}, {Vicente}, {Vogt}, {Voss}, {Votruba},
  {Voutsinas}, {Walmsley}, {Weiler}, {Weingrill}, {Werner}, {Wevers},
  {Whitehead}, {Wyrzykowski}, {Yoldas}, {{\v{Z}}erjal}, {Zucker}, {Zurbach},
  {Zwitter}, {Alecu}, {Allen}, {Allende Prieto}, {Amorim},
  {Anglada-Escud{\'e}}, {Arsenijevic}, {Azaz}, {Balm}, {Beck}, {Bernstein},
  {Bigot}, {Bijaoui}, {Blasco}, {Bonfigli}, {Bono}, {Boudreault}, {Bressan},
  {Brown}, {Brunet}, {Bunclark}, {Buonanno}, {Butkevich}, {Carret}, {Carrion},
  {Chemin}, {Ch{\'e}reau}, {Corcione}, {Darmigny}, {de Boer}, {de Teodoro}, {de
  Zeeuw}, {Delle Luche}, {Domingues}, {Dubath}, {Fodor}, {Fr{\'e}zouls},
  {Fries}, {Fustes}, {Fyfe}, {Gallardo}, {Gallegos}, {Gardiol}, {Gebran},
  {Gomboc}, {G{\'o}mez}, {Grux}, {Gueguen}, {Heyrovsky}, {Hoar}, {Iannicola},
  {Isasi Parache}, {Janotto}, {Joliet}, {Jonckheere}, {Keil}, {Kim},
  {Klagyivik}, {Klar}, {Knude}, {Kochukhov}, {Kolka}, {Kos}, {Kutka}, {Lainey},
  {LeBouquin}, {Liu}, {Loreggia}, {Makarov}, {Marseille}, {Martayan},
  {Martinez-Rubi}, {Massart}, {Meynadier}, {Mignot}, {Munari}, {Nguyen},
  {Nordlander}, {Ocvirk}, {O'Flaherty}, {Olias Sanz}, {Ortiz}, {Osorio},
  {Oszkiewicz}, {Ouzounis}, {Palmer}, {Park}, {Pasquato}, {Peltzer}, {Peralta},
  {P{\'e}turaud}, {Pieniluoma}, {Pigozzi}, {Poels}, {Prat}, {Prod'homme},
  {Raison}, {Rebordao}, {Risquez}, {Rocca-Volmerange}, {Rosen}, {Ruiz-Fuertes},
  {Russo}, {Sembay}, {Serraller Vizcaino}, {Short}, {Siebert}, {Silva},
  {Sinachopoulos}, {Slezak}, {Soffel}, {Sosnowska}, {Strai{\v{z}}ys}, {ter
  Linden}, {Terrell}, {Theil}, {Tiede}, {Troisi}, {Tsalmantza}, {Tur},
  {Vaccari}, {Vachier}, {Valles}, {Van Hamme}, {Veltz}, {Virtanen}, {Wallut},
  {Wichmann}, {Wilkinson}, {Ziaeepour}, \& {Zschocke}}]{Gaia2016}
{Gaia Collaboration}, {Prusti}, T., {de Bruijne}, J.~H.~J., {et~al.} 2016,
  \aap, 595, A1, \dodoi{10.1051/0004-6361/201629272}

\bibitem[{{Gaidos} \& {Williams}(2004)}]{Gaidos2004}
{Gaidos}, E., \& {Williams}, D.~M. 2004, \na, 10, 67,
  \dodoi{10.1016/j.newast.2004.04.009}

\bibitem[{{Gardner} {et~al.}(2006){Gardner}, {Mather}, {Clampin}, {Doyon},
  {Greenhouse}, {Hammel}, {Hutchings}, {Jakobsen}, {Lilly}, {Long}, {Lunine},
  {McCaughrean}, {Mountain}, {Nella}, {Rieke}, {Rieke}, {Rix}, {Smith},
  {Sonneborn}, {Stiavelli}, {Stockman}, {Windhorst}, \& {Wright}}]{Gardner2006}
{Gardner}, J.~P., {Mather}, J.~C., {Clampin}, M., {et~al.} 2006, \ssr, 123,
  485, \dodoi{10.1007/s11214-006-8315-7}

\bibitem[{{Goldreich} \& {Peale}(1970)}]{Goldreich1970}
{Goldreich}, P., \& {Peale}, S.~J. 1970, \aj, 75, 273, \dodoi{10.1086/110975}

\bibitem[{{Hansen} \& {Naoz}(2020)}]{Hansen2020}
{Hansen}, B. M.~S., \& {Naoz}, S. 2020, \mnras, 499, 1682,
  \dodoi{10.1093/mnras/staa2602}

\bibitem[{{Harrington}(1968)}]{Harrington1968}
{Harrington}, R.~S. 1968, \aj, 73, 190, \dodoi{10.1086/110614}

\bibitem[{{Huang} \& {Ji}(2022)}]{Huang2022}
{Huang}, X., \& {Ji}, J. 2022, \aj, 164, 177, \dodoi{10.3847/1538-3881/ac8f4c}

\bibitem[{{Hut}(1981)}]{Hut1981}
{Hut}, P. 1981, \aap, 99, 126

\bibitem[{{Ji} {et~al.}(2022){Ji}, {Li}, {Zhang}, {Fang}, {Li}, {Wang}, {Cao},
  {Deng}, {Li}, {Xian}, {Gao}, {Zhang}, {Li}, {Liu}, {Qi}, {Jin}, {Liu},
  {Chen}, {Li}, {Dong}, {Zhu}, \& {CHES Consortium}}]{Ji2022}
{Ji}, J.-H., {Li}, H.-T., {Zhang}, J.-B., {et~al.} 2022, Research in Astronomy
  and Astrophysics, 22, 072003, \dodoi{10.1088/1674-4527/ac77e4}

\bibitem[{{Kawahara}(2016)}]{Kawahara2016}
{Kawahara}, H. 2016, \apj, 822, 112, \dodoi{10.3847/0004-637X/822/2/112}

\bibitem[{{Kozai}(1962)}]{Kozai1962}
{Kozai}, Y. 1962, \aj, 67, 591, \dodoi{10.1086/108790}

\bibitem[{{Kozai}(1965)}]{Kozai1965}
---. 1965, \pasj, 17, 395

\bibitem[{{Lagrange} {et~al.}(2006){Lagrange}, {Beust}, {Udry}, {Chauvin}, \&
  {Mayor}}]{Lagrange2006}
{Lagrange}, A.~M., {Beust}, H., {Udry}, S., {Chauvin}, G., \& {Mayor}, M. 2006,
  arXiv e-prints, astro, \dodoi{10.48550/arXiv.astro-ph/0606167}

\bibitem[{{Lai}(2012)}]{Lai2012}
{Lai}, D. 2012, \mnras, 423, 486, \dodoi{10.1111/j.1365-2966.2012.20893.x}

\bibitem[{{Lai} {et~al.}(2018){Lai}, {Anderson}, \& {Pu}}]{Lai2018}
{Lai}, D., {Anderson}, K.~R., \& {Pu}, B. 2018, \mnras, 475, 5231,
  \dodoi{10.1093/mnras/sty133}

\bibitem[{{Laskar} \& {Robutel}(1993)}]{Laskar1993}
{Laskar}, J., \& {Robutel}, P. 1993, \nat, 361, 608, \dodoi{10.1038/361608a0}

\bibitem[{{Leconte} {et~al.}(2011){Leconte}, {Chabrier}, \&
  {Baraffe}}]{Leconte2011}
{Leconte}, J., {Chabrier}, G., \& {Baraffe}, I. 2011, in The Astrophysics of
  Planetary Systems: Formation, Structure, and Dynamical Evolution, ed.
  A.~{Sozzetti}, M.~G. {Lattanzi}, \& A.~P. {Boss}, Vol. 276, 248--251,
  \dodoi{10.1017/S1743921311020266}

\bibitem[{{Lee} \& {Peale}(2003)}]{Lee2003}
{Lee}, M.~H., \& {Peale}, S.~J. 2003, \apj, 592, 1201, \dodoi{10.1086/375857}

\bibitem[{{Li} {et~al.}(2014){Li}, {Naoz}, {Holman}, \& {Loeb}}]{Li2014}
{Li}, G., {Naoz}, S., {Holman}, M., \& {Loeb}, A. 2014, \apj, 791, 86,
  \dodoi{10.1088/0004-637X/791/2/86}

\bibitem[{{Lidov}(1962)}]{Lidov1962}
{Lidov}, M.~L. 1962, \planss, 9, 719, \dodoi{10.1016/0032-0633(62)90129-0}

\bibitem[{{Liu} {et~al.}(2015){Liu}, {Lai}, \& {Yuan}}]{Liu2015}
{Liu}, B., {Lai}, D., \& {Yuan}, Y.-F. 2015, \prd, 92, 124048,
  \dodoi{10.1103/PhysRevD.92.124048}

\bibitem[{{Liu} {et~al.}(2013){Liu}, {Guillochon}, {Lin}, \&
  {Ramirez-Ruiz}}]{Liu2013}
{Liu}, S.-F., {Guillochon}, J., {Lin}, D. N.~C., \& {Ramirez-Ruiz}, E. 2013,
  \apj, 762, 37, \dodoi{10.1088/0004-637X/762/1/37}

\bibitem[{{Millholland} \& {Spalding}(2020)}]{Millholland2020}
{Millholland}, S.~C., \& {Spalding}, C. 2020, \apj, 905, 71,
  \dodoi{10.3847/1538-4357/abc4e5}

\bibitem[{{Mugrauer} \& {Neuh{\"a}user}(2005)}]{Mugrauer2005}
{Mugrauer}, M., \& {Neuh{\"a}user}, R. 2005, \mnras, 361, L15,
  \dodoi{10.1111/j.1745-3933.2005.00055.x}

\bibitem[{{Naoz}(2016)}]{Naoz2016}
{Naoz}, S. 2016, \araa, 54, 441, \dodoi{10.1146/annurev-astro-081915-023315}

\bibitem[{{Naoz} {et~al.}(2011){Naoz}, {Farr}, {Lithwick}, {Rasio}, \&
  {Teyssandier}}]{Naoz2011}
{Naoz}, S., {Farr}, W.~M., {Lithwick}, Y., {Rasio}, F.~A., \& {Teyssandier}, J.
  2011, \nat, 473, 187, \dodoi{10.1038/nature10076}

\bibitem[{{Naoz} {et~al.}(2013){Naoz}, {Farr}, {Lithwick}, {Rasio}, \&
  {Teyssandier}}]{Naoz2013}
---. 2013, \mnras, 431, 2155, \dodoi{10.1093/mnras/stt302}

\bibitem[{{Naoz} {et~al.}(2012){Naoz}, {Farr}, \& {Rasio}}]{Naoz2012}
{Naoz}, S., {Farr}, W.~M., \& {Rasio}, F.~A. 2012, \apjl, 754, L36,
  \dodoi{10.1088/2041-8205/754/2/L36}

\bibitem[{{Nikolov} \& {Sainsbury-Martinez}(2015)}]{Nikolov2015}
{Nikolov}, N., \& {Sainsbury-Martinez}, F. 2015, \apj, 808, 57,
  \dodoi{10.1088/0004-637X/808/1/57}

\bibitem[{{Peale}(1969)}]{Peale1969}
{Peale}, S.~J. 1969, \aj, 74, 483, \dodoi{10.1086/110825}

\bibitem[{{Perryman} {et~al.}(1997){Perryman}, {Lindegren}, {Kovalevsky},
  {Hoeg}, {Bastian}, {Bernacca}, {Cr{\'e}z{\'e}}, {Donati}, {Grenon},
  {Grewing}, {van Leeuwen}, {van der Marel}, {Mignard}, {Murray}, {Le Poole},
  {Schrijver}, {Turon}, {Arenou}, {Froeschl{\'e}}, \&
  {Petersen}}]{Perryman1997}
{Perryman}, M.~A.~C., {Lindegren}, L., {Kovalevsky}, J., {et~al.} 1997, \aap,
  323, L49

\bibitem[{{Petrovich}(2015{\natexlab{a}})}]{Petrovich2015bb}
{Petrovich}, C. 2015{\natexlab{a}}, \apj, 805, 75,
  \dodoi{10.1088/0004-637X/805/1/75}

\bibitem[{{Petrovich}(2015{\natexlab{b}})}]{Petrovich2015aa}
---. 2015{\natexlab{b}}, \apj, 799, 27, \dodoi{10.1088/0004-637X/799/1/27}

\bibitem[{{Queloz} {et~al.}(2000){Queloz}, {Mayor}, {Weber}, {Bl{\'e}cha},
  {Burnet}, {Confino}, {Naef}, {Pepe}, {Santos}, \& {Udry}}]{Queloz2000}
{Queloz}, D., {Mayor}, M., {Weber}, L., {et~al.} 2000, \aap, 354, 99

\bibitem[{{Rauer} {et~al.}(2014){Rauer}, {Catala}, {Aerts}, {Appourchaux},
  {Benz}, {Brandeker}, {Christensen-Dalsgaard}, {Deleuil}, {Gizon}, {Goupil},
  {G{\"u}del}, {Janot-Pacheco}, {Mas-Hesse}, {Pagano}, {Piotto}, {Pollacco},
  {Santos}, {Smith}, {Su{\'a}rez}, {Szab{\'o}}, {Udry}, {Adibekyan}, {Alibert},
  {Almenara}, {Amaro-Seoane}, {Eiff}, {Asplund}, {Antonello}, {Barnes},
  {Baudin}, {Belkacem}, {Bergemann}, {Bihain}, {Birch}, {Bonfils}, {Boisse},
  {Bonomo}, {Borsa}, {Brand{\~a}o}, {Brocato}, {Brun}, {Burleigh}, {Burston},
  {Cabrera}, {Cassisi}, {Chaplin}, {Charpinet}, {Chiappini}, {Church},
  {Csizmadia}, {Cunha}, {Damasso}, {Davies}, {Deeg}, {D{\'\i}az}, {Dreizler},
  {Dreyer}, {Eggenberger}, {Ehrenreich}, {Eigm{\"u}ller}, {Erikson}, {Farmer},
  {Feltzing}, {de Oliveira Fialho}, {Figueira}, {Forveille}, {Fridlund},
  {Garc{\'\i}a}, {Giommi}, {Giuffrida}, {Godolt}, {Gomes da Silva}, {Granzer},
  {Grenfell}, {Grotsch-Noels}, {G{\"u}nther}, {Haswell}, {Hatzes},
  {H{\'e}brard}, {Hekker}, {Helled}, {Heng}, {Jenkins}, {Johansen},
  {Khodachenko}, {Kislyakova}, {Kley}, {Kolb}, {Krivova}, {Kupka}, {Lammer},
  {Lanza}, {Lebreton}, {Magrin}, {Marcos-Arenal}, {Marrese}, {Marques},
  {Martins}, {Mathis}, {Mathur}, {Messina}, {Miglio}, {Montalban}, {Montalto},
  {Monteiro}, {Moradi}, {Moravveji}, {Mordasini}, {Morel}, {Mortier},
  {Nascimbeni}, {Nelson}, {Nielsen}, {Noack}, {Norton}, {Ofir}, {Oshagh},
  {Ouazzani}, {P{\'a}pics}, {Parro}, {Petit}, {Plez}, {Poretti}, {Quirrenbach},
  {Ragazzoni}, {Raimondo}, {Rainer}, {Reese}, {Redmer}, {Reffert},
  {Rojas-Ayala}, {Roxburgh}, {Salmon}, {Santerne}, {Schneider}, {Schou},
  {Schuh}, {Schunker}, {Silva-Valio}, {Silvotti}, {Skillen}, {Snellen}, {Sohl},
  {Sousa}, {Sozzetti}, {Stello}, {Strassmeier}, {{\v{S}}vanda}, {Szab{\'o}},
  {Tkachenko}, {Valencia}, {Van Grootel}, {Vauclair}, {Ventura}, {Wagner},
  {Walton}, {Weingrill}, {Werner}, {Wheatley}, \& {Zwintz}}]{Rauer2014}
{Rauer}, H., {Catala}, C., {Aerts}, C., {et~al.} 2014, Experimental Astronomy,
  38, 249, \dodoi{10.1007/s10686-014-9383-4}

\bibitem[{{Ricker} {et~al.}(2015){Ricker}, {Winn}, {Vanderspek}, {Latham},
  {Bakos}, {Bean}, {Berta-Thompson}, {Brown}, {Buchhave}, {Butler}, {Butler},
  {Chaplin}, {Charbonneau}, {Christensen-Dalsgaard}, {Clampin}, {Deming},
  {Doty}, {De Lee}, {Dressing}, {Dunham}, {Endl}, {Fressin}, {Ge}, {Henning},
  {Holman}, {Howard}, {Ida}, {Jenkins}, {Jernigan}, {Johnson}, {Kaltenegger},
  {Kawai}, {Kjeldsen}, {Laughlin}, {Levine}, {Lin}, {Lissauer}, {MacQueen},
  {Marcy}, {McCullough}, {Morton}, {Narita}, {Paegert}, {Palle}, {Pepe},
  {Pepper}, {Quirrenbach}, {Rinehart}, {Sasselov}, {Sato}, {Seager},
  {Sozzetti}, {Stassun}, {Sullivan}, {Szentgyorgyi}, {Torres}, {Udry}, \&
  {Villasenor}}]{Ricker2015}
{Ricker}, G.~R., {Winn}, J.~N., {Vanderspek}, R., {et~al.} 2015, Journal of
  Astronomical Telescopes, Instruments, and Systems, 1, 014003,
  \dodoi{10.1117/1.JATIS.1.1.014003}

\bibitem[{{Schwartz} {et~al.}(2016){Schwartz}, {Sekowski}, {Haggard},
  {Pall{\'e}}, \& {Cowan}}]{Schwartz2016}
{Schwartz}, J.~C., {Sekowski}, C., {Haggard}, H.~M., {Pall{\'e}}, E., \&
  {Cowan}, N.~B. 2016, \mnras, 457, 926, \dodoi{10.1093/mnras/stw068}

\bibitem[{{Snellen} {et~al.}(2014){Snellen}, {Brandl}, {de Kok}, {Brogi},
  {Birkby}, \& {Schwarz}}]{Snellen2014}
{Snellen}, I. A.~G., {Brandl}, B.~R., {de Kok}, R.~J., {et~al.} 2014, \nat,
  509, 63, \dodoi{10.1038/nature13253}

\bibitem[{{Stephan} {et~al.}(2018){Stephan}, {Naoz}, \& {Gaudi}}]{Stephan2018}
{Stephan}, A.~P., {Naoz}, S., \& {Gaudi}, B.~S. 2018, \aj, 156, 128,
  \dodoi{10.3847/1538-3881/aad6e5}

\bibitem[{{Storch} \& {Lai}(2014)}]{Storch2014b}
{Storch}, N.~I., \& {Lai}, D. 2014, \mnras, 438, 1526,
  \dodoi{10.1093/mnras/stt2292}

\bibitem[{{Storch} \& {Lai}(2015)}]{Storch2015}
---. 2015, \mnras, 448, 1821, \dodoi{10.1093/mnras/stv119}

\bibitem[{{Su} \& {Lai}(2020)}]{Su2020}
{Su}, Y., \& {Lai}, D. 2020, \apj, 903, 7, \dodoi{10.3847/1538-4357/abb6f3}

\bibitem[{{Su} \& {Lai}(2022{\natexlab{a}})}]{Su2022b}
---. 2022{\natexlab{a}}, \mnras, 513, 3302, \dodoi{10.1093/mnras/stac1096}

\bibitem[{{Su} \& {Lai}(2022{\natexlab{b}})}]{Su2022a}
---. 2022{\natexlab{b}}, \mnras, 509, 3301, \dodoi{10.1093/mnras/stab3172}

\bibitem[{{Tan} {et~al.}(2020){Tan}, {Hou}, {Liao}, {Wang}, \&
  {Tang}}]{Tan2020}
{Tan}, P., {Hou}, X., {Liao}, X., {Wang}, W., \& {Tang}, J. 2020, \aj, 160,
  139, \dodoi{10.3847/1538-3881/aba89c}

\bibitem[{{Teyssandier} {et~al.}(2013){Teyssandier}, {Naoz}, {Lizarraga}, \&
  {Rasio}}]{Teyssandier2013}
{Teyssandier}, J., {Naoz}, S., {Lizarraga}, I., \& {Rasio}, F.~A. 2013, \apj,
  779, 166, \dodoi{10.1088/0004-637X/779/2/166}

\bibitem[{{Tinetti} {et~al.}(2021){Tinetti}, {Eccleston}, {Haswell}, {Lagage},
  {Leconte}, {L{\"u}ftinger}, {Micela}, {Min}, {Pilbratt}, {Puig}, {Swain},
  {Testi}, {Turrini}, {Vandenbussche}, {Rosa Zapatero Osorio}, {Aret},
  {Beaulieu}, {Buchhave}, {Ferus}, {Griffin}, {Guedel}, {Hartogh}, {Machado},
  {Malaguti}, {Pall{\'e}}, {Rataj}, {Ray}, {Ribas}, {Szab{\'o}}, {Tan},
  {Werner}, {Ratti}, {Scharmberg}, {Salvignol}, {Boudin}, {Halain}, {Haag},
  {Crouzet}, {Kohley}, {Symonds}, {Renk}, {Caldwell}, {Abreu}, {Alonso},
  {Amiaux}, {Berth{\'e}}, {Bishop}, {Bowles}, {Carmona}, {Coffey},
  {Colom{\'e}}, {Crook}, {D{\'e}sjonqueres}, {D{\'\i}az}, {Drummond},
  {Focardi}, {G{\'o}mez}, {Holmes}, {Krijger}, {Kovacs}, {Hunt}, {Machado},
  {Morgante}, {Ollivier}, {Ottensamer}, {Pace}, {Pagano}, {Pascale}, {Pearson},
  {M{\o}ller Pedersen}, {Pniel}, {Roose}, {Savini}, {Stamper}, {Szirovicza},
  {Szoke}, {Tosh}, {Vilardell}, {Barstow}, {Borsato}, {Casewell}, {Changeat},
  {Charnay}, {Civi{\v{s}}}, {Coud{\'e} du Foresto}, {Coustenis}, {Cowan},
  {Danielski}, {Demangeon}, {Drossart}, {Edwards}, {Gilli}, {Encrenaz}, {Kiss},
  {Kokori}, {Ikoma}, {Morales}, {Mendon{\c{c}}a}, {Moneti}, {Mugnai},
  {Garc{\'\i}a Mu{\~n}oz}, {Helled}, {Kama}, {Miguel}, {Nikolaou}, {Pagano},
  {Panic}, {Rengel}, {Rickman}, {Rocchetto}, {Sarkar}, {Selsis}, {Tennyson},
  {Tsiaras}, {Venot}, {Vida}, {Waldmann}, {Yurchenko}, {Szab{\'o}}, {Zellem},
  {Al-Refaie}, {Perez Alvarez}, {Anisman}, {Arhancet}, {Ateca}, {Baeyens},
  {Barnes}, {Bell}, {Benatti}, {Biazzo}, {B{\l}{\k{e}}cka}, {Bonomo}, {Bosch},
  {Bossini}, {Bourgalais}, {Brienza}, {Brucalassi}, {Bruno}, {Caines},
  {Calcutt}, {Campante}, {Canestrari}, {Cann}, {Casali}, {Casas}, {Cassone},
  {Cara}, {Carmona}, {Carone}, {Carrasco}, {Changeat}, {Chioetto},
  {Cortecchia}, {Czupalla}, {Chubb}, {Ciaravella}, {Claret}, {Claudi},
  {Codella}, {Garcia Comas}, {Cracchiolo}, {Cubillos}, {Da Peppo}, {Decin},
  {Dejabrun}, {Delgado-Mena}, {Di Giorgio}, {Diolaiti}, {Dorn}, {Doublier},
  {Doumayrou}, {Dransfield}, {Dumaye}, {Dunford}, {Jimenez Escobar}, {Van
  Eylen}, {Farina}, {Fedele}, {Fern{\'a}ndez}, {Fleury}, {Fonte}, {Fontignie},
  {Fossati}, {Funke}, {Galy}, {Garai}, {Garc{\'\i}a}, {Garc{\'\i}a-Rigo},
  {Garufi}, {Germano Sacco}, {Giacobbe}, {G{\'o}mez}, {Gonzalez},
  {Gonzalez-Galindo}, {Grassi}, {Griffith}, {Guarcello}, {Goujon}, {Gressier},
  {Grzegorczyk}, {Guillot}, {Guilluy}, {Hargrave}, {Hellin}, {Herrero},
  {Hills}, {Horeau}, {Ito}, {Jessen}, {Kabath}, {K{\'a}lm{\'a}n}, {Kawashima},
  {Kimura}, {Kn{\'\i}{\v{z}}ek}, {Kreidberg}, {Kruid}, {Kruijssen},
  {Kubel{\'\i}k}, {Lara}, {Lebonnois}, {Lee}, {Lefevre}, {Lichtenberg},
  {Locci}, {Lombini}, {Sanchez Lopez}, {Lorenzani}, {MacDonald}, {Magrini},
  {Maldonado}, {Marcq}, {Migliorini}, {Modirrousta-Galian}, {Molaverdikhani},
  {Molinari}, {Molli{\`e}re}, {Moreau}, {Morello}, {Morinaud}, {Morvan},
  {Moses}, {Mouzali}, {Nakhjiri}, {Naponiello}, {Narita}, {Nascimbeni},
  {Nikolaou}, {Noce}, {Oliva}, {Palladino}, {Papageorgiou}, {Parmentier},
  {Peres}, {P{\'e}rez}, {Perez-Hoyos}, {Perger}, {Cecchi Pestellini},
  {Petralia}, {Philippon}, {Piccialli}, {Pignatari}, {Piotto}, {Podio},
  {Polenta}, {Preti}, {Pribulla}, {Lopez Puertas}, {Rainer}, {Reess}, {Rimmer},
  {Robert}, {Rosich}, {Rossi}, {Rust}, {Saleh}, {Sanna}, {Schisano},
  {Schreiber}, {Schwartz}, {Scippa}, {Seli}, {Shibata}, {Simpson}, {Shorttle},
  {Skaf}, {Skup}, {Sobiecki}, {Sousa}, {Sozzetti}, {{\v{S}}poner}, {Steiger},
  {Tanga}, {Tackley}, {Taylor}, {Tecza}, {Terenzi}, {Tremblin}, {Tozzi},
  {Triaud}, {Trompet}, {Tsai}, {Tsantaki}, {Valencia}, {Carine Vandaele}, {Van
  der Swaelmen}, {Adibekyan}, {Vasisht}, {Vazan}, {Del Vecchio}, {Waltham},
  {Wawer}, {Widemann}, {Wolkenberg}, {Hou Yip}, {Yung}, {Zilinskas},
  {Zingales}, \& {Zuppella}}]{Tinetti2021}
{Tinetti}, G., {Eccleston}, P., {Haswell}, C., {et~al.} 2021, arXiv e-prints,
  arXiv:2104.04824, \dodoi{10.48550/arXiv.2104.04824}

\bibitem[{{Valente} \& {Correia}(2022)}]{Valente2022}
{Valente}, E. F.~S., \& {Correia}, A. C.~M. 2022, \aap, 665, A130,
  \dodoi{10.1051/0004-6361/202244010}

\bibitem[{{Vervoort} {et~al.}(2022){Vervoort}, {Horner}, {Kane}, {Kirtland
  Turner}, \& {Gilmore}}]{Vervoort2022}
{Vervoort}, P., {Horner}, J., {Kane}, S.~R., {Kirtland Turner}, S., \&
  {Gilmore}, J.~B. 2022, \aj, 164, 130, \dodoi{10.3847/1538-3881/ac87fd}

\bibitem[{{Vick} {et~al.}(2019){Vick}, {Lai}, \& {Anderson}}]{Vick2019}
{Vick}, M., {Lai}, D., \& {Anderson}, K.~R. 2019, \mnras, 484, 5645,
  \dodoi{10.1093/mnras/stz354}

\bibitem[{{von Zeipel}(1910)}]{vonZeipel1910}
{von Zeipel}, H. 1910, Astronomische Nachrichten, 183, 345,
  \dodoi{10.1002/asna.19091832202}

\bibitem[{{Wang} {et~al.}(2019){Wang}, {Xu}, \& {Liao}}]{Wang2019}
{Wang}, W.-L., {Xu}, X.-Q., \& {Liao}, X.-H. 2019, Research in Astronomy and
  Astrophysics, 19, 130, \dodoi{10.1088/1674-4527/19/9/130}

\bibitem[{{Winn} \& {Holman}(2005)}]{Winn2005}
{Winn}, J.~N., \& {Holman}, M.~J. 2005, \apjl, 628, L159,
  \dodoi{10.1086/432834}

\bibitem[{{Wu}(2018)}]{Wu2018}
{Wu}, Y. 2018, \aj, 155, 118, \dodoi{10.3847/1538-3881/aaa970}

\bibitem[{{Zeng} {et~al.}(2022){Zeng}, {Brandt}, {Li}, {Dupuy}, {Li}, {Brandt},
  {Farihi}, {Horner}, {Wittenmyer}, {Butler}, {Tinney}, {Carter}, {Wright},
  {Jones}, \& {O'Toole}}]{Zeng2022}
{Zeng}, Y., {Brandt}, T.~D., {Li}, G., {et~al.} 2022, \aj, 164, 188,
  \dodoi{10.3847/1538-3881/ac8ff7}

\end{thebibliography}
\bibliographystyle{aasjournal}

\end{CJK*}
\end{document}